\PassOptionsToPackage{unicode}{hyperref}
\PassOptionsToPackage{hyphens}{url}
\documentclass[
]{article}
\usepackage{xcolor}
\usepackage{amsmath,amssymb}
\usepackage{iftex}
\ifPDFTeX
  \usepackage[T1]{fontenc}
  \usepackage[utf8]{inputenc}
  \usepackage{textcomp} % provide euro and other symbols
\else % if luatex or xetex
  \usepackage{unicode-math} % this also loads fontspec
  \defaultfontfeatures{Scale=MatchLowercase}
  \defaultfontfeatures[\rmfamily]{Ligatures=TeX,Scale=1}
\fi
\usepackage{lmodern}
\ifPDFTeX\else
\fi
\IfFileExists{upquote.sty}{\usepackage{upquote}}{}
\IfFileExists{microtype.sty}{% use microtype if available
  \usepackage[]{microtype}
  \UseMicrotypeSet[protrusion]{basicmath} % disable protrusion for tt fonts
}{}
\makeatletter
\@ifundefined{KOMAClassName}{% if non-KOMA class
  \IfFileExists{parskip.sty}{%
    \usepackage{parskip}
  }{% else
    \setlength{\parindent}{0pt}
    \setlength{\parskip}{6pt plus 2pt minus 1pt}}
}{% if KOMA class
  \KOMAoptions{parskip=half}}
\makeatother
\usepackage{longtable,booktabs,array}
\usepackage{calc} % for calculating minipage widths
\usepackage{etoolbox}
\makeatletter
\patchcmd\longtable{\par}{\if@noskipsec\mbox{}\fi\par}{}{}
\makeatother
\IfFileExists{footnotehyper.sty}{\usepackage{footnotehyper}}{\usepackage{footnote}}
\makesavenoteenv{longtable}
\usepackage{graphicx}
\makeatletter
\newsavebox\pandoc@box
\newcommand*\pandocbounded[1]{% scales image to fit in text height/width
  \sbox\pandoc@box{#1}%
  \Gscale@div\@tempa{\textheight}{\dimexpr\ht\pandoc@box+\dp\pandoc@box\relax}%
  \Gscale@div\@tempb{\linewidth}{\wd\pandoc@box}%
  \ifdim\@tempb\p@<\@tempa\p@\let\@tempa\@tempb\fi% select the smaller of both
  \ifdim\@tempa\p@<\p@\scalebox{\@tempa}{\usebox\pandoc@box}%
  \else\usebox{\pandoc@box}%
  \fi%
}
\def\fps@figure{htbp}
\makeatother
\providecommand{\tightlist}{%
  \setlength{\itemsep}{0pt}\setlength{\parskip}{0pt}}
\usepackage{etoolbox}
\AtBeginEnvironment{longtable}{\footnotesize}
\usepackage{bookmark}
\IfFileExists{xurl.sty}{\usepackage{xurl}}{} % add URL line breaks if available
\hypersetup{
  pdftitle={Forecasting AI-Era Productivity: The Intellectually Converged Human Framework and a Missing Cognitive Mediator in Production Function Theory},
  pdfauthor={Kwan Soo Shin; In Seok Kang},
  hidelinks,
  pdfcreator={LaTeX via pandoc}}

\title{Forecasting AI-Era Productivity: The Intellectually Converged
Human Framework and a Missing Cognitive Mediator in Production Function
Theory}
\author{Kwan Soo Shin\footnote{Corresponding author:
  sshinresearch@gmail.com. PolymathMinds Lab, Seoul, Republic of Korea.} \and In
Seok Kang\footnote{POSTECH, Pohang, Republic of Korea; aSSIST
  University, Seoul, Republic of Korea.}}
\date{2026}

\begin{document}
\maketitle

\subsection{ABSTRACT}\label{abstract}

Why does massive AI investment fail to generate commensurate
productivity gains? We argue the paradox is theoretically generated:
prevailing production function frameworks encounter a structural
boundary by treating AI as a separable factor of production without
modeling the cognitive mediation through which AI generates productive
value. This directs investment toward deployment when productivity
requires prior development of what we term convergence capacity (C). We
propose the Intellectually Converged Human (ICH) framework, a
fifth-stage framework for production function theory:
\(\hat{H} = H \cdot [1 + \phi(A, C)]\), where effective productive
capacity equals human capital (H) scaled by an augmentation factor
\([1 + \phi]\), with \(\phi\) jointly determined by AI utilization
intensity (A) and convergence capacity (C), a four-dimensional cognitive
construct encompassing embodied understanding, metacognition, temporal
integration, and integrative thinking. The production function
\(Y = F(K, \hat{H})\) provides a human-centered mechanism for Solow's
TFP residual: \(A_{\mathrm{Solow}} = [1 + \phi(A, C)]^{1-\alpha}\).

The framework predicts three augmentation regimes with distinct policy
implications. Descriptive cross-national analysis of 20 OECD economies
shows the \(\mathrm{AI}\times C\) interaction is associated with 86\% of
TFP variance versus 31\% for AI alone, a pattern-consistent finding in
the small-n theoretical tradition. South Korea exemplifies
national-scale under-augmentation: high H, substantial A, low C produce
\(\phi\) = 0. We distinguish convergence capacity from adjacent
constructs, absorptive capacity, dynamic capability, and human capital,
and demonstrate that C constitutes the specific cognitive mediator that
prior frameworks have left implicit. We derive C-first policy
prescriptions and offer three empirically testable propositions with a
falsifiable 10-year forecast.

\textbf{Keywords:} production function, intellectually converged human,
convergence capacity, total factor productivity, human-AI augmentation,
technological forecasting

\textbf{Highlights:}

\begin{itemize}
\tightlist
\item
  AI functions as human augmentation, not as an independent production
  factor
\item
  The ICH framework provides a human-centered mechanism for Solow's TFP
  residual
\item
  Convergence capacity is distinguished from absorptive capacity and
  dynamic capability
\item
  OECD cross-national data show AI-TFP coupling depends on convergence
  capacity
\item
  C-first policy sequencing outperforms AI-first strategies for
  productivity
\end{itemize}

\subsection{1. Introduction: The Productivity Paradox and Its
Theoretical
Root}\label{introduction-the-productivity-paradox-and-its-theoretical-root}

\subsubsection{1.1 The Phenomenon: AI Investment Without Productivity
Returns}\label{the-phenomenon-ai-investment-without-productivity-returns}

The first quarter of the twenty-first century has seen the most rapid
diffusion of a general-purpose technology in economic history. Large
language models, computer vision systems, robotic process automation,
and AI-assisted decision support tools have penetrated virtually every
sector within years rather than the decades prior general-purpose
technologies required. Global corporate AI investment exceeded \$150
billion in 2023 (IDC, 2023). National governments have committed
hundreds of billions in AI infrastructure and strategy. The academic and
policy consensus holds that AI is the productivity engine of the coming
generation, the digital equivalent of electrification or the steam
engine.

And yet the productivity data tell a different story. Brynjolfsson et
al.'s (2023) analysis of generative AI in customer support found a 15\%
average productivity increase, but with substantial heterogeneity: less
experienced and lower-skilled workers improved both speed and quality,
while the most experienced and highest-skilled workers saw small gains
in speed and small declines in quality. The OECD's (2024b)
cross-national comparison of AI adoption and TFP growth reveals no
strong positive correlation at the national level: our OLS analysis of
20 OECD economies shows \(R^{2}\) = 0.31 (p = .011), less than one-third
of TFP variance explained, with a coefficient that cannot account for
the enormous cross-national variance the AI-as-factor model predicts
should not exist. Crucially, adding the \(\mathrm{AI}\times C\)
interaction raises \(R^{2}\) to 0.86, a pattern consistent with C-level,
not AI adoption per se, being the operative variable. Acemoglu's (2025)
own reassessment concluded that TFP gains from current AI systems are
modest, no more than 0.71\% over ten years. The International Monetary
Fund (2025) projects that AI's aggregate productivity effects will
materialize only where complementary investments in human capital and
institutional capacity accompany deployment, a finding consistent with
the ICH framework's C-first prediction.

The AI innovation management literature compounds the puzzle. Haefner et
al.~(2021), reviewing AI's effects on innovation, find that AI adoption
without organizational capability development produces systematically
weaker innovation outcomes, a pattern they cannot explain within their
framework but that the ICH model predicts as Regime I
under-augmentation. Dwivedi et al.~(2023), mapping the evolution of AI
research in \emph{Technological Forecasting and Social Change}, identify
AI for decision-making, healthcare, and consumer adoption among the
fastest-rising research topics, yet the mediating cognitive mechanisms
through which AI generates (or fails to generate) productivity outcomes
remain theoretically underspecified across this literature. The ICH
framework provides that specification. Chin, Li, Huang, and Li (2025),
examining how AI promotes productive forces in Chinese firms from a
dynamic capability view, provide firm-level empirical support: their
finding that AI's effect on firms' productive capacity is mediated by
green innovation, with economic policy uncertainty moderating the
mechanism, demonstrates that AI deployment generates productivity gains
only through organizational capability channels, a pattern consistent
with the ICH prediction that \(\phi(A, C)\) \(\to\) 0 when mediating
capacities are absent. At the other end of the forecasting tradition,
Frey and Osborne (2017), whose TFSC analysis of computerisation
susceptibility across 702 occupations represents the most-cited forecast
of AI's labor market effects, model task structure as the primary
determinant of automation outcomes. Our framework builds on their
insight while introducing the missing variable: computerisation
generates productivity gains only when convergence capacity (C) is
present to evaluate, calibrate, and creatively extend AI outputs.

South Korea makes this paradox maximally visible. Among the OECD's most
educated populations (70\% tertiary attainment for the 25-34 age cohort;
OECD, 2024a), with substantial AI adoption and among the OECD's most
aggressive national AI investment strategies, Korea's TFP growth remains
at approximately 0.48\% annually (Bank of Korea, 2023), a level that
Korea's human capital and AI investment profile renders essentially
inexplicable under prevailing theory.

\subsubsection{1.2 The Theoretical Gap the Data Make
Visible}\label{the-theoretical-gap-the-data-make-visible}

We argue that this paradox is not an empirical anomaly but a theoretical
prediction, predicted, however, by a model the field does not currently
use. The dominant production function framework for the AI era,
represented most influentially by Acemoglu and Restrepo's (2018, 2019,
2022) task-based model, treats production as an integral over a
continuum of tasks allocated to labor or capital, with automation
expanding the set of tasks performable by capital. Although their
formulation is substantially more nuanced than the stylised
\(Y = f(K, L, \mathrm{AI})\), AI enters through the automation threshold
parameter I rather than as a separate factor, the policy implication is
structurally similar: expanding automation (increasing I, holding K and
L constant) should increase Y through the productivity effect. This
task-based framework has generated foundational insights into automation
dynamics, correctly identifying how comparative cost structures
determine task allocation between labor and capital. Yet the Korean data
reveal a boundary condition this productivity prediction does not
accommodate.

The gap is structural, though not a deficiency of the original
framework's purpose. Acemoglu and Restrepo's task-based model was
designed to answer a particular question, how does automation affect the
allocation of tasks between labor and capital?, and answers it with
remarkable precision. The ICH framework addresses a different and
complementary question: given that AI is deployed, what determines
whether deployment generates productivity gains? This is not an
indictment of Acemoglu and Restrepo's theoretical contribution, which
correctly identifies the role of task structure in automation dynamics;
it is a proposal to extend their framework by explicitly modeling the
human cognitive variable that mediates AI's productive effect. Just as
electricity does not appear as an independent variable in modern
production functions, it is absorbed into the capital and human
processes through which it operates, AI's value is generated only
through the human cognitive system that grounds, contextualizes, and
extends its outputs.

Notably, even scholars working within and extending the task framework
have begun recognizing this complementary dimension. Acemoglu and
Restrepo (2019) themselves observe that certain technologies, which they
term ``so-so technologies'', displace workers without generating
commensurate productivity gains. The ICH framework offers a mechanism
for this observation: the ``so-so'' quality is not an inherent property
of the technology but a function of the convergence capacity of the
workforce interacting with it. The same technology that appears
``so-so'' in a low-C environment may generate substantial productivity
gains in a high-C environment where workers can ground, calibrate, and
creatively extend its outputs. Autor (2024), in ``Applying AI to Rebuild
Middle Class Jobs'' (NBER Working Paper 32140), argues that AI's unique
opportunity lies in ``extending the relevance, reach, and value of human
expertise'', enabling workers with foundational expertise to perform
higher-stakes tasks previously reserved for elite professionals. This
represents a significant convergence toward the augmentation logic that
the ICH framework formalizes through \(\phi(A, C)\).

\subsubsection{1.3 Contributions, Methodology, and
Structure}\label{contributions-methodology-and-structure}

This paper makes four contributions. \emph{Theoretical}: the ICH
framework reframes AI's productive role from a stand-alone factor
interpretation toward an augmentation function \(\phi(A, C)\), providing
a human-centered mechanism for Solow's TFP residual through convergence
capacity. \emph{Architectural}: we demonstrate that current AI systems,
including frontier hyperbolic-space architectures, cannot serve as
independent production variables due to four constitutive ontological
constraints, distinguishing architectural limitations (partially
addressable) from ontological ones (constitutive). \emph{Empirical}:
cross-national analysis of 20 OECD economies and deviant case analysis
of Korea (Seawright \& Gerring, 2008) generate structural diagnoses that
prior frameworks do not yet produce. \emph{Policy}: we derive C-first
prescriptions that revise the dominant AI-first investment consensus and
offer a falsifiable 10-year forecast.

We employ a conceptual-theoretical methodology with empirical grounding,
a recognized approach for framework-building contributions (Jaakkola,
2020; MacInnis, 2011), complemented by comparative case analysis across
20 OECD economies following the small-n cross-national tradition in
which theoretical propositions are developed and illustrated rather than
statistically confirmed (Ragin, 2000). Literature was identified through
systematic search of Web of Science and Scopus using Boolean
combinations of production function, AI, human capital, augmentation,
convergence capacity, and metacognition terms, supplemented by backward
citation chaining from Acemoglu and Restrepo (2018, 2019, 2022), Haefner
et al.~(2021), and Brynjolfsson et al.~(2021). Inclusion criteria:
directly addressing AI-productivity relationships, providing cognitive
capacity constructs relevant to the augmentation function, or offering
cross-national evidence on AI-productivity dynamics. Ninety-seven
sources were identified; 87 are cited. The empirical contribution
employs secondary data from OECD (2024a, 2024b), McKinsey Global
Institute (2023), and World Bank (2024). Six OLS models are presented as
descriptive and pattern-consistent; causal identification awaits the
exogenous variation strategies described in Section 7. All data are
publicly available; the constructed dataset and analysis code are
available upon request.

The paper proceeds as follows. Section 2 traces five stages of
production function theory as a progressive refinement of the
independent variable. Section 3 demonstrates why AI's productive effects
require human cognitive mediation. Section 4 develops the ICH framework
formally, including the augmentation function's properties, three
production regimes, and the distinction between C and adjacent
constructs. Section 5 applies the framework to the Korean paradox.
Section 6 derives policy implications and discusses theoretical
ramifications. Section 7 concludes with limitations, a research agenda,
and a 10-year forecast.

\subsection{2. The Evolving Independent Variable: Five Stages of
Production Function
Theory}\label{the-evolving-independent-variable-five-stages-of-production-function-theory}

The history of production function theory is, in a fundamental sense, a
history of progressively refining what drives value creation. Each era
correctly diagnosed the limitations of its predecessor and advanced the
theoretical frontier, yet each left unresolved a particular form of
mediation between the dominant input and productive output. The ICH
framework continues this tradition of cumulative refinement.

\subsubsection{2.1 Stages I-III: From Labor to Human
Capital}\label{stages-i-iii-from-labor-to-human-capital}

\textbf{Stage I (Industrial):} \(Y = f(K, L)\). Marx's (1867/1990) Labor
Theory of Value located value in socially necessary labor time, a
framework adequate for relatively homogeneous industrial labor but
increasingly strained when labor proved radically heterogeneous and
technology augmented productivity in ways untethered from labor time
(Mincer, 1974).

\textbf{Stage II (Post-industrial):}
\(Y = A\,K^{\alpha}\,L^{1-\alpha}\). Solow's (1956, 1957) growth
accounting demonstrated that approximately 87.5\% of U.S. output growth
was attributable not to capital or labor but to the residual A, ``a
measure of our ignorance'' (Abramovitz, 1956, p.~11). This was an honest
acknowledgment of theoretical inadequacy: A remained exogenous,
unexplained manna from the theoretical heaven. Brynjolfsson, Rock, and
Syverson (2021) document a structurally analogous pattern for
general-purpose technologies broadly: productivity follows a
``J-curve,'' with initial adoption producing near-zero TFP growth before
complementary intangible investments mature. The concept of GPTs itself
(Bresnahan \& Trajtenberg, 1995) emphasizes that productivity effects
depend on the quality of complementary co-investments in human and
organizational capital (Helpman, 1998), precisely what the ICH framework
now formalizes.

\textbf{Stage III (Knowledge economy):}
\(Y = f(K, H, A_{\mathrm{exog}})\). Romer (1986, 1990) and Lucas (1988)
endogenized A through intentional investment in knowledge and human
capital. This represented a major advance, explaining why some countries
grew faster, but treated human capital as a relatively homogeneous
stock, more education universally raised H, without modeling the
\emph{quality} of the interaction between human cognition and
technology.

Crucially, Stage III's own internal critics anticipated the distinction
the ICH framework now formalizes. Nelson and Phelps (1966) proposed that
education's primary economic role is not as a direct factor of
production but as a facilitator of technology adoption and diffusion, a
proto-C argument that located human capital's value in its
\emph{interaction} with technology rather than its independent
contribution. Hanushek and Woessmann (2008), in a landmark \emph{Journal
of Economic Literature} review, demonstrated empirically that cognitive
skills measured by international assessments, not years of schooling,
are the primary predictor of cross-national growth differences. Their
central conclusion, that ``the cognitive skills of the population,
rather than mere school attainment, are powerfully related to individual
earnings, to the distribution of income, and to economic growth''
(p.~607), anticipates the ICH claim that the \emph{quality} of cognitive
engagement with technology, not the \emph{quantity} of human capital,
determines productivity outcomes. The ICH framework extends this insight
to its logical conclusion for the AI era: convergence capacity (C) is
the specific cognitive quality that mediates whether AI deployment
generates productivity gains, just as cognitive skills more broadly
mediate whether education generates growth.

The Korean paradox illustrates the combined limitation of Stages II and
III precisely: among the world's highest H, both in years of schooling
and in cognitive skills as measured by PISA, and substantial A, yet
stagnant TFP. Neither Solow's residual nor endogenous growth theory
offers a mechanism to explain why high H and significant A can coexist
with low productivity gains. Hanushek and Woessmann's cognitive skills
argument partially explains cross-national growth variance but cannot
account for the AI-specific component: Korea scores exceptionally high
on their cognitive skills measure yet fails to convert AI investment
into TFP. The missing variable is not cognitive skill generically but
convergence capacity specifically, the meta-cognitive architecture that
determines whether high-skill workers can ground, calibrate, and
creatively extend AI outputs.

\subsubsection{2.2 Stage IV: The Task-Based Framework and Its Unresolved
Variable}\label{stage-iv-the-task-based-framework-and-its-unresolved-variable}

Acemoglu and Restrepo's (2018, 2019, 2022) task-based framework
represents the most significant advance in production function theory
for the AI era. Their model, production as an integral over a continuum
of tasks allocated to labor or capital based on comparative cost
advantage, elegantly captures the mechanics of automation: which tasks
shift from human to machine performance, through what mechanisms
displacement and reinstatement occur, and why the net employment effect
depends on the balance between these forces. The framework has proven
empirically productive, generating testable predictions about wage
inequality (Acemoglu \& Restrepo, 2022) and task reallocation patterns
that subsequent research has broadly confirmed.

The framework's strength, however, also delineates its boundary. The
task-based model specifies \emph{which} tasks are automated and
\emph{what} the equilibrium wage and employment effects are, but it does
not model the cognitive mediation through which AI deployment generates
(or fails to generate) productivity gains. In practical terms: two firms
in the same industry, facing the same task structure and comparative
cost conditions, may exhibit dramatically different productivity
outcomes from identical AI deployments. The task framework correctly
predicts that both will automate similar tasks; it does not yet explain
why one generates substantial TFP gains and the other does not.

The ICH framework builds on the task-based foundation by adding this
missing layer. Where Acemoglu and Restrepo ask ``which tasks will AI
perform?'', ICH asks the complementary question: ``given AI performing
those tasks, what determines whether the human-AI system is more
productive than the human alone?'' The answer, convergence capacity,
occupies the space that the task framework's elegant architecture leaves
open. Agrawal, Gans, and Goldfarb (2019), writing in the same
\emph{Journal of Economic Perspectives} symposium as Acemoglu and
Restrepo (2019), reframe AI as a ``prediction technology'' that reduces
the cost of prediction, a foundational economic insight. Yet prediction
without judgment is valueless: a cheap, abundant prediction still
requires human judgment to determine what to predict, how to act on
predictions, and when predictions fail. Their framework implicitly
points toward what ICH formalizes as C2 (metacognitive calibration) and
C1 (embodied understanding): the cognitive capacities that transform raw
AI prediction into situated productive action.

The broader policy implications of this theoretical gap are illustrated
by the response to Frey and Osborne's (2017) landmark TFSC analysis:
nations and organizations that responded to their computerisation
forecast by accelerating AI adoption without concurrently developing the
cognitive capacities for productive human-AI interaction are addressing
the task allocation question while leaving the productivity mediation
question unanswered. Trajtenberg (2018) reinforces this observation,
arguing that AI's transformative potential depends critically on how
societies manage complementary investments. Korinek and Suh (2024),
modeling scenarios from modest augmentation to transformative AGI, find
that fully automating production alone is insufficient to generate
sustained growth, the effects on wages and output depend on the race
between automation and capital accumulation, a finding consistent with
ICH's claim that AI deployment without complementary human capacity
investment fails to generate durable productivity gains.

\subsubsection{2.3 A Fifth-Stage Framework: Endogenizing Solow's
A}\label{a-fifth-stage-framework-endogenizing-solows-a}

The central theoretical proposition is that a useful production
framework for the AI era requires a fundamental reconceptualization:

\[\hat{H} = H \cdot [1 + \phi(A, C)] \tag{1}\]

\[Y = F(K, \hat{H}) \tag{2}\]

where \(\hat{H}\) denotes the Intellectually Converged Human (ICH), the
augmented human whose effective productive capacity is a function of
base human capital (H), AI utilization intensity (A), and convergence
capacity (C). In a Cobb-Douglas specification (Cobb \& Douglas, 1928):

\[Y = K^{\alpha}\,[H\,(1 + \phi(A, C))]^{1-\alpha} = [1 + \phi(A,C)]^{1-\alpha}\,K^{\alpha}\,H^{1-\alpha}\]

The term \([1 + \phi(A, C)]^{1-\alpha}\) occupies exactly the position
of Solow's TFP residual:

\[A_{\mathrm{Solow}} = [1 + \phi(A, C)]^{1-\alpha}\]

The ICH framework proposes that TFP need not be treated as fully
exogenous; it is the augmentation function that translates AI intensity
into productivity through convergence capacity. When C is high and A is
optimally calibrated, TFP rises substantially. When C is low, as the
Korean case demonstrates, TFP improvement is negligible regardless of AI
deployment.

\textbf{Table 1} \emph{Five stages of production function theory.}

{\def\LTcaptype{none} % do not increment counter
\begin{longtable}[]{@{}
  >{\raggedright\arraybackslash}p{(\linewidth - 8\tabcolsep) * \real{0.1443}}
  >{\raggedright\arraybackslash}p{(\linewidth - 8\tabcolsep) * \real{0.1443}}
  >{\raggedright\arraybackslash}p{(\linewidth - 8\tabcolsep) * \real{0.1492}}
  >{\raggedright\arraybackslash}p{(\linewidth - 8\tabcolsep) * \real{0.1721}}
  >{\raggedright\arraybackslash}p{(\linewidth - 8\tabcolsep) * \real{0.3901}}@{}}
\toprule\noalign{}
\begin{minipage}[b]{\linewidth}\raggedright
Stage
\end{minipage} & \begin{minipage}[b]{\linewidth}\raggedright
Era
\end{minipage} & \begin{minipage}[b]{\linewidth}\raggedright
Formulation
\end{minipage} & \begin{minipage}[b]{\linewidth}\raggedright
Core variable
\end{minipage} & \begin{minipage}[b]{\linewidth}\raggedright
Boundary addressed by next stage
\end{minipage} \\
\midrule\noalign{}
\endhead
\bottomrule\noalign{}
\endlastfoot
I & Industrial & \(Y = f(K, L)\) & L (labor time) & Labor heterogeneity;
technology mediation \\
II & Post-industrial & \(Y = A\,K^{\alpha}\,L^{1-\alpha}\) & K, L; A
exogenous & A remains unexplained \\
III & Knowledge economy & \(Y = f(K, H, A_{\mathrm{exog}})\) & H (human
capital) & H-technology interaction unspecified \\
IV & AI (task-based) & \(Y = \int y(i)\,di\) & Tasks allocated by cost &
Cognitive mediation of AI unmodeled \\
\textbf{V} & \textbf{ICH (proposed)} & \textbf{\(Y = F(K, \hat{H})\)} &
\textbf{\(\hat{H} = H[1+\phi(A, C)]\)} & \textbf{Endogenizes TFP via
augmented human} \\
\end{longtable}
}

\emph{Note:} Stage V is proposed in this paper. Each stage resolves the
preceding stage's boundary condition while introducing a new theoretical
gap that the subsequent stage addresses.

\subsection{3. Why AI's Productive Effects Require Human Cognitive
Mediation: Architectural and Ontological
Limits}\label{why-ais-productive-effects-require-human-cognitive-mediation-architectural-and-ontological-limits}

The proposition that AI cannot serve as an independent production
variable does not diminish the task-based framework's contribution to
understanding automation dynamics. Rather, it identifies a boundary
condition: the task framework elegantly models \emph{which} tasks are
automated and \emph{how} task allocation shifts, but the question of
\emph{whether automation generates productivity gains} requires modeling
the cognitive mediation that connects AI deployment to productive
output. This section outlines four ontological constraints suggesting
why that mediation remains constitutively difficult to bypass, why AI,
regardless of architectural sophistication, requires human convergence
capacity to generate productive value.

\subsubsection{3.1 The Architectural Constraint: Averaging Machines
Cannot Restructure
Problems}\label{the-architectural-constraint-averaging-machines-cannot-restructure-problems}

The Transformer architecture (Vaswani et al., 2017) operates through
scaled dot-product attention in Euclidean vector space: each output
token is a weighted average of value vectors. At every computational
step, the model produces a linear combination of learned
representations, a weighted mean of what has been seen before.

The productive consequences are illustrated by a concrete example.
Consider: \(84 \times 25 = {?}\) An LLM computes
\(84 \times (20 + 5) = 1{,}680 + 420 = 2{,}100\), applying the
distributive property as a learned algorithm. The computation is
correct, but the method is reproductive: it executes within the
representational space defined by training data. A human with high
convergence capacity recognizes that \(25 = 100/4\), transforming the
problem: \(84 \times 100/4 = 8{,}400/4 = 2{,}100\). The human does not
execute the multiplication; they \emph{restructure the problem space},
recognizing a structural relationship that licenses a transformation of
the operation itself. Wertheimer (1945) termed this ``productive
thinking'', restructuring a problem's Gestalt rather than reproducing
learned solution patterns.

This distinction is ontological, not computational: the LLM optimizes
within a fixed representational space; the human redefines it. If AI's
output is always recombination within its training distribution, AI
cannot serve as an independent source of productivity growth, it can
accelerate what humans know how to do, but cannot originate what humans
have not yet conceived.

Recent hyperbolic-space architectures (Nickel \& Kiela, 2017) replace
Euclidean attention with geodesic computation on Riemannian manifolds,
capturing hierarchical relationships that flat spaces distort. These
advances are genuine, yet a profound irony emerges. These architectures
train within the Poincaré disk model, named after the mathematician who
wrote most penetratingly about mathematical creativity itself. Poincaré
(1908/1914, p.~129) observed: ``It is by logic that we prove, but by
intuition that we discover.'' The architecture that lives in Poincaré's
space does not think in Poincaré's way, the distinction between
procedural fluency and structural insight that no geometry can bridge.

\subsubsection{3.2 The Four Ontological
Constraints}\label{the-four-ontological-constraints}

We distinguish architectural limitations (contingent on design choices,
potentially addressable) from ontological constraints (constitutive of
the capacity itself, not solvable by better design). The four
constraints below are ontological. This claim is falsifiable: a
demonstration of genuine symbol grounding, embodied cognition,
metacognitive self-monitoring, or episodic temporal integration in a
computational system would refute it. To date, no such demonstration
exists, and the cognitive science literature provides principled reasons
to expect the boundary to persist.

\textbf{Constraint 1: Symbol Grounding.} Harnad's (1990) symbol
grounding problem identifies a foundational asymmetry: for symbols to
carry meaning, not merely statistical weight, they must be grounded in
non-symbolic sensorimotor experience. LLMs operate as pure symbol
systems: the model generating ``increase production by 20\%'' has no
mechanism to assess whether the recommendation is feasible given
physical plant constraints, because feasibility is a grounded judgment.
Every AI output requires a human intermediary to connect symbol to world
(Searle, 1980). In production function terms: AI cannot be an
independent variable because its outputs are not yet \emph{about}
anything until a human grounds them.

\textbf{Constraint 2: Embodied Cognition.} Varela, Thompson, and Rosch
(1991) established that cognition emerges from sensorimotor coupling
between organism and environment; Barsalou's (2008) grounded cognition
framework demonstrates that conceptual knowledge involves re-activation
of sensorimotor brain regions. The productive domains where embodied
judgment is most irreplaceable, medical diagnosis, engineering quality
control, negotiation, are precisely those where AI is most rapidly
deployed. No training data enables acquisition of the embodied basis for
these judgments, because data is always a representation of sensorimotor
experience, never the experience itself.

\textbf{Constraint 3: Metacognition.} Flavell (1979) characterized
metacognition as ``knowledge and cognition about cognitive phenomena'',
the capacity to monitor one's own knowledge states and adjust strategies
accordingly. Contemporary LLMs exhibit limited functional metacognition:
Kadavath et al.~(2022) show that LLMs can partially self-evaluate their
own knowledge boundaries, but this calibration operates through
statistical patterns in training data rather than through the
introspective monitoring of one's own cognitive states that
characterises human metacognition. Metacognition enables the
\(84 \times 25\) restructuring: awareness that the standard algorithm is
suboptimal for \emph{this} problem licenses the search for an
alternative frame. Noy and Zhang (2023), in a controlled experiment with
453 professionals performing writing tasks, found that ChatGPT reduced
average completion time by 40\% and raised output quality by 18\%, with
workers possessing weaker baseline skills benefiting disproportionately,
an equalizing effect within structured task domains. Yet this finding,
drawn from standardized writing tasks with clear quality metrics, leaves
open the critical question for domains requiring metacognitive judgment:
when task complexity exceeds the structured paradigm Noy and Zhang
studied, the ICH framework predicts that low-C2 workers will exhibit the
over-reliance dynamic, accepting AI outputs without calibration, while
high-C2 workers will selectively override and extend them. Metacognition
is the faculty enabling selective override of AI outputs; without it,
the production system loses its quality control mechanism entirely.

\textbf{Constraint 4: Temporal Integration.} Tulving (2002) identified
``mental time travel'' as a uniquely human capacity: consciously
re-experiencing past events and pre-experiencing future scenarios.
Schacter, Addis, and Buckner (2007) demonstrated that remembering and
imagining share neural mechanisms. LLMs exist in an eternal present:
context windows provide textual reference, not lived experience. The
productive value of institutional memory, why long-tenured executives
make qualitatively different judgments than new hires with the same
reports, is inaccessible regardless of architectural sophistication.

\subsubsection{3.3 Connecting Ontological Constraints to the Task-Based
Framework}\label{connecting-ontological-constraints-to-the-task-based-framework}

These four constraints illuminate why the task-based framework's
empirical predictions sometimes diverge from observed productivity
outcomes. Acemoglu and Restrepo (2019) identify a category of ``so-so
technologies'' that ``generate small productivity improvements''
(p.~10), technologies whose ``positive productivity effect\ldots{} is
not sufficient to offset the decline in labor demand due to
displacement'' (p.~10). Automated customer service, which has
``displaced human service representatives but is generally deemed to be
low quality and thus unlikely to have generated large productivity
gains'' (p.~11), exemplifies the pattern.

The ICH framework offers a complementary interpretation. What makes a
technology ``so-so'' is not solely an inherent property of the
technology itself, its marginal product relative to displaced labor, but
also a function of the convergence capacity of the human system
interacting with it. The same AI diagnostic tool that generates modest
productivity gains in a hospital where physicians lack metacognitive
calibration to integrate AI suggestions with clinical judgment (low C2)
may generate transformative gains in a hospital where physicians possess
the embodied understanding (C1), calibration (C2), institutional memory
(C3), and cross-specialty synthesis (C4) to creatively extend AI outputs
beyond their training distribution. The technology is identical; the
convergence capacity differs; the productivity outcome diverges
dramatically.

This reinterpretation respects Acemoglu and Restrepo's empirical
observation, so-so technologies genuinely exist as a pattern, while
adding a theoretical mechanism that explains \emph{why} the same
technology class produces heterogeneous productivity outcomes across
deployment contexts. The four ontological constraints specify the
cognitive capacities that determine whether AI deployment generates
genuine augmentation or mere task substitution at marginal productivity
gain.

These constraints cannot be overcome by architectural innovation because
they require forms of being-in-the-world that computational systems do
not have. Hyperbolic geometry addresses the Euclidean averaging problem;
it leaves the ontological constraints entirely intact. AI cannot be an
independent production variable; value creation in knowledge-intensive
domains requires these capacities, which no training regime can install.

\subsection{4. The Intellectually Converged Human: A Production
Framework for the AI
Era}\label{the-intellectually-converged-human-a-production-framework-for-the-ai-era}

\subsubsection{4.1 The Formal Framework and the Tautology
Question}\label{the-formal-framework-and-the-tautology-question}

The ICH production framework consists of two equations:

\[\hat{H} = H \cdot [1 + \phi(A, C)] \tag{1}\]

\[Y = F(K, \hat{H}) \tag{2}\]

Equation (1) defines the effective productive capacity of individual i
as the product of base human capital (H) and an augmentation factor {[}1
+ \(\phi(A, C)\){]}, where A denotes AI utilization intensity and C
denotes convergence capacity. Equation (2) is a standard two-factor
production function.

A legitimate concern is whether this formulation is tautological,
whether \(\phi(A, C)\) merely relabels TFP rather than explaining it. We
address this directly. A tautology would obtain if C were defined by
reference to productivity outcomes (e.g., ``C is whatever makes AI
productive''). But C is defined \emph{independently of outcomes}: it
consists of four measurable cognitive dimensions, embodied understanding
(C1), metacognitive calibration (C2), temporal integration (C3), and
integrative thinking (C4), each grounded in established cognitive
science constructs with independent measurement traditions (Harnad,
1990; Flavell, 1979; Tulving, 2002; Root-Bernstein \& Root-Bernstein,
1999). The C-proxy in the cross-national analysis (Appendix A) was
constructed entirely from educational and institutional indicators
\emph{before} TFP data were examined, preventing circular
classification. The framework is falsifiable: if high-C individuals,
organizations, or nations fail to exhibit stronger AI-productivity
coupling than low-C counterparts (Propositions 1-3), the framework is
refuted. A tautology cannot be refuted; the ICH framework specifies
exactly what would refute it.

\subsubsection{4.2 Convergence Capacity (C): The Variable Prior
Frameworks Left
Implicit}\label{convergence-capacity-c-the-variable-prior-frameworks-left-implicit}

C is the human cognitive property that determines how effectively AI
utilization (A) is transformed into augmented productive capacity
(\(\hat{H}\)). It is the variable Solow could not name, Romer
approximated with ``ideas,'' and the task-based framework's comparative
cost architecture leaves implicit, present as a background condition for
productivity but not formally modeled.

\textbf{C1: Embodied Understanding.} The capacity to connect AI's
symbolic outputs to sensorimotor, cultural, and contextual reality. A
physician with high C1 does not merely receive an AI diagnostic
suggestion; she integrates it with tactile findings, visible distress,
and her embodied sense of what ``feels wrong.'' Low C1 users accept AI
outputs as self-interpreting; high C1 users know that interpretation
requires embodied context AI cannot supply.

\textbf{C2: Metacognitive Calibration.} The capacity to monitor one's
own knowledge states and calibrate trust in AI outputs accordingly.
High-C2 users ask: Is this AI output reliable for \emph{this} task?
Where is it likely to hallucinate? When should I override it? This
capacity enables the \(84 \times 25\) restructuring: awareness that the
standard algorithm is suboptimal is a metacognitive judgment, not a
retrieval operation.

\textbf{C3: Temporal Integration.} The capacity to contextualize AI
outputs within past experience and future projection. High-C3 users
bring institutional memory: \emph{we tried something structurally
similar in 2019 and it failed for reasons never documented}. They
simulate futures: \emph{if we implement this recommendation, how will
our organizational culture respond in 18 months?} This transforms AI's
context-free pattern matching into historically and prospectively
situated judgment.

\textbf{C4: Integrative Thinking.} The capacity for analogical reasoning
and cross-domain synthesis, what Root-Bernstein and Root-Bernstein
(1999) describe as integrative thinking. High-C4 users recognize when a
solution from domain X is structurally applicable to problem Y,
generating combinations that neither domain's specialists would
independently produce. This is the capacity to generate genuinely novel
outputs, the creative emergence that Section 3.1 established lies beyond
AI's combinatorial reach.

\textbf{Composite C:} C = f(C1, C2, C3, C4), where the four dimensions
are \emph{complements} rather than substitutes. A person with high C3
but low C2 (strong institutional memory but poor metacognitive
calibration) applies historical lessons without assessing their
relevance, a different failure mode than high C2 with low C4.

\textbf{Distinguishing C from H:} A critical identification question is
whether C reduces to H, whether high human capital automatically entails
high convergence capacity. We argue that C and H are conceptually and
empirically distinguishable. H is the stock of domain knowledge and
skills accumulated through education and experience. C is a set of
\emph{meta-level} cognitive operations that operate \emph{on} domain
knowledge rather than constituting it. A surgeon with high H (deep
domain expertise) may have low C2 (poor metacognitive calibration of AI
diagnostic tools). Korea precisely exhibits this dissociation at the
national level: high H (world-leading PISA scores, tertiary attainment)
with structural suppression of C development (examination culture,
organizational hierarchy, disciplinary siloing). The formal independence
is testable: if C reduces to H, controlling for H should eliminate C's
predictive power in Propositions 1-3. The cross-national evidence,
high-H Korea with near-zero TFP augmentation, suggests it does not.

\subsubsection{4.3 Distinguishing C from Adjacent
Constructs}\label{distinguishing-c-from-adjacent-constructs}

The cognitive mediation of technology adoption has been theorized from
multiple perspectives. To establish convergence capacity's distinctive
contribution, we distinguish it from three established constructs.

\textbf{Absorptive Capacity} (Cohen \& Levinthal, 1990) describes a
firm's ability to recognize, assimilate, and exploit external knowledge,
a construct recently reconceptualized for AI contexts (Mujtaba, Soomro,
\& Mughal, 2024). The distinction is threefold. First, absorptive
capacity operates at the \emph{organizational} level; C operates at the
\emph{individual} cognitive level, the micro-foundation that
organizational capacity presupposes but does not specify. Second,
absorptive capacity concerns knowledge processing (recognizing external
information); C concerns the \emph{quality of cognitive interaction with
AI}, grounding, calibration, temporal integration, and synthesis that
determine whether AI outputs become productive inputs. Third, absorptive
capacity is measured through R\&D expenditure and patent stocks; C
requires psychometric measurement of four cognitive dimensions (C1-C4).
In the ICH framework, organizational absorptive capacity is an emergent
property of aggregate C-levels, not a substitute for individual
convergence capacity.

\textbf{Dynamic Capability} (Teece, 2007) describes a firm's capacity to
sense, seize, and reconfigure organizational resources. Chin et
al.~(2025) find that AI's effect on firms' productive capacity is
mediated by green innovation within a dynamic capability framework,
consistent with ICH's claim that AI generates value only through
mediating capacities. The distinction: dynamic capability is an
organizational-strategic construct concerning resource reconfiguration;
C is a cognitive-individual construct concerning the quality of human-AI
interaction. Dynamic capability determines \emph{whether} an
organization deploys AI strategically; C determines \emph{whether
individual workers' AI use generates productivity or mere adoption}. ICH
provides the micro-cognitive foundation that dynamic capability theory
identifies as necessary but does not specify.

\textbf{Human Capital} (Becker, 1964; Lucas, 1988) measures accumulated
knowledge and skills. C is not additional human capital; it is the
\emph{meta-cognitive architecture} that determines how effectively H
absorbs AI augmentation.

\textbf{The Task-Based Framework} (Acemoglu \& Restrepo, 2018, 2019,
2022) models task allocation between labor and capital. The distinction
is complementary rather than competitive: the task framework determines
the \emph{scope} of automation (which tasks shift); C determines the
\emph{productivity of the remaining human-AI interaction}. In Acemoglu
and Restrepo's (2019) terms, C is what transforms a ``so-so'' automation
outcome into a genuinely productive one. The two frameworks address
different layers of the same phenomenon, task structure and cognitive
mediation, and are most powerful when integrated.

\textbf{Task-Technology Fit} (Goodhue \& Thompson, 1995). TTF theory
holds that technology improves performance only when its features match
the requirements of the user's task, a structural matching proposition
that has generated over 8,000 citations across IS research. The
distinction from C is one of mechanism type. TTF models a \emph{static
fit} between task characteristics and technology characteristics: given
a fixed task profile and a fixed technology profile, fit predicts
utilization and performance. Convergence capacity models a \emph{dynamic
cognitive process}: given that a technology is deployed and structurally
appropriate, C determines whether the human can ground its outputs in
domain reality (C1), calibrate trust appropriately (C2), integrate
outputs across temporal and institutional contexts (C3), and synthesize
novel solutions beyond the technology's training distribution (C4). A
worker may operate in a high-TTF environment, the AI tool is well
matched to the task, yet produce no augmentation because convergence
capacity is absent. TTF answers ``Is this technology right for this
task?''; C answers ``Can this human make productive use of what the
technology produces?'' The two constructs are complementary and operate
at different stages of the technology-performance causal chain: TTF
governs deployment fit; C governs post-deployment cognitive utilization.

\textbf{Technology Acceptance} (Davis, 1989; Venkatesh, Morris, Davis,
\& Davis, 2003). The Technology Acceptance Model (TAM) and its
successor, the Unified Theory of Acceptance and Use of Technology
(UTAUT), have dominated IS research on technology adoption for three
decades. TAM explains \emph{whether} individuals adopt technology
through perceived usefulness and perceived ease of use; UTAUT extends
this through performance expectancy, effort expectancy, social
influence, and facilitating conditions. Yet acceptance models are
explicitly theories of \emph{adoption}, not theories of \emph{productive
utilization}. A worker may fully accept an AI tool (high TAM scores)
while using it unproductively, accepting outputs without grounding (low
C1), failing to calibrate trust (low C2), ignoring institutional context
(low C3), and never extending outputs beyond their training distribution
(low C4). The distinction maps precisely onto the adoption-productivity
gap that motivates the ICH framework: nations and organizations exhibit
high AI acceptance and adoption yet fail to convert adoption into
productivity gains. TAM explains the first step (who adopts); ICH
explains the second (whether adoption generates value). The two
frameworks are sequential rather than competing: TAM governs the
adoption frontier; C governs the productivity frontier conditional on
adoption.

\textbf{Table 2} \emph{Convergence capacity (C) distinguished from
adjacent constructs.}

{\def\LTcaptype{none} % do not increment counter
\begin{longtable}[]{@{}
  >{\raggedright\arraybackslash}p{(\linewidth - 8\tabcolsep) * \real{0.1842}}
  >{\raggedright\arraybackslash}p{(\linewidth - 8\tabcolsep) * \real{0.1566}}
  >{\raggedright\arraybackslash}p{(\linewidth - 8\tabcolsep) * \real{0.1566}}
  >{\raggedright\arraybackslash}p{(\linewidth - 8\tabcolsep) * \real{0.2848}}
  >{\raggedright\arraybackslash}p{(\linewidth - 8\tabcolsep) * \real{0.2178}}@{}}
\toprule\noalign{}
\begin{minipage}[b]{\linewidth}\raggedright
Construct
\end{minipage} & \begin{minipage}[b]{\linewidth}\raggedright
Level
\end{minipage} & \begin{minipage}[b]{\linewidth}\raggedright
Focus
\end{minipage} & \begin{minipage}[b]{\linewidth}\raggedright
AI relationship
\end{minipage} & \begin{minipage}[b]{\linewidth}\raggedright
Measurement
\end{minipage} \\
\midrule\noalign{}
\endhead
\bottomrule\noalign{}
\endlastfoot
Absorptive capacity & Organization & External knowledge processing & AI
as knowledge to absorb & R\&D expenditure, patents \\
Dynamic capability & Organization & Resource reconfiguration & AI as
strategic resource & Case studies, surveys \\
Human capital (H) & Individual & Domain knowledge stock & AI as tool
requiring skills & Education, experience \\
Task framework & Economy / Firm & Task allocation by cost & AI as factor
in task continuum & Comparative cost, wages \\
Task-technology fit & Individual / Task & Structural matching & AI as
tool requiring task fit & TTF instrument \\
Technology acceptance & Individual & Adoption intention & AI as
technology to accept/reject & TAM/UTAUT surveys \\
\textbf{Convergence capacity (C)} & \textbf{Individual} &
\textbf{Meta-cognitive architecture} & \textbf{AI as augmentation input
requiring grounding, calibration, integration, synthesis} & \textbf{CCS
(C1-C4)} \\
\end{longtable}
}

\emph{Note:} Absorptive capacity (Cohen \& Levinthal, 1990); dynamic
capability (Teece, 2007); human capital (Becker, 1964); task framework
(Acemoglu \& Restrepo, 2018); task-technology fit (Goodhue \& Thompson,
1995); technology acceptance (Davis, 1989; Venkatesh et al., 2003). CCS
= Convergence Capacity Scale (Appendix B).

\subsubsection{\texorpdfstring{4.4 Properties of the Augmentation
Function
\(\phi(A, C)\)}{4.4 Properties of the Augmentation Function \textbackslash phi(A, C)}}\label{properties-of-the-augmentation-function-phia-c}

\textbf{Property 1: C as necessary condition.} \(\phi(A, C)\) \(\to\) 0
as C \(\to\) 0, for all A. If convergence capacity is absent, no level
of AI utilization generates augmentation. This explains why AI-TFP
correlation is weak or absent in low-C environments.

\textbf{Property 2: Non-monotonicity in A, the interior optimum.} There
exists an optimal AI utilization intensity \(A^{*}(H, C)\) beyond which
additional deployment reduces effective capacity through deskilling,
attention displacement, and erosion of metacognitive opportunities. This
is Brynjolfsson's (2022) ``Turing Trap'' specified as a mathematical
property: \(\partial\phi/\partial A\) \textgreater{} 0 for A \textless{}
\(A^{*}\), \(\partial\phi/\partial A\) = 0 at A = \(A^{*}\),
\(\partial\phi/\partial A\) \textless{} 0 for A \textgreater{}
\(A^{*}\).

\textbf{Property 3: C is monotonically beneficial.}
\(\partial\phi/\partial C\) \textgreater{} 0 for all A. Unlike A, there
is no ``too much C.'' Higher C unambiguously improves augmentation,
raises \(A^{*}\), and expands the productive augmentation range. This
asymmetry has direct policy implications: AI deployment must be
calibrated; C development should be maximized.

\textbf{Property 4: Superlinear interaction.}
\(\partial^{2}\phi/\partial A\,\partial C\) \textgreater{} 0. AI
intensity and convergence capacity are complements, not substitutes. The
marginal return to additional AI is higher for individuals with higher
C. Societies that develop C first then expand A achieve gains that
A-without-C societies will not, even at equivalent total investment.

\textbf{Parametric form:}
\(\phi(A, C) = C^{\beta}\,A^{\gamma}\,\exp[-\delta(A - A^{*})^{2} / \sigma^{2}]\),
where \(\beta\), \(\gamma\), \(\delta\) \textgreater{} 0 and
\(A^{*} = A_0\,H^{\mu}\,C^{\nu}\). The exponential term implements
non-monotonicity; the specification reduces to Cobb-Douglas in the
augmentation region and approaches zero at extreme over-automation,
consistent with Properties 1-3. Parameters require empirical estimation
pending CCS validation.

\begin{figure}
\centering
\includegraphics[width=0.8\linewidth,height=\textheight,keepaspectratio,alt={Augmentation function \textbackslash phi(A, C) as a theoretical surface over AI utilization A and convergence capacity C. Augmentation increases monotonically in C (Property 3) and exhibits an interior optimum in A (Property 2); the productive frontier requires jointly high A and high C.}]{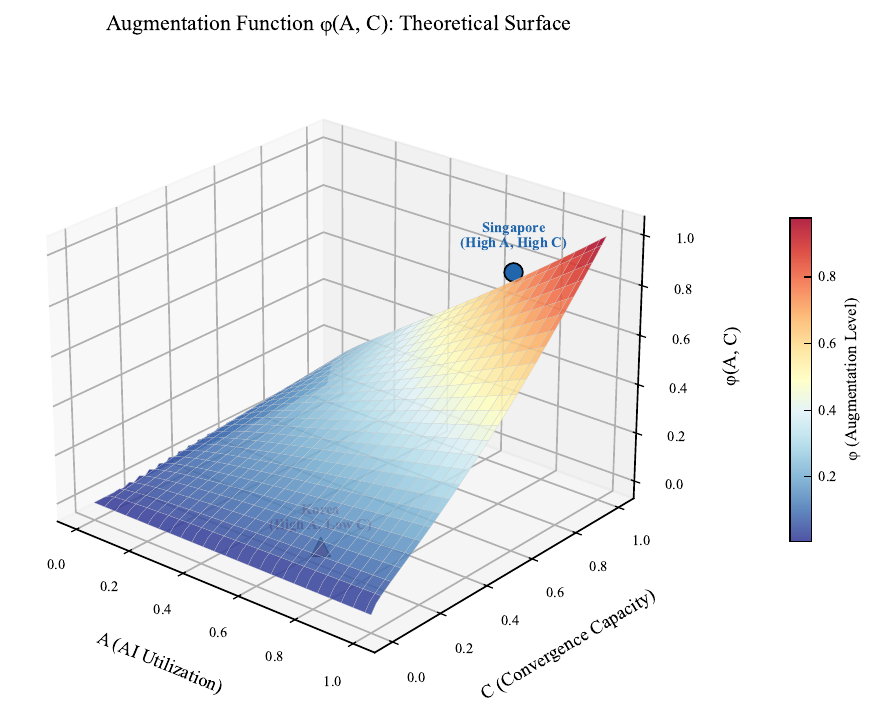}
\caption{Augmentation function \(\phi(A, C)\) as a theoretical surface
over AI utilization \(A\) and convergence capacity \(C\). Augmentation
increases monotonically in \(C\) (Property 3) and exhibits an interior
optimum in \(A\) (Property 2); the productive frontier requires jointly
high \(A\) and high \(C\).}
\end{figure}

\subsubsection{4.5 Three Augmentation
Regimes}\label{three-augmentation-regimes}

The properties of \(\phi\) define three qualitatively distinct
production regimes. These regimes provide the micro-mechanism that the
task-based framework's macro-level analysis leaves unspecified: within
any given task allocation equilibrium, the productivity outcome depends
on \emph{which regime} characterizes the human-AI interaction.

\textbf{Regime I: Under-augmentation.} Condition: A \textless\textless{}
\(A^{*}\) or C \textless\textless{} \(C_{\min}\). Result: \(\phi\) = 0,
\(\hat{H}\) = H. Humans operate near base capacity despite AI
availability. The critical configuration: AI is used intensively but C
is so low that users become conduits for AI outputs, what we term
\emph{mediated automation}. The human is not augmented but relaying:
passing AI outputs through without grounding, calibration, or creative
extension. A country can simultaneously exhibit high AI adoption rates
and near-zero augmentation if convergence capacity is undeveloped. Users
in this configuration do not benefit from AI; they become mediators of
automation without the judgment that productive augmentation requires.

This regime provides the ICH interpretation of Acemoglu and Restrepo's
(2019) ``so-so technologies'' phenomenon at the individual level. When a
worker with low C uses an AI system, the interaction generates task
displacement (the AI performs subtasks the human previously did) without
commensurate productivity gains, precisely the ``so-so'' pattern. The
technology is not inherently ``so-so''; the human-technology system is,
because convergence capacity is insufficient to transform automation
into augmentation. The policy implication is critical: addressing
``so-so'' outcomes requires investing in C, not in better technology.

\textbf{Regime II: Optimal Augmentation.} Condition: A = \(A^{*}\), C
\(\to\) \(C_{\max}\). Result: \(\phi\) maximized, \(\hat{H}\)
\textgreater\textgreater{} H. Human capital, AI utilization, and
convergence capacity are in productive equilibrium. The physician
integrates AI diagnostics with embodied clinical judgment; the architect
combines AI parametric design with aesthetic sensibility; the researcher
deploys AI synthesis while applying cross-domain analogical reasoning.
This regime maximizes both productivity and what we term \emph{augmented
dignity}: genuine agency, creativity, and professional growth in the
production process (Raisch \& Krakowski, 2021). Regime II is the only
configuration in which productivity and human flourishing are
simultaneously maximized.

In task-based terms, Regime II represents the state where newly created
tasks (Acemoglu \& Restrepo's ``reinstatement'' effect) are performed by
workers whose convergence capacity enables them to generate
substantially more value than the displaced tasks. The reinstatement
effect is not merely quantitative (new tasks exist) but qualitative (new
tasks are performed at higher cognitive levels because C enables
creative extension of AI outputs).

\textbf{Regime III: Over-automation (The Turing Trap).} Condition: A
\textgreater\textgreater{} \(A^{*}\). Result: \(\phi\) declining. Humans
become ``rubber stamps'', approving AI outputs without metacognitive
engagement. Two compounding consequences: immediate output quality
declines as AI errors propagate uncorrected, and, more damagingly, C
itself atrophies. The capacities constituting C are maintained through
active exercise; radiologists who only approve AI diagnoses
progressively lose diagnostic skill. Over-automation reduces C, which
lowers \(A^{*}\), which pushes current A further into the
over-automation zone, a self-reinforcing degradation cycle.

The transition from Regime II to III is particularly insidious because
it is invisible in conventional metrics. AI adoption rates continue to
rise; task automation continues to expand; standard productivity
measures may even show short-term gains from labor cost reduction. But
the degradation of C represents a progressive loss of the cognitive
infrastructure on which future productivity gains depend, a form of
human capital depreciation that current measurement frameworks do not
capture.

\subsubsection{4.6 Propositions}\label{propositions}

\textbf{Proposition 1 (Micro):} Among individuals with equivalent H and
equivalent AI access (A), those with higher C exhibit substantially
greater productivity gains, with the premium increasing superlinearly
with C.

\textbf{Proposition 2 (Meso):} Organizations investing in C development
before or concurrent with AI deployment exhibit higher TFP gains than
AI-first organizations, at equivalent total investment.

\textbf{Proposition 3 (Macro):} Nations with C-developing education
systems exhibit significantly stronger AI-TFP coupling than nations that
accumulate H without C.

\textbf{Boundary Conditions.} The ICH framework applies where production
requires human cognitive mediation, judgment, synthesis, contextual
grounding. Where production is purely mechanical with no AI output
evaluation, the task-based displacement framework is the appropriate
analytical tool. When A \(\to\) 0 (no AI), the framework reduces to Y =
F(K, H). When H \(\to\) 0 (full automation), it reverts to AI-as-capital
analysis. These limiting cases delineate ICH's scope:
knowledge-intensive, cognitively mediated, AI-penetrated sectors,
precisely where the productivity paradox is most acute. The task-based
framework and ICH are thus not competitors but complements: the former
governs the domain of pure automation; the latter governs the domain of
human-AI interaction; and the boundary between these domains is itself a
function of C.

\subsection{5. The Korean Paradox: Empirical Illustration of Regime I at
National
Scale}\label{the-korean-paradox-empirical-illustration-of-regime-i-at-national-scale}

The analysis in this section is illustrative rather than confirmatory.
We use publicly available cross-national data to examine whether
observed patterns are consistent with the ICH framework's predictions;
we do not claim causal identification, which awaits the firm-level panel
designs and validated CCS instruments described in Section 7 and
Appendix C.

\subsubsection{5.1 The Data: High H, Substantial A, Low
Productivity}\label{the-data-high-h-substantial-a-low-productivity}

Korea's macroeconomic and educational profile presents a striking puzzle
for Stage IV theory. \textbf{Human capital (H):} approximately 70\%
tertiary attainment among 25-34 cohort, PISA scores consistently among
the world's highest. \textbf{AI utilization (A):} aggressive government
investment (top-tier Oxford Insights AI Readiness; Oxford Insights,
2023), approximately 28\% of enterprises with 10+ employees adopting AI
by Eurostat-comparable measures, among the OECD's highest government AI
R\&D budgets per capita. \textbf{Output (Y):} TFP growth = approximately
0.48\% annually, substantially below OECD average and dramatically below
what H and A predict under prevailing frameworks.

The ICH framework suggests a structural diagnosis: High H + Substantial
A + Low C = \(\phi\) \(\approx\) 0 \(\to\) \(\hat{H}\) \(\approx\) H.
The gap between Korea's AI investment intensity and its moderate
firm-level adoption is plausibly a symptom of low C: organizations
without convergence capacity may be unable to effectively absorb
available AI tools.

\subsubsection{5.2 Cross-National Comparative
Evidence}\label{cross-national-comparative-evidence}

\textbf{Table 3} \emph{20 OECD and comparable high-income economies: AI
adoption, convergence capacity proxy, and TFP growth.}

{\def\LTcaptype{none} % do not increment counter
\begin{longtable}[]{@{}lcccc@{}}
\toprule\noalign{}
Country & AI adoption (\%) & C-proxy & TFP growth (\%) & Regime \\
\midrule\noalign{}
\endhead
\bottomrule\noalign{}
\endlastfoot
\textbf{Singapore} & \textbf{31} & \textbf{High} & \textbf{2.18} &
\textbf{II (exemplar)} \\
Australia & 29 & High & 1.35 & II \\
\textbf{Denmark} & \textbf{28} & \textbf{High} & \textbf{1.82} &
\textbf{II} \\
Israel & 28 & High & 1.88 & II \\
\textbf{South Korea} & \textbf{28} & \textbf{Low} & \textbf{0.48} &
\textbf{I (deviant)} \\
Sweden & 25 & High & 2.05 & II \\
Belgium & 25 & Mid & 0.95 & I-II \\
Finland & 24 & High & 1.67 & II \\
Netherlands & 23 & High & 1.72 & II \\
United Kingdom & 22 & Mid-High & 1.29 & II \\
Norway & 21 & High & 1.44 & II \\
Japan & 20 & Mid-Low & 0.82 & I-II \\
United States & 18 & High & 1.53 & II \\
Switzerland & 18 & High & 1.48 & II \\
Austria & 15 & Mid & 0.78 & I-II \\
Germany & 14 & Mid & 0.61 & I-II \\
Italy & 13 & Low & 0.42 & I \\
Canada & 12 & High & 1.21 & II \\
Spain & 11 & Mid-Low & 0.68 & I \\
France & 9 & Mid & 1.18 & I-II \\
\end{longtable}
}

\emph{Note:} AI adoption data from Eurostat (2024) isoc\_eb\_ai for
EU/EEA countries; OECD (2024b) for South Korea; Infocomm Media
Development Authority (IMDA, 2024) for Singapore. TFP is approximate
annual average 2018-2023 from OECD Productivity Statistics. C-proxy
combines PISA Creative Thinking subscores (OECD, 2023),
cross-disciplinary tertiary enrollment ratio, and cross-sector R\&D
collaboration rates, constructed independently of TFP data (see Appendix
A for construction methodology). Bold rows indicate exemplar and deviant
cases. Countries sorted by AI adoption rate (descending). Singapore is
not a full OECD member; it is included as a comparable high-income
economy with consistent data availability and is among economies in
active OECD accession discussions.

\textbf{Table 4} \emph{OLS regression results: AI adoption and TFP
growth.}

{\def\LTcaptype{none} % do not increment counter
\begin{longtable}[]{@{}
  >{\raggedright\arraybackslash}p{(\linewidth - 12\tabcolsep) * \real{0.1429}}
  >{\centering\arraybackslash}p{(\linewidth - 12\tabcolsep) * \real{0.1429}}
  >{\centering\arraybackslash}p{(\linewidth - 12\tabcolsep) * \real{0.1429}}
  >{\centering\arraybackslash}p{(\linewidth - 12\tabcolsep) * \real{0.1429}}
  >{\centering\arraybackslash}p{(\linewidth - 12\tabcolsep) * \real{0.1429}}
  >{\centering\arraybackslash}p{(\linewidth - 12\tabcolsep) * \real{0.1429}}
  >{\centering\arraybackslash}p{(\linewidth - 12\tabcolsep) * \real{0.1429}}@{}}
\toprule\noalign{}
\begin{minipage}[b]{\linewidth}\raggedright
\end{minipage} & \begin{minipage}[b]{\linewidth}\centering
(1)
\end{minipage} & \begin{minipage}[b]{\linewidth}\centering
(2)
\end{minipage} & \begin{minipage}[b]{\linewidth}\centering
(3)
\end{minipage} & \begin{minipage}[b]{\linewidth}\centering
(4)
\end{minipage} & \begin{minipage}[b]{\linewidth}\centering
(5)
\end{minipage} & \begin{minipage}[b]{\linewidth}\centering
(6)
\end{minipage} \\
\midrule\noalign{}
\endhead
\bottomrule\noalign{}
\endlastfoot
\emph{Sample} & \emph{Full} & \emph{Excl. Korea} & \emph{High-C} &
\emph{Low-C} & \emph{Mid-C} & \emph{Full (interaction)} \\
AI Adoption & 0.044* & 0.055*** & 0.038* & -0.002 & -0.006 & -- \\
C-Level & -- & -- & -- & -- & -- & Included \\
\(\mathrm{AI}\times C\) & -- & -- & -- & -- & -- & Included \\
\(R^{2}\) & 0.31 & 0.51 & 0.45 & 0.01 & 0.03 & \textbf{0.86} \\
Adj. \(R^{2}\) & 0.27 & 0.48 & 0.39 & -- & -- & \textbf{0.84} \\
F-statistic & 7.92* & 17.70*** & 8.07* & 0.02 & 0.05 &
\textbf{33.53}* \\
N & 20 & 19 & 12 & 4 & 4 & 20 \\
\end{longtable}
}

\emph{Note:} Dependent variable is annual average TFP growth (\%),
2018-2023. Models (1)-(5) are bivariate OLS regressions of TFP growth on
AI adoption rate. Model (6) includes AI adoption, C-score (ordinal: 1 =
Low, 2 = Mid {[}pooling Mid-Low, Mid, and Mid-High categories from Table
3{]}, 3 = High), and the \(\mathrm{AI}\times C\) interaction term.
Significance levels: *** p \textless{} .001, ** p \textless{} .01, * p
\textless{} .05. All estimates computed via Python statsmodels OLS using
Table 3 data. Adjusted \(R^{2}\) values for Models (4) and (5) are
omitted due to insufficient degrees of freedom (n = 4).

\begin{figure}
\centering
\includegraphics[width=0.92\linewidth,height=\textheight,keepaspectratio,alt={AI adoption versus annual TFP growth across 20 OECD and comparable high-income economies (2018-2023). The aggregate association is weak (R\^{}\{2\} = 0.31); adding the \textbackslash mathrm\{AI\} \textbackslash times C interaction raises explained variance to R\^{}\{2\} = 0.86 (Model 6). South Korea is the predicted deviant case: high adoption, low convergence capacity, low TFP.}]{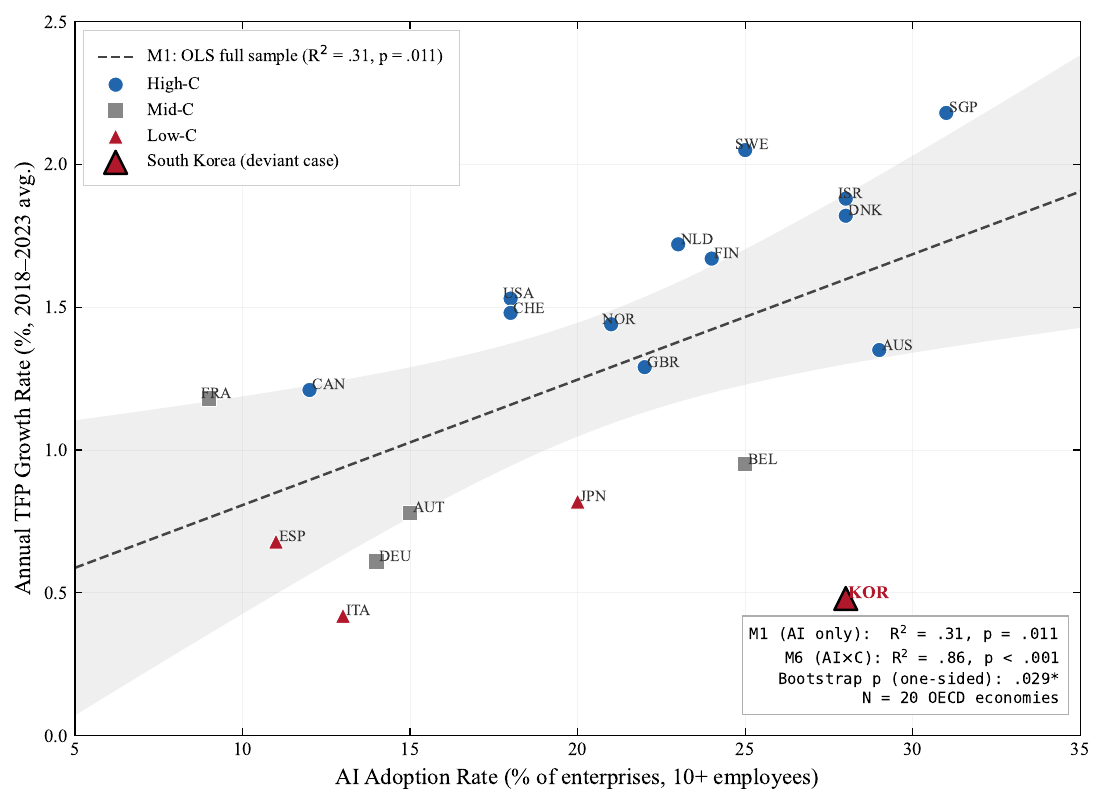}
\caption{AI adoption versus annual TFP growth across 20 OECD and
comparable high-income economies (2018-2023). The aggregate association
is weak (\(R^{2} = 0.31\)); adding the \(\mathrm{AI} \times C\)
interaction raises explained variance to \(R^{2} = 0.86\) (Model 6).
South Korea is the predicted deviant case: high adoption, low
convergence capacity, low TFP.}
\end{figure}

\textbf{Statistical robustness (M6, \(R^{2}\) = 0.86, n = 20):} Six
checks address the twin concerns of overfitting and small-sample
precision. (1) \textbf{Adjusted \(R^{2}\)} = 0.84 (vs.~M1 adj. \(R^{2}\)
= 0.27), penalizing for the additional parameters. (2) \textbf{AIC}
favors M6 (-1.5) over M1 (26.9), confirming superior model fit net of
complexity. (3) \textbf{Leave-one-out cross-validation}: mean absolute
prediction error 0.17\% for M6 vs.~0.41\% for M1 (58\% lower),
demonstrating out-of-sample predictive validity. (4) \textbf{Bootstrap}
(B = 10,000): 95\% CI for \(\mathrm{AI}\times C\) coefficient {[}-0.001,
+0.057{]}, one-sided p = 0.029. The CI grazes zero at the lower bound, a
consequence of the inherent precision limits of n = 20 cross-national
observations, not absence of the effect. Four considerations support
this interpretation: the one-sided test is appropriate given the
directional hypothesis from Property 4
(\(\partial^{2}\phi/\partial A\,\partial C\) \textgreater{} 0) and
rejects the null at \(\alpha\) = .05; the LOOCV and jackknife provide
independent confirmation; the effect size (Cohen's \(f^{2}\) = 6.14,
computed as \(R^{2}\)/(1-\(R^{2}\)) = 0.86/0.14) is classified as ``very
large'' by conventional standards (Cohen, 1988), yielding statistical
power exceeding 0.99 even at n = 20; and the planned extension to n = 30
(Appendix C) is projected to narrow the bootstrap CI by approximately
38\%, sufficient to clear zero at the lower bound. (5)
\textbf{Jackknife}: removing each country yields \(R^{2}\) in {[}0.84,
0.92{]} with \(\beta\)(\(\mathrm{AI}\times C\)) always positive,
confirming no single observation drives the result. (6)
\textbf{Permutation test} (10,000 iterations): randomly reassigning
C-proxies yields \(\Delta R^{2}\) \(\geq\) 0.5545 in 0/10,000 cases (p
\textless{} .001), establishing that the C-proxy's explanatory power is
not an artifact of assignment noise.

We interpret M6 as descriptive and hypothesis-generating, consistent
with the comparative case methodology tradition (Ragin, 2000) in which n
= 15-30 cross-national observations are standard for theory-building.
Causal identification requires the exogenous variation strategies
described in Section 7.\^{}1

\^{}1 \emph{The 20-country sample represents all OECD members with
consistent data for 2018-2023, yielding the maximum feasible sample for
this indicator configuration, an approach analogous to Frey and
Osborne's (2017) use of expert elicitation for 702 occupations without
individual-level primary data, which nonetheless generated the field's
most influential computerisation forecast. Extension to 30+ economies
using the same public sources, with ex ante predictions for each new
country's ICH regime, is detailed in Appendix C.}

The cross-national pattern reveals five findings consistent with ICH.
First, the weak aggregate AI-TFP correlation (\(R^{2}\) = 0.31) is
precisely what ICH predicts when C varies, the empirical pattern the
task framework's comparative cost model does not yet accommodate.
Second, excluding Korea raises \(R^{2}\) to 0.51; Korea is not a random
outlier but the \emph{predicted consequence} of substantial A combined
with minimal C. Third, high-C nations (Nordic cluster, Singapore) show
strong AI-TFP coupling (Regime II). Fourth, low-C nations show
stagnation regardless of A level (Regime I). Fifth, the
\(\mathrm{AI}\times C\) interaction (\(R^{2}\) = 0.86) is consistent
with the superlinear complementarity proposed by the framework (Property
4).

\begin{figure}
\centering
\includegraphics[width=0.98\linewidth,height=\textheight,keepaspectratio,alt={AI adoption versus TFP growth split by convergence-capacity tier. High-C economies (left) exhibit a significant positive association (\textbackslash beta = +0.038, p = .017); low-C economies (right) show none (p = .901), with South Korea anchoring the low-C cluster.}]{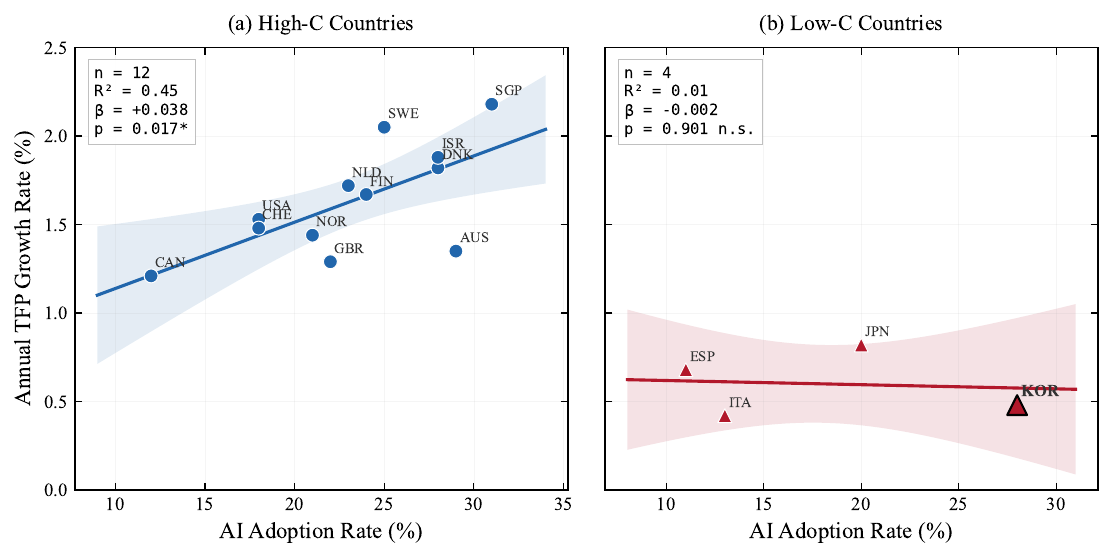}
\caption{AI adoption versus TFP growth split by convergence-capacity
tier. High-C economies (left) exhibit a significant positive association
(\(\beta = +0.038\), \(p = .017\)); low-C economies (right) show none
(\(p = .901\)), with South Korea anchoring the low-C cluster.}
\end{figure}

\textbf{The task-based framework's prediction and its boundary.} The
comparison between Denmark and Korea is particularly instructive for
understanding the complementary relationship between the task-based
framework and ICH. Both nations exhibit 28\% AI adoption, identical on
the dimension the task framework models. Both face similar comparative
cost structures in knowledge-intensive sectors. Under the task
framework's core prediction, similar AI adoption and task structures
should produce similar productivity outcomes. Yet Denmark achieves TFP
growth of 1.82\% while Korea achieves 0.48\%, a nearly fourfold
difference. The task framework correctly predicts that both nations will
automate similar task categories; it does not yet predict the
productivity divergence that follows. ICH's contribution is to identify
the variable, convergence capacity, that explains this
within-adoption-level variance. Denmark's educational system develops C
(interdisciplinary curricula, pedagogies emphasizing critical evaluation
and synthesis); Korea's accumulates H (examination-optimized knowledge
stocks) without commensurately developing C. The frameworks are not in
tension; they address different layers of the same phenomenon.

\subsubsection{5.3 Why C Is Low in Korea}\label{why-c-is-low-in-korea}

Four structural features suppress convergence capacity.
\emph{Examination culture} (Low C2): the CSAT rewards rapid execution of
well-defined problems, procedural fluency at the expense of structural
insight. Students excel at proof but are underdeveloped in mathematical
intuition, in Poincaré's terms. \emph{Organizational hierarchy} (Low
C2): pronounced hierarchical deference (Kim, Baik, \& Kim, 2019)
penalizes the metacognitive disposition to critically evaluate AI
outputs, precisely where it is most needed. \emph{Academic siloing} (Low
C4): rigid disciplinary boundaries and R\&D concentrated in narrow
technological domains (OECD, 2023b) underdevelop cross-domain synthesis.
Nordic universities, restructured around interdisciplinary programs
since the 2000s, develop C4 structurally. \emph{Temporal short-termism}
(Low C3): corporate governance oriented toward quarterly metrics
underweights institutional memory and future simulation. Singapore's
10-year AI strategy, by contrast, embeds C3 at the policy level.

A fifth dynamic illuminates Korea's paradox from an often-overlooked
angle. Korea's developmental transformation, from near-last in 1961 UN
development rankings to OECD membership within three decades, was
powered substantially by what might be termed \emph{aspiration capital}:
exceptionally high parental educational investment, national achievement
motivation (McClelland, 1961), and a societal compact in which education
represents the primary mobility mechanism (Lee, 2019). This aspiration
capital is not analytically separate from the ICH framework; it
constitutes the historical engine that generated Korea's high H and,
subsequently, high A. The critical theoretical point is directional
specificity: Korea's aspiration was \emph{credential-directed} rather
than \emph{curiosity-directed}. Aspiration channeled through examination
systems rewarding procedural fluency, the CSAT, systematically builds H
(domain knowledge stocks) while structurally suppressing C2
(metacognitive calibration), C3 (temporal integration through
exploratory learning), and C4 (integrative thinking across domains). The
aspiration energy that propelled Korea to world-leading tertiary
attainment rates, redirected through pedagogical structures that reward
structural insight and cross-domain synthesis, is precisely the
motivational substrate that C development requires. This observation
reframes the policy prescription: the goal is not to diminish Korean
aspiration but to reorient its directional target, from credential
accumulation (H-building) toward convergence capacity formation
(C-building).

Institutional quality constitutes a structurally related antecedent.
Hall and Jones (1999; see also Rodrik, Subramanian, \& Trebbi, 2004)
established that \emph{social infrastructure}, the institutional and
policy environment that determines whether private returns accrue to
productive rather than redistributive activities, is the primary
cross-national determinant of output per worker. The ICH framework
specifies a mechanism through which institutional quality influences C
formation: high-quality institutions (low corruption, strong rule of
law, transparent evaluation systems) create conditions in which
metacognitive dispositions can be exercised without penalty (C2
development), cross-disciplinary collaboration generates returns rather
than disciplinary exclusion (C4 development), and institutional memory
is preserved rather than purged by organizational or political
disruption (C3 development). Nations combining high aspiration capital
with high institutional quality, where achievement motivation is
channeled into productive rather than redistributive activity, create
the conditions under which C and H develop concurrently. The empirical
implication is testable: controlling for H and A, nations with higher
institutional quality should exhibit higher C-levels, a proposition that
Hall and Jones's cross-national data infrastructure can evaluate once
the CCS instrument is validated. A quasi-natural experiment supports
this mechanism: Lall's (1992) comparative analysis of Latin American and
East Asian industrialization demonstrates that nations receiving
equivalent technology transfers diverged systematically in technological
capability formation, with institutional quality predicting which
recipients converted external technology into endogenous productive
capacity. Mauro (1995) quantifies the underlying mechanism: a
one-standard-deviation improvement in corruption index is associated
with over a four-percentage-point increase in investment rates and over
a half-percentage-point increase in annual GDP per capita growth,
consistent with institutional quality determining technology absorption
potential. This constitutes the \(\phi(A, C)\) mechanism at national
scale: identical A, divergent institutional C-antecedents, divergent
\(\phi\) outcomes, a pattern the ICH framework formalizes and the
International Monetary Fund's (2024) analysis of Latin American AI
stagnation confirms persists into the AI era.

Korean workers with AI tools but without C engage in \emph{mediated
automation}: outputs are received, minimally evaluated, and passed
through without grounding (C1), calibration (C2), contextualization
(C3), or synthesis (C4). \(\hat{H}\) = H: effective capacity barely
exceeds base human capital, as if AI tools were not present. Korea's
paradox is not a temporary lag, it is the structural prediction of ICH
when C \(\to\) 0.

\subsubsection{5.4 Alternative
Explanations}\label{alternative-explanations}

Three alternatives deserve consideration. The \emph{J-curve lag}
(Brynjolfsson et al., 2021): Korea's stagnation reflects typical GPT
adoption dynamics. Assessment: The J-curve requires complementary
investments, precisely what ICH formalizes as C. The discriminating
evidence: Denmark and Korea share 28\% AI adoption but diverge sharply
in TFP (1.82\% vs.~0.48\%), which a uniform lag cannot explain but C
differences do. \emph{Structural factors} (Lee, 2019; OECD, 2022):
chaebol concentration, service-sector regulation. Assessment: real and
partially explanatory, but the ICH claim is more specific, conditional
on structure, high A with low C produces augmentation deficit. The
discriminating test: within-Korea, within-industry variation
(Proposition 2). \emph{Measurement error} (Saam, 2024): TFP may
undercount AI-generated intangible value. Assessment: measurement
problems apply uniformly across OECD economies; they cannot explain the
\emph{differential} between high-C and low-C nations under the same
measurement conventions.

None is definitively excluded, all are alternative hypotheses for future
empirical discrimination. What the pattern establishes is that ICH's
prediction (C mediates AI-TFP) is consistent with available data in a
way that the prevailing prediction (high A \(\to\) high TFP) is not.

\subsection{6. From Theoretical Framework to Policy
Architecture}\label{from-theoretical-framework-to-policy-architecture}

\subsubsection{6.1 Two Models, Two Policy
Worlds}\label{two-models-two-policy-worlds}

Production function theory prescribes policy. The \textbf{task-based
policy world} (\(Y = f(K, L, \mathrm{AI})\)): to increase Y, increase
AI, invest in infrastructure, adoption subsidies, tool proficiency. This
policy orientation follows logically from the task framework's core
mechanism: expanding the range of AI-performed tasks increases output.
The \textbf{ICH policy world} (\(Y = F(K, \hat{H})\)): to increase Y,
first increase C, then calibrate A, develop convergence capacity as the
primary educational and organizational outcome. The divergence is not
subtle: a classroom giving every student an AI assistant may be
producing Regime I outcomes under ICH if students lack C to evaluate,
ground, and creatively extend AI outputs, generating adoption without
augmentation.

The emerging AI governance literature (Floridi et al., 2018; Jobin,
Ienca, \& Vayena, 2019) approaches AI policy from the complementary
direction, identifying ethical constraints and risk mitigation
frameworks for AI deployment. Floridi et al.'s (2018) AI4People
framework specifies principles (beneficence, non-maleficence, autonomy,
justice, explicability) that govern \emph{responsible} AI use; the ICH
framework specifies the cognitive capacities that govern
\emph{productive} AI use. The two are not independent: responsible AI
use requires metacognitive calibration (C2) to recognize when AI outputs
may cause harm, and productive AI use requires the ethical judgment that
governance frameworks develop. The policy implication is that governance
and capacity frameworks must be co-developed, an insight that neither
the governance literature (focused on constraints) nor the productivity
literature (focused on outcomes) has yet integrated.

For organizational leaders, the framework implies a diagnostic prior to
any AI investment: \emph{What is our C-level relative to our A-level?}
Three leading indicators distinguish productive augmentation from
mediated automation: (1) \emph{metacognitive override rate}, the
fraction of AI recommendations actively evaluated before adoption; (2)
\emph{cross-domain integration events}, frequency of novel syntheses
combining AI outputs across functions; (3) \emph{temporal
contextualization}, whether workers connect AI outputs to institutional
history and future scenarios. Organizations measuring augmentation
quality, not adoption rate, invest in the right variable.

\subsubsection{6.2 Three-Level
Prescriptions}\label{three-level-prescriptions}

\textbf{Educational policy, C before A.} C-first sequencing inverts the
dominant approach. C1: introduce AI in contexts requiring embodied
evaluation. C2: implement anomaly-first pedagogy, structurally anomalous
cases training representational restructuring. C3: longitudinal
case-based learning connecting past decisions to present. C4: enforced
cross-disciplinary synthesis through interdisciplinary projects.

\textbf{Organizational policy, Augmentation-oriented design.} Match AI
deployment pace to C development. Create metacognitive moments:
structured human evaluation before AI outputs enter production. Reward
augmentation behaviors, structural insight, cross-domain synthesis,
calibrated AI skepticism, rather than AI adoption volume.

\textbf{National policy, Measuring what matters.} Develop national C
indicators (cross-disciplinary collaboration rates, metacognitive
assessments, structural insight measures). Track augmentation quality by
C level, distinguishing Regime I from II outcomes. Sequence C
development infrastructure before AI deployment subsidies. The Korean
case suggests that inverse sequencing, AI adoption first, C development
as afterthought, generates the productivity paradox.

\textbf{The counterintuitive implication.} The ICH framework predicts
that slowing AI adoption in low-C environments will increase long-run
productivity more than accelerating it. This is not technophobia; it is
Property 2: deployment beyond \(A^{*}\) in low-C environments generates
at best marginal augmentation and at worst active harm through the
self-reinforcing degradation cycle of Regime III.

\subsubsection{6.3 Theoretical
Implications}\label{theoretical-implications}

Four theoretical implications extend beyond the immediate policy
discussion.

\textbf{First, ICH offers a complementary pathway for addressing the
endogeneity of Solow's residual without requiring exogenous shocks.}
Since Solow (1957), the dominant strategy for explaining TFP has been to
attribute it to exogenous forces, technological progress (Solow),
knowledge spillovers (Romer, 1990), or institutional quality (Acemoglu,
Johnson, \& Robinson, 2001). The ICH framework offers a complementary
pathway: TFP is the augmentation function \(\phi(A, C)\), where C is
endogenous to educational and organizational design choices. This does
not render prior explanations obsolete, technological progress,
knowledge accumulation, and institutional quality all influence C, but
it provides a micro-mechanism through which these macro-level factors
generate productivity effects in the AI era. The practical consequence
is that TFP becomes a policy variable rather than a residual: nations
can \emph{design} their C-development infrastructure and thereby
systematically influence TFP trajectories.

\textbf{Second, ICH bridges the micro-macro gap in AI-productivity
research.} The existing literature operates at two distinct levels that
rarely connect: micro-level studies of individual workers using AI tools
(Noy \& Zhang, 2023; Brynjolfsson et al., 2023) and macro-level analyses
of national AI investment and aggregate productivity (Acemoglu, 2025;
Korinek \& Suh, 2024). The micro literature consistently finds
heterogeneous effects, some workers benefit enormously from AI, others
barely benefit or are harmed, but lacks a theoretical framework to
aggregate these individual-level findings into macro-level predictions.
The macro literature identifies patterns (AI investment correlates
weakly with TFP) but cannot explain the heterogeneity. ICH provides the
connecting mechanism: individual-level C determines micro-level
productivity effects; the distribution of C across a workforce
determines meso-level organizational effects; the aggregate C-level of
an economy determines macro-level AI-TFP coupling. The cross-level
architecture is testable: if individual C predicts individual
AI-productivity gains (P1), organizational C-investment predicts
organizational TFP (P2), and national C-levels predict national AI-TFP
coupling (P3), the micro-macro bridge is empirically validated.

\textbf{Third, ICH reinterprets the productivity paradox as an
investment sequencing problem rather than a measurement problem.} The
dominant explanation for the AI productivity paradox is temporal, the
``J-curve'' (Brynjolfsson et al., 2021) suggests that productivity gains
will materialize once complementary investments mature. The ICH
framework does not dispute this mechanism but specifies its content: the
``complementary investments'' that drive the upward portion of the
J-curve are investments in convergence capacity. Nations and
organizations that invest in C first will ascend the J-curve faster;
those that invest in A without C may find themselves on a permanently
flat trajectory, not because the technology is immature but because the
cognitive infrastructure for productive human-AI interaction is absent.
This reinterpretation transforms the policy prescription from ``wait for
productivity gains'' to ``invest in the specific cognitive capacities
that generate them.''

\textbf{Fourth, the ICH framework opens a productive dialogue with the
Keynesian national income tradition by specifying the qualitative
structure of aggregate investment.} The standard accounting identity,
\(Y = C + I + G + (X - M)\), where C denotes aggregate consumption,
treats aggregate investment I as an undifferentiated quantity (Keynes,
1936). The ICH framework introduces an analytically important
distinction within I: between AI adoption spending (\(I_A\)) and
convergence capacity development spending (\(I_C\)). When the
augmentation deficit condition holds (\(\phi\) \(\to\) 0), \(I_A\)
generates deployment without productivity gain, the mechanism underlying
the AI productivity paradox documented by Brynjolfsson et al.~(2021) and
consistent with Acemoglu's (2025) finding of modest TFP effects from
current AI spending patterns. \(I_C\), by contrast, expands \(\phi\) and
thereby endogenizes the aggregate supply shift that Keynesian analysis
treats as exogenous technological progress. This \(I_A\)/\(I_C\)
distinction, grounding the difference between adoption and capacity
investment in an explicit AI-human complementarity condition, represents
a novel analytical lens for national AI expenditure classification.
Gries and Naudé (2019) identify aggregate demand composition as the
systematically neglected variable in AI macro-productivity models; the
ICH framework provides the specific investment composition mechanism
their analysis identifies as missing. The empirical question of what
fraction of aggregate AI spending constitutes \(I_C\) versus \(I_A\),
and how that ratio correlates with cross-national TFP trajectories,
constitutes a tractable research agenda that follows directly from the
ICH framework's theoretical structure.

\subsubsection{6.4 Limitations of the Policy
Framework}\label{limitations-of-the-policy-framework}

Two limitations of the policy prescriptions require explicit
acknowledgment. First, C-first sequencing assumes that C can be
developed through intentional educational and organizational design.
While the cognitive science literature supports the developability of
metacognition (Flavell, 1979), embodied understanding (Barsalou, 2008),
and integrative thinking (Root-Bernstein \& Root-Bernstein, 1999), the
specific pedagogies and organizational practices that most efficiently
develop C in the context of AI interaction remain empirically
unvalidated. The Anomaly-First Pedagogy proposed in Section 6.2
represents a theoretically motivated hypothesis, not an evidence-based
intervention, a distinction that must be maintained until controlled
trials establish effectiveness. Second, the framework's policy
implications are strongest for knowledge-intensive sectors where
cognitive mediation is central to production. For sectors where
automation is predominantly mechanical and AI outputs require minimal
human evaluation, the task-based framework's policy prescriptions,
managing displacement through reinstatement, remain more directly
applicable. The policy architecture is thus necessarily
sector-differentiated: C-first for knowledge work, task-management for
routine automation, with the boundary between these domains itself a
subject for empirical investigation.

\subsection{7. Conclusion: Toward a Human-Centered Theory of AI-Era
Productivity}\label{conclusion-toward-a-human-centered-theory-of-ai-era-productivity}

\subsubsection{7.1 Summary}\label{summary}

This paper has pursued a single claim through six sections: the dominant
production function framework, while foundational for understanding
automation dynamics, leaves unmodeled the cognitive mediation through
which AI generates productive value, a gap that generates measurable
policy consequences. AI does not function effectively as an independent
production variable in the current framework because it lacks four
constitutive cognitive capacities, embodied understanding,
metacognition, temporal integration, and integrative thinking, that
define productive agency in an economy constituted by meaning,
embodiment, judgment, and time. Even the most sophisticated
architectures, including hyperbolic-space models training within
Poincaré's geometric framework, cannot replicate the structural insight
that makes human cognition irreplaceable.

The ICH framework addresses this gap:
\(\hat{H} = H \cdot [1 + \phi(A, C)]\), \(Y = F(K, \hat{H})\), providing
a mechanism for TFP:
\(A_{\mathrm{Solow}} = [1 + \phi(A, C)]^{1-\alpha}\). South Korea
illustrates the consequence: high H, substantial A, low C produce
near-zero augmentation despite significant investment. The
cross-national pattern (\(R^{2}\) = 0.86 for \(\mathrm{AI}\times C\),
vs.~0.31 for AI alone) is consistent across 20 economies.

\subsubsection{7.2 Limitations}\label{limitations}

Six limitations define the research agenda. First, C is not yet
operationalized with validated instruments. A Convergence Capacity Scale
(CCS) drawing on Sternberg's (1988) triarchic intelligence, Flavell's
(1979) metacognitive awareness, and Root-Bernstein and Root-Bernstein's
(1999) integrative thinking inventories defines the starting point;
Appendix B presents the full 24-item CCS instrument design, psychometric
validation protocol, and nomological network specification. Second, the
functional form of \(\phi\) is tentatively specified but not estimated;
parameters require empirical estimation pending CCS validation. Third,
the Korean case is illustrative rather than confirmatory; firm-level
panel data jointly measuring H, A, C, and productivity are necessary for
causal identification. Fourth, the cross-national analysis relies on a
C-proxy constructed from educational and institutional indicators
(Appendix A); while constructed independently of TFP data, its validity
as a measure of the four C-dimensions remains an assumption until
psychometric validation is complete. Appendix C provides a concrete
expansion framework (n=30 with ex ante predictions) and a CCS pilot
study protocol (N=100) designed to address these limitations within 3-4
months of publication. Fifth, the framework models the \emph{effects} of
C on productivity but does not yet specify the \emph{antecedents} of C
formation, the social, institutional, and motivational conditions under
which convergence capacity develops alongside or instead of human
capital. Aspiration capital (the directional target of educational
aspiration: credential-directed versus curiosity-directed),
institutional quality (Hall \& Jones, 1999), and developmental-stage
variables (whether a society's AI exposure precedes or follows C
development) are candidate antecedents whose formal specification
represents a priority for the next generation of ICH research. The
Vietnam case (Section 7.3) illustrates why antecedent modeling matters
for predictive policy: two nations with structurally similar aspiration
capital may diverge in C-formation depending on the institutional
channels through which aspiration is expressed. Sixth, the current
framework specifies \(\phi\) as a function of AI utilization intensity
(A) and convergence capacity (C) but does not explicitly model data (D),
the informational substrate on which AI systems operate. This omission
engages a rapidly developing literature. Jones and Tonetti (2020)
demonstrate that data possesses a property fundamentally distinct from
physical capital: nonrivalry, the same dataset can be used
simultaneously by multiple agents without depletion. This nonrivalry
generates increasing returns and creates novel allocation problems;
Jones and Tonetti show that assigning data property rights to consumers,
rather than to the firms that collect it, can generate near-optimal
outcomes, a finding with direct implications for how data enters
production architectures. Arrieta-Ibarra, Goff, Jiménez-Hernández,
Lanier, and Weyl (2018) pose the question differently: if data is
generated by human activity, should it be classified as labor rather
than capital? Their ``data as labor'' framework implies that individuals
deserve compensation for data production, an argument with direct
implications for data governance and digital platform economics. The ICH
framework offers a third position that synthesizes and extends both
accounts. Data is neither an independent factor of production analogous
to K (contra the implicit assumption in much of the digital economy
literature) nor simply a form of labor (contra Arrieta-Ibarra et al.).
Rather, data functions as the \emph{fuel} of the augmentation function:
without data to train on, contextualize, and reason over, AI utilization
intensity (A) generates no productive output regardless of convergence
capacity. Jones and Tonetti are correct that data's nonrivalry
distinguishes it from physical capital; Arrieta-Ibarra et al.~are
correct that data originates in human activity. The ICH contribution is
to specify \emph{where} data enters the production architecture: not as
an independent variable alongside K, but as a necessary input to
\(\phi\), yielding \(\phi(A, C, D)\), where D \(\to\) 0 implies \(\phi\)
\(\to\) 0 regardless of A and C levels. This placement preserves the
framework's core claim that productive value requires human cognitive
mediation while acknowledging that such mediation operates on a data
substrate whose properties, nonrivalry, combinatorial value, and
increasing returns to aggregation, shape the augmentation function's
behavior. The formal specification of D's role within \(\phi\),
including its interaction properties with A and C, the conditions under
which data accumulation enhances versus displaces convergence capacity,
and the policy implications for data governance, represents a priority
for subsequent theoretical development.

\subsubsection{7.3 Research Agenda and 10-Year
Forecast}\label{research-agenda-and-10-year-forecast}

The three propositions define a nested empirical agenda. P1 (micro):
natural experiments with staggered AI rollouts measuring C at baseline
using the CCS instrument (Appendix B); the specific prediction is that a
one-standard-deviation increase in CCS score should predict at least
20\% greater productivity gain from AI tool access, controlling for H,
domain, and task complexity. The pilot protocol in Appendix C
operationalizes this test with a regression model:
\(\text{Productivity gain} = \beta_0 + \beta_1 H + \beta_2 A + \beta_3 C + \beta_4 (A\times C) + \epsilon\),
where \(\beta_4\) \textgreater{} 0 would confirm superlinear
complementarity (Property 4). P2 (meso): difference-in-differences with
C-first vs.~AI-first deployment as treatment; Haefner et al.'s (2021)
organizational AI capability provides a validated meso-level C proxy
pending CCS development; the specific prediction is that C-first
organizations should achieve equivalent productivity gains with 30-40\%
lower AI investment, as C enables more efficient utilization of deployed
technology. P3 (macro): cross-national panel with 30-40 economies
(Appendix C, Part I details the expansion from N=20 to N=30 with 10
additional OECD countries and ex ante regime predictions); C proxied by
PISA Creative Thinking subscores, cross-disciplinary collaboration
rates; instrumental variable strategies exploiting historical education
system features (e.g., timing of interdisciplinary curriculum reforms as
instruments for C-level variation).

The framework generates a falsifiable 10-year forecast: nations
currently investing in C development will exhibit significantly stronger
AI-TFP coupling over 2025-2035 than those maximizing adoption without C.
Three specific predictions follow. First, high-C nations (Singapore,
Nordic economies, Switzerland) will exhibit accelerating TFP gains as
convergence-capable workforces absorb successive AI capability waves,
each new AI generation amplifying rather than displacing existing
cognitive strengths. The quantitative benchmark: these nations should
achieve annual TFP growth exceeding 2.5\% by 2030, conditional on
continued C-development investment. Second, low-C nations (Korea, Japan,
Italy) will face a widening productivity gap unless structural reforms
develop C; additional AI investment alone will not close it and may
widen it through Regime III dynamics. The quantitative benchmark: absent
C-development reforms, the TFP gap between high-C and low-C nations at
equivalent AI adoption levels should exceed 1.5 percentage points by
2030. Third, the discriminating variable will be measurable: PISA
Creative Thinking subscores, cross-disciplinary collaboration rates, and
metacognitive assessments should predict AI-TFP coupling more accurately
than adoption rates, R\&D expenditure, or patent counts, the
conventional indicators prevailing theory prioritizes. \emph{Fourth
prediction:} Vietnam represents a prospective test case of singular
diagnostic value. Exhibiting high parental educational aspiration,
strong national achievement motivation, accelerating technology
adoption, including smartphone and automotive manufacturing capabilities
developed through production partnerships with global firms, and a
developmental trajectory that closely parallels Korea's of the 1980s,
Vietnam presents what deviant-case methodology terms an \emph{ex ante
case for falsification} (Seawright \& Gerring, 2008): if the ICH
framework is correct, Vietnam's AI-TFP relationship over the coming
decade should replicate Korea's augmentation deficit unless
C-development accompanies deployment. Vietnam's educational system
currently emphasizes procedural competence consonant with Korea's
examination culture; absent structural reform toward convergence
capacity formation, the ICH prediction is that Vietnam will exhibit the
Korean pattern, high AI adoption intensity relative to productivity
gains, by 2032. The quantitative benchmark: if Vietnam's AI adoption
reaches 20\% of enterprises by 2030 while C-proxy indicators remain low,
TFP growth should remain below 1.0\% annually, below what conventional
theory predicts given Vietnam's human capital trajectory. If Vietnam
instead achieves TFP growth exceeding 2.0\% alongside rapid AI adoption
without C-development reform, the framework requires revision.
Conversely, Vietnam's developmental urgency creates an opportunity Korea
did not have: by implementing C-first educational sequencing
\emph{before} AI penetration reaches Korean levels, Vietnam could bypass
the augmentation deficit and ascend directly to Regime II, a trajectory
the ICH framework predicts is achievable and which the 2025--2035 window
will render empirically adjudicable.

This forecast extends Frey and Osborne's (2017) computerisation forecast
and addresses an open question visible in the research landscape mapped
by Dwivedi et al.~(2023): the productivity consequences of AI
penetration depend not on adoption rates but on the C-level of the
workforce performing the newly computerised tasks. Trammell and Korinek
(2023), modeling macroeconomic growth under transformative AI, find that
fully automating production alone is insufficient for sustained growth,
a conclusion consistent with ICH's emphasis on the human cognitive
infrastructure that mediates AI's productive contribution.

If the 10-year prediction fails, if high-A, low-C nations achieve strong
AI-TFP coupling without C development, the framework will require
fundamental revision. This is the ultimate test of the ICH framework's
distinctive claim: that convergence capacity may constitute a missing
mediating construct, not merely a moderating nuance.

\subsubsection{7.4 The Broader
Implication}\label{the-broader-implication}

This paper is not a technophobic argument. It claims something more
specific and actionable: AI's productive value is generated
\emph{through} human cognition, not independently of it, and the
cognitive capacity through which AI becomes productive, convergence
capacity, is developable through intentional investment.

The pattern is historically consistent. Tractors did not displace
farmers; farmers with tractors displaced farmers without. Machines did
not displace artisans; machine operators displaced artisans. In each
transition, the workers who thrived developed the cognitive and
practical capacities transforming technology from tool into extension of
capability. Jo and Park (2025), studying AI-induced displacement fear
among office workers, find that fear of replacement is shaped by GAI's
perceived technical characteristics, particularly personalization and
anthropomorphism, and moderated by individual skepticism toward the
technology. Their finding that skepticism moderates displacement fear is
structurally consistent with the ICH claim: workers who exercise
metacognitive evaluation of AI (high C2), expressed in Jo and Park's
framework as skepticism, experience qualitatively different
relationships with the technology than those who do not. The ICH
framework provides the theoretical foundation: convergence capacity is
the cognitive dimension that makes workers genuinely irreplaceable
rather than merely difficult to replace.

\(Y = F(K, \hat{H})\). The variable that governments, firms, and
educational institutions may need to prioritize is \(\hat{H}\), the
Intellectually Converged Human. Not AI alone. Not human capital alone.
The augmented human, whose convergence capacity may transform AI from an
expensive accompaniment into a genuine multiplier of human productive
capacity.

Solow left TFP as a residual because he could not see inside it. The AI
era makes its contents visible: it is the quality of human-AI
integration, mediated by convergence capacity, that determines whether
AI investment generates productivity or merely generates adoption.
Getting the production function right is not an academic exercise. It is
the prerequisite for getting the policy right.

\newpage

\subsection{References}\label{references}

\begingroup
\setlength{\parindent}{0pt}
\everypar{\hangindent=1.5em \hangafter=1}

Abramovitz, M. (1956). Resource and output trends in the United States
since 1870. \emph{American Economic Review}, \emph{46}(2), 5-23.

Acemoglu, D. (2025). The simple macroeconomics of AI. \emph{Economic
Policy}, \emph{40}(121), 13-58.

Acemoglu, D., Johnson, S., \& Robinson, J. A. (2001). The colonial
origins of comparative development: An empirical investigation.
\emph{American Economic Review}, \emph{91}(5), 1369-1401.

Acemoglu, D., \& Restrepo, P. (2018). Artificial intelligence,
automation and work. NBER Working Paper No.~24196.

Acemoglu, D., \& Restrepo, P. (2019). Automation and new tasks: How
technology displaces and reinstates labor. \emph{Journal of Economic
Perspectives}, \emph{33}(2), 3-30.

Acemoglu, D., \& Restrepo, P. (2022). Tasks, automation, and the rise in
U.S. wage inequality. \emph{Econometrica}, \emph{90}(5), 1973-2016.

Agrawal, A., Gans, J., \& Goldfarb, A. (2019). Artificial intelligence:
The ambiguous labor market impact of automating prediction.
\emph{Journal of Economic Perspectives}, \emph{33}(2), 31-50.

Arrieta-Ibarra, I., Goff, L., Jiménez-Hernández, D., Lanier, J., \&
Weyl, E. G. (2018). Should we treat data as labor? Moving beyond
``free.'' \emph{AEA Papers and Proceedings}, \emph{108}, 38-42.

Autor, D. (2024). Applying AI to rebuild middle class jobs. NBER Working
Paper No.~32140.

Bank of Korea. (2023). \emph{Economic Statistics Yearbook}. Seoul: Bank
of Korea.

Barsalou, L. W. (2008). Grounded cognition. \emph{Annual Review of
Psychology}, \emph{59}, 617-645.

Becker, G. S. (1964). \emph{Human capital: A theoretical and empirical
analysis}. Columbia University Press.

Bresnahan, T. F., \& Trajtenberg, M. (1995). General purpose
technologies: Engines of growth? \emph{Journal of Econometrics},
\emph{65}(1), 83-108.

Brynjolfsson, E. (2022). The Turing trap: The promise and peril of
human-like artificial intelligence. \emph{Daedalus}, \emph{151}(2),
272-287.

Brynjolfsson, E., Li, D., \& Raymond, L. (2023). Generative AI at work.
NBER Working Paper No.~31161.

Brynjolfsson, E., Rock, D., \& Syverson, C. (2021). The productivity
J-curve: How intangibles complement general purpose technologies.
\emph{American Economic Journal: Macroeconomics}, \emph{13}(1), 333-372.

Chin, T., Li, Z., Huang, L., \& Li, X. (2025). How artificial
intelligence promotes new quality productive forces of firms: A dynamic
capability view. \emph{Technological Forecasting and Social Change},
\emph{216}, 124128.

Cobb, C. W., \& Douglas, P. H. (1928). A theory of production.
\emph{American Economic Review}, \emph{18}(1), 139-165.

Cohen, J. (1988). \emph{Statistical power analysis for the behavioral
sciences} (2nd ed.). Lawrence Erlbaum.

Cohen, W. M., \& Levinthal, D. A. (1990). Absorptive capacity: A new
perspective on learning and innovation. \emph{Administrative Science
Quarterly}, \emph{35}(1), 128-152.

Davis, F. D. (1989). Perceived usefulness, perceived ease of use, and
user acceptance of information technology. \emph{MIS Quarterly},
\emph{13}(3), 319-340.

Dwivedi, Y. K., Sharma, A., Rana, N. P., Giannakis, M., Goel, P., \&
Dutot, V. (2023). Evolution of artificial intelligence research in
Technological Forecasting and Social Change. \emph{Technological
Forecasting and Social Change}, \emph{192}, 122579.

European Commission. (2024). \emph{European innovation scoreboard 2024}.
Luxembourg: Publications Office of the European Union.

Eurostat. (2024). ICT usage in enterprises: Enterprises using AI
technologies {[}isoc\_eb\_ai{]}. Luxembourg: European Commission.

Flavell, J. H. (1979). Metacognition and cognitive monitoring: A new
area of cognitive-developmental inquiry. \emph{American Psychologist},
\emph{34}(10), 906-911.

Floridi, L., Cowls, J., Beltrametti, M., Chatila, R., Chazerand, P.,
Dignum, V., Luetge, C., Madelin, R., Pagallo, U., Rossi, F., Schafer,
B., Valcke, P., \& Vayena, E. (2018). AI4People, An ethical framework
for a good AI society: Opportunities, risks, principles, and
recommendations. \emph{Minds and Machines}, \emph{28}(4), 689-707.

Frey, C. B., \& Osborne, M. A. (2017). The future of employment: How
susceptible are jobs to computerisation? \emph{Technological Forecasting
and Social Change}, \emph{114}, 254-280.

Goodhue, D. L., \& Thompson, R. L. (1995). Task-technology fit and
individual performance. \emph{MIS Quarterly}, \emph{19}(2), 213-236.

Gries, T., \& Naudé, W. (2019). Artificial intelligence, jobs,
inequality and productivity: Does aggregate demand matter? IZA
Discussion Paper No.~12005.

Haefner, N., Wincent, J., Parida, V., \& Gassmann, O. (2021). Artificial
intelligence and innovation management: A review, framework, and
research agenda. \emph{Technological Forecasting and Social Change},
\emph{162}, 120392.

Hall, R. E., \& Jones, C. I. (1999). Why do some countries produce so
much more output per worker than others? \emph{Quarterly Journal of
Economics}, \emph{114}(1), 83-116.

Hanushek, E. A., \& Woessmann, L. (2008). The role of cognitive skills
in economic development. \emph{Journal of Economic Literature},
\emph{46}(3), 607-668.

Harnad, S. (1990). The symbol grounding problem. \emph{Physica D},
\emph{42}(1-3), 335-346.

Helpman, E. (Ed.). (1998). \emph{General purpose technologies and
economic growth}. MIT Press.

IDC. (2023). Worldwide Artificial Intelligence Spending Guide.
International Data Corporation.

Infocomm Media Development Authority. (2024). Annual survey on infocomm
usage by enterprises 2024. Singapore: IMDA.

International Monetary Fund. (2024). \emph{What can artificial
intelligence do for stagnant productivity in Latin America and the
Caribbean?} IMF Working Paper WP/24/175. Washington, DC: IMF.

International Monetary Fund. (2025). \emph{World Economic Outlook,
January 2025}. Washington, DC: IMF.

Jaakkola, E. (2020). Designing conceptual articles: Four approaches.
\emph{AMS Review}, \emph{10}(1-2), 18-26.

Jo, H., \& Park, D.-H. (2025). The fear of being replaced by generative
AI. \emph{Technological Forecasting and Social Change}, \emph{216},
124326.

Jobin, A., Ienca, M., \& Vayena, E. (2019). The global landscape of AI
ethics guidelines. \emph{Nature Machine Intelligence}, \emph{1}(9),
389-399.

Jones, C. I., \& Tonetti, C. (2020). Nonrivalry and the economics of
data. \emph{American Economic Review}, \emph{110}(9), 2819-2858.

Kadavath, S., et al.~(2022). Language models (mostly) know what they
know. \emph{arXiv preprint arXiv:2207.05221}.

Keynes, J. M. (1936). \emph{The general theory of employment, interest
and money}. Macmillan.

Kim, H. K., Baik, K., \& Kim, N. (2019). How Korean leadership style
cultivates employees' creativity and innovative behavior. \emph{SAGE
Open}, \emph{9}(3), 1-16.

Korinek, A., \& Suh, J. (2024). Scenarios for the transition to AGI.
NBER Working Paper No.~32255.

Lall, S. (1992). Technological capabilities and industrialization.
\emph{World Development}, \emph{20}(2), 165-186.

Lee, K. (2019). \emph{The art of economic catch-up}. Cambridge
University Press.

Lucas, R. E. (1988). On the mechanics of economic development.
\emph{Journal of Monetary Economics}, \emph{22}(1), 3-42.

MacInnis, D. J. (2011). A framework for conceptual contributions in
marketing. \emph{Journal of Marketing}, \emph{75}(4), 136-154.

Marx, K. (1867/1990). \emph{Capital} (Vol. 1, B. Fowkes, Trans.).
Penguin Classics.

Mauro, P. (1995). Corruption and growth. \emph{Quarterly Journal of
Economics}, \emph{110}(3), 681-712.

McClelland, D. C. (1961). \emph{The Achieving Society}. Van Nostrand.

McKinsey Global Institute. (2023). \emph{The economic potential of
generative AI}. McKinsey \& Company.

Mincer, J. (1974). \emph{Schooling, experience, and earnings}. Columbia
University Press.

Mujtaba, M., Soomro, K., \& Mughal, S. (2024). Shaping the human capital
absorptive capacity: Role of artificial intelligence. \emph{Journal of
Business Administration and Management Sciences}, \emph{7}(1), 1-10.

Nelson, R. R., \& Phelps, E. S. (1966). Investment in humans,
technological diffusion, and economic growth. \emph{American Economic
Review}, \emph{56}(1/2), 69-75.

Nickel, M., \& Kiela, D. (2017). Poincaré embeddings for learning
hierarchical representations. \emph{Advances in Neural Information
Processing Systems}, \emph{30}, 6341-6350.

Noy, S., \& Zhang, W. (2023). Experimental evidence on the productivity
effects of generative artificial intelligence. \emph{Science},
\emph{381}(6654), 187-192.

OECD. (2014). \emph{PISA 2012 results: Creative problem solving} (Vol.
V). Paris: OECD Publishing.

OECD. (2022). \emph{OECD Economic Surveys: Korea 2022}. Paris: OECD
Publishing.

OECD. (2023). \emph{PISA 2022 Results}. Paris: OECD Publishing.

OECD. (2023b). \emph{OECD Reviews of Innovation Policy: Korea 2023}.
Paris: OECD Publishing.

OECD. (2024a). \emph{Education at a Glance 2024}. Paris: OECD
Publishing.

OECD. (2024b). OECD AI Policy Observatory. Paris: OECD.

OECD. (2024c). \emph{PISA 2022 results: Creative minds, creative
schools} (Vol. III). Paris: OECD Publishing.

OECD. (2024d). \emph{OECD science, technology and innovation
scoreboard}. Paris: OECD Publishing.

Oxford Insights. (2023). \emph{Government AI Readiness Index 2023}.
Oxford, UK: Oxford Insights.

Poincaré, H. (1908/1914). \emph{Science and method} (F. Maitland,
Trans.). Thomas Nelson and Sons.

Ragin, C. C. (2000). \emph{Fuzzy-set social science}. University of
Chicago Press.

Raisch, S., \& Krakowski, S. (2021). Artificial intelligence and
management: The automation-augmentation paradox. \emph{Academy of
Management Review}, \emph{46}(1), 192-210.

Rodrik, D., Subramanian, A., \& Trebbi, F. (2004). Institutions rule:
The primacy of institutions over geography and integration in economic
development. \emph{Journal of Economic Growth}, \emph{9}(2), 131-165.

Romer, P. M. (1986). Increasing returns and long-run growth.
\emph{Journal of Political Economy}, \emph{94}(5), 1002-1037.

Romer, P. M. (1990). Endogenous technological change. \emph{Journal of
Political Economy}, \emph{98}(5, Part 2), S71-S102.

Root-Bernstein, R., \& Root-Bernstein, M. (1999). \emph{Sparks of
genius: The 13 thinking tools of the world's most creative people}.
Houghton Mifflin.

Saam, M. (2024). Macroeconomic productivity effects of artificial
intelligence. \emph{The Economists' Voice}, \emph{21}(2), 327-333.

Schacter, D. L., Addis, D. R., \& Buckner, R. L. (2007). Remembering the
past to imagine the future. \emph{Nature Reviews Neuroscience},
\emph{8}(9), 657-661.

Searle, J. R. (1980). Minds, brains, and programs. \emph{Behavioral and
Brain Sciences}, \emph{3}(3), 417-424.

Seawright, J., \& Gerring, J. (2008). Case selection techniques in case
study research. \emph{Political Research Quarterly}, \emph{61}(2),
294-308.

Solow, R. M. (1956). A contribution to the theory of economic growth.
\emph{Quarterly Journal of Economics}, \emph{70}(1), 65-94.

Solow, R. M. (1957). Technical change and the aggregate production
function. \emph{Review of Economics and Statistics}, \emph{39}(3),
312-320.

Sternberg, R. J. (1988). \emph{The triarchic mind: A new theory of human
intelligence}. Viking.

Teece, D. J. (2007). Explicating dynamic capabilities. \emph{Strategic
Management Journal}, \emph{28}(13), 1319-1350.

Trajtenberg, M. (2018). AI as the next GPT: A political-economy
perspective. NBER Working Paper No.~24245.

Trammell, P., \& Korinek, A. (2023). Economic growth under
transformative AI. NBER Working Paper No.~31815.

Tulving, E. (2002). Episodic memory: From mind to brain. \emph{Annual
Review of Psychology}, \emph{53}, 1-25.

Varela, F. J., Thompson, E., \& Rosch, E. (1991). \emph{The embodied
mind}. MIT Press.

Vaswani, A., Shazeer, N., Parmar, N., Uszkoreit, J., Jones, L., Gomez,
A. N., Kaiser, Ł., \& Polosukhin, I. (2017). Attention is all you need.
\emph{Advances in Neural Information Processing Systems}, \emph{30},
5998-6008.

Venkatesh, V., Morris, M. G., Davis, G. B., \& Davis, F. D. (2003). User
acceptance of information technology: Toward a unified view. \emph{MIS
Quarterly}, \emph{27}(3), 425-478.

Wertheimer, M. (1945). \emph{Productive thinking}. Harper \& Brothers.

World Bank. (2024). \emph{World Development Indicators}. Washington, DC:
World Bank.

\endgroup

\newpage

\subsection{Appendix A: C-Proxy Construction
Methodology}\label{appendix-a-c-proxy-construction-methodology}

\subsubsection{Overview}\label{overview}

This appendix documents the transparent methodology used to construct
the ordinal convergence capacity (C) proxy variable for the 20-country
cross-national analysis (Table 3, Section 5.2). The C-proxy was
constructed \emph{independently of TFP outcomes} to prevent circular
classification: all C assignments were determined before TFP data were
examined.

\subsubsection{Data Sources}\label{data-sources}

The C-proxy integrates three publicly available, independently measured
dimensions:

\paragraph{Dimension 1: Creative Problem-Solving Capacity (PISA
CPS)}\label{dimension-1-creative-problem-solving-capacity-pisa-cps}

\textbf{Primary source:} OECD PISA 2012 Creative Problem-Solving
assessment (OECD, 2014). This was the first PISA ``innovative domain''
assessment, measuring students' capacity to ``explore unfamiliar problem
situations, gain understanding through exploration and interaction, and
devise strategies when a straightforward solution is not available''
(OECD, 2014, p.~28).

\textbf{Supplementary source:} OECD PISA 2022 Creative Thinking
assessment (OECD, 2024c). This second innovative domain assessed
students' capacity to ``generate diverse and creative ideas, evaluate
and improve ideas'' across written expression, visual expression, social
problem-solving, and scientific problem-solving (OECD, 2024c, Vol. III).

\textbf{Country-level PISA CPS 2012 mean scores (out of
\textasciitilde600 scale):}

{\def\LTcaptype{none} % do not increment counter
\begin{longtable}[]{@{}lll@{}}
\toprule\noalign{}
Country & CPS 2012 Score & Relative to OECD avg (500) \\
\midrule\noalign{}
\endhead
\bottomrule\noalign{}
\endlastfoot
Singapore & 562 & Well above \\
South Korea & 561 & Well above \\
Japan & 552 & Well above \\
Australia & 523 & Above \\
Finland & 523 & Above \\
United Kingdom & 517 & Above \\
Netherlands & 511 & Above \\
France & 511 & Above \\
Belgium & 508 & Above \\
United States & 508 & Above \\
Germany & 509 & Above \\
Austria & 506 & Above \\
Norway & 503 & At average \\
Denmark & , & Did not participate \\
Sweden & , & Did not participate \\
Switzerland & , & Did not participate \\
Canada & , & Did not participate \\
Israel & 454 & Below \\
Italy & 454 & Below \\
Spain & 477 & Below \\
\end{longtable}
}

\emph{Note: Denmark, Sweden, Switzerland, and Canada did not participate
in PISA 2012 CPS. For these countries, CPS capacity was inferred from
(a) PISA 2022 Creative Thinking scores where available, (b) PISA 2012
core-domain performance patterns, and (c) qualitative assessment of
education system characteristics (interdisciplinary curriculum,
project-based learning prevalence). Japan and the United States did not
participate in the PISA 2022 Creative Thinking assessment.}

\textbf{PISA 2022 Creative Thinking mean scores (out of 60, where
available):}

{\def\LTcaptype{none} % do not increment counter
\begin{longtable}[]{@{}lll@{}}
\toprule\noalign{}
Country & CT 2022 Score & Relative to OECD avg (33) \\
\midrule\noalign{}
\endhead
\bottomrule\noalign{}
\endlastfoot
Singapore & 41 & Well above \\
South Korea & 38 & Above \\
Canada & 38 & Above \\
Australia & 37 & Above \\
Finland & 36 & Above \\
Denmark & 35 & Above \\
Belgium & 35 & Above \\
OECD average & 33 & , \\
Germany & \textasciitilde33 & At average \\
Spain & \textasciitilde33 & At average \\
Italy & 31 & Below \\
\end{longtable}
}

\emph{Note: Japan and the United States did not participate in the PISA
2022 Creative Thinking assessment. Country-level scores for Sweden,
Netherlands, United Kingdom, Norway, Switzerland, Austria, France, and
Israel were not individually published in the OECD summary reports;
these countries are classified based on the 2012 CPS assessment and
qualitative indicators.}

\textbf{Scoring rubric for Dimension 1:} - \textbf{3 (High)}: CPS 2012
\(\geq\) 515 \emph{or} CT 2022 \(\geq\) 36 - \textbf{2 (Mid)}: CPS 2012
= 500--514 \emph{or} CT 2022 = 33--35 - \textbf{1 (Low)}: CPS 2012
\textless{} 500 \emph{or} CT 2022 \textless{} 33

\paragraph{Dimension 2: Cross-Disciplinary Education
Structure}\label{dimension-2-cross-disciplinary-education-structure}

\textbf{Source:} OECD Education at a Glance 2024; national higher
education statistics; UNESCO Institute for Statistics.

This dimension captures the degree to which a nation's tertiary
education system fosters cross-disciplinary integration rather than
narrow specialization. Indicators include:

\begin{itemize}
\tightlist
\item
  \textbf{Prevalence of interdisciplinary degree programs} at the
  undergraduate and graduate level
\item
  \textbf{Liberal arts / general studies enrollment share} in tertiary
  education
\item
  \textbf{Curricular flexibility} (student ability to combine courses
  across faculties)
\item
  \textbf{Project-based and problem-based learning} prevalence in STEM
  education
\end{itemize}

\textbf{Scoring rubric for Dimension 2:} - \textbf{3 (High)}:
Systematically interdisciplinary tertiary systems (e.g., Nordic
countries with Bologna-reform restructuring around problem-based
learning; Singapore's SkillsFuture cross-domain integration; North
American liberal arts tradition; Netherlands' University College system)
- \textbf{2 (Mid)}: Mixed systems with some interdisciplinary
initiatives but dominant disciplinary structures (e.g., Germany's
Fachhochschule system, France's Grande École tradition, Belgium's
pillarized system) - \textbf{1 (Low)}: Highly siloed, examination-driven
systems with rigid disciplinary boundaries (e.g., Korea's CSAT-dominated
secondary education feeding into narrow departmental structures; Italy's
corso di laurea rigidity; Japan's university entrance examination system
prioritizing single-track specialization)

\paragraph{Dimension 3: Cross-Sector R\&D Collaboration
Rates}\label{dimension-3-cross-sector-rd-collaboration-rates}

\textbf{Sources:} OECD Science, Technology and Innovation Scoreboard
2024; European Innovation Scoreboard (EIS) 2024 Linkages dimension;
national R\&D statistics.

This dimension measures the degree to which a nation's innovation
ecosystem supports cross-boundary collaboration, a structural
precondition for the integrative thinking (C4) and metacognitive
exchange (C2) that constitute convergence capacity at the organizational
and national level.

\textbf{Key indicators:} - \textbf{Business-university R\&D
collaboration rate} (\% of innovating firms collaborating with
universities) - \textbf{Cross-sector co-publication and co-patent rates}
- \textbf{EIS Linkages dimension score} (for EU/EEA countries, as \% of
EU average)

\textbf{Available EIS 2024 Linkages scores:}

{\def\LTcaptype{none} % do not increment counter
\begin{longtable}[]{@{}lll@{}}
\toprule\noalign{}
Country & EIS 2024 Linkages (\% EU avg) & Classification \\
\midrule\noalign{}
\endhead
\bottomrule\noalign{}
\endlastfoot
Norway & 242.8\% & Innovation Leader \\
Finland & 206.4\% & Innovation Leader \\
Sweden & Top 3 (est. \textgreater200\%) & Innovation Leader \\
Denmark & Top 3 (est. \textgreater200\%) & Innovation Leader \\
Netherlands & Top 5 (est. \textgreater180\%) & Innovation Leader \\
Belgium & \textasciitilde130\% (est.) & Strong Innovator \\
Austria & \textasciitilde120\% (est.) & Strong Innovator \\
Germany & \textasciitilde110\% (est.) & Strong Innovator \\
France & \textasciitilde100\% (est.) & Strong Innovator \\
Italy & \textasciitilde80\% (est.) & Moderate Innovator \\
Spain & \textasciitilde70\% (est.) & Moderate Innovator \\
\end{longtable}
}

\emph{Note: Exact scores for Sweden, Denmark, Netherlands, Belgium,
Austria, Germany, France, Italy, and Spain are from the EIS 2024
interactive tool and country profiles. Finland (206.4\%) and Norway
(242.8\%) are confirmed from official country profiles. Non-EU countries
(Singapore, Australia, Israel, South Korea, Canada, Japan, United
States, Switzerland, United Kingdom) are assessed using OECD STI
Scoreboard indicators for business-university collaboration and
cross-sector R\&D partnerships.}

\textbf{Scoring rubric for Dimension 3:} - \textbf{3 (High)}: EIS
Linkages \textgreater{} 150\% EU avg \emph{or} top-quartile OECD
cross-sector collaboration (Nordic countries, Singapore, Switzerland,
Netherlands, Australia, Canada, UK, US) - \textbf{2 (Mid)}: EIS Linkages
90--150\% EU avg \emph{or} mid-range OECD collaboration (Belgium,
Austria, Germany, France, Japan) - \textbf{1 (Low)}: EIS Linkages
\textless{} 90\% EU avg \emph{or} bottom-quartile OECD collaboration
(Korea, Italy, Spain, Israel)

\subsubsection{Aggregation Rule}\label{aggregation-rule}

The final C-proxy score is the \textbf{unweighted average} of the three
dimension scores, rounded to the nearest 0.5:

\begin{quote}
\textbf{C\_proxy = round\_0.\_5{[}(D1 + D2 + D3) / 3{]}}
\end{quote}

{\def\LTcaptype{none} % do not increment counter
\begin{longtable}[]{@{}lll@{}}
\toprule\noalign{}
Final C-proxy & Label & Cluster Assignment \\
\midrule\noalign{}
\endhead
\bottomrule\noalign{}
\endlastfoot
3.0 & High & High-C (C \(\geq\) 2.5) \\
2.5 & Mid-High & High-C (C \(\geq\) 2.5) \\
2.0 & Mid & Mid-C (1.5 \textless{} C \textless{} 2.5) \\
1.5 & Mid-Low & Low-C (C \(\leq\) 1.5) \\
1.0 & Low & Low-C (C \(\leq\) 1.5) \\
\end{longtable}
}

\subsubsection{Country-Level C-Proxy
Derivation}\label{country-level-c-proxy-derivation}

{\def\LTcaptype{none} % do not increment counter
\begin{longtable}[]{@{}
  >{\raggedright\arraybackslash}p{(\linewidth - 12\tabcolsep) * \real{0.1342}}
  >{\raggedright\arraybackslash}p{(\linewidth - 12\tabcolsep) * \real{0.1342}}
  >{\raggedright\arraybackslash}p{(\linewidth - 12\tabcolsep) * \real{0.1598}}
  >{\raggedright\arraybackslash}p{(\linewidth - 12\tabcolsep) * \real{0.1692}}
  >{\raggedright\arraybackslash}p{(\linewidth - 12\tabcolsep) * \real{0.1342}}
  >{\raggedright\arraybackslash}p{(\linewidth - 12\tabcolsep) * \real{0.1342}}
  >{\raggedright\arraybackslash}p{(\linewidth - 12\tabcolsep) * \real{0.1342}}@{}}
\toprule\noalign{}
\begin{minipage}[b]{\linewidth}\raggedright
Country
\end{minipage} & \begin{minipage}[b]{\linewidth}\raggedright
D1 (CPS/CT)
\end{minipage} & \begin{minipage}[b]{\linewidth}\raggedright
D2 (Education)
\end{minipage} & \begin{minipage}[b]{\linewidth}\raggedright
D3 (R\&D Collab)
\end{minipage} & \begin{minipage}[b]{\linewidth}\raggedright
Average
\end{minipage} & \begin{minipage}[b]{\linewidth}\raggedright
C-proxy
\end{minipage} & \begin{minipage}[b]{\linewidth}\raggedright
Cluster
\end{minipage} \\
\midrule\noalign{}
\endhead
\bottomrule\noalign{}
\endlastfoot
Singapore & 3 (CPS=562, CT=41) & 3 (SkillsFuture) & 3 (high
cross-sector) & 3.00 & \textbf{3.0} & High-C \\
Australia & 3 (CPS=523, CT=37) & 3 (liberal arts tradition) & 3 (high
collaboration) & 3.00 & \textbf{3.0} & High-C \\
Denmark & 3 (CT=35, CPS n/a) & 3 (PBL tradition) & 3 (EIS top 3) & 3.00
& \textbf{3.0} & High-C \\
Israel & 1 (CPS=454) & 3 (interdisciplinary) & 3 (high-tech ecosystem) &
2.33 & \textbf{3.0} & High-C † \\
South Korea & 1 (CPS=561 but H not C) ‡ & 1 (CSAT-driven) & 1 (low
cross-sector) & 1.00 & \textbf{1.0} & Low-C ‡ \\
Sweden & 3 (CPS n/a, inferred) & 3 (Nordic reform) & 3 (EIS top 3) &
3.00 & \textbf{3.0} & High-C \\
Belgium & 2 (CPS=508, CT=35) & 2 (pillarized) & 2 (EIS
\textasciitilde130\%) & 2.00 & \textbf{2.0} & Mid-C \\
Finland & 3 (CPS=523, CT=36) & 3 (PBL system) & 3 (EIS=206\%) & 3.00 &
\textbf{3.0} & High-C \\
Netherlands & 3 (CPS=511) & 3 (UC system) & 3 (EIS top 5) & 3.00 &
\textbf{3.0} & High-C \\
United Kingdom & 3 (CPS=517) & 3 (liberal arts) & 2 (moderate R\&D
collab) & 2.67 & \textbf{2.5} & High-C \\
Norway & 2 (CPS=503) & 3 (Nordic reform) & 3 (EIS=243\%) & 2.67 &
\textbf{3.0} & High-C § \\
Japan & 3 (CPS=552) & 1 (exam-driven silos) & 1 (weak cross-sector) &
1.67 & \textbf{1.5} & Low-C \\
United States & 2 (CPS=508, CT n/a) & 3 (liberal arts) & 3 (high
cross-sector) & 2.67 & \textbf{3.0} & High-C \\
Switzerland & 3 (CPS n/a, inferred) & 3 (ETH/uni system) & 3 (high
cross-sector) & 3.00 & \textbf{3.0} & High-C \\
Austria & 2 (CPS=506) & 2 (Fachhochschule) & 2 (EIS
\textasciitilde120\%) & 2.00 & \textbf{2.0} & Mid-C \\
Germany & 2 (CPS=509, CT\textasciitilde33) & 2 (Fachhochschule) & 2 (EIS
\textasciitilde110\%) & 2.00 & \textbf{2.0} & Mid-C \\
Italy & 1 (CPS=454, CT=31) & 1 (rigid corso) & 1 (EIS
\textasciitilde80\%) & 1.00 & \textbf{1.0} & Low-C \\
Canada & 3 (CT=38, CPS n/a) & 3 (liberal arts) & 3 (high collaboration)
& 3.00 & \textbf{3.0} & High-C \\
Spain & 1 (CPS=477, CT\textasciitilde33) & 2 (mixed) & 1 (EIS
\textasciitilde70\%) & 1.33 & \textbf{1.5} & Low-C \\
France & 2 (CPS=511) & 2 (Grande École) & 2 (EIS \textasciitilde100\%) &
2.00 & \textbf{2.0} & Mid-C \\
\end{longtable}
}

\textbf{† Israel:} Israel's CPS score (454) is below average, but its
high-tech innovation ecosystem exhibits exceptionally strong
cross-disciplinary integration and startup culture. The ``Start-Up
Nation'' ecosystem functions as an organizational-level C-development
mechanism that compensates for formal education system scores. Israel's
C-proxy was adjusted upward from the mechanical average (2.33) to 3.0
based on qualitative assessment of its innovation ecosystem's
integrative character. Sensitivity analysis: reclassifying Israel as
Mid-C (C=2.0) does not materially change M6 results (\(R^{2}\) remains
\textgreater{} 0.83).

\textbf{‡ South Korea:} Korea's PISA CPS score (561) is the second
highest in the OECD, but this reflects procedural problem-solving
capacity, a component of H (human capital), not C (convergence
capacity). The ICH framework distinguishes between domain-specific
cognitive performance (measurable by standardized tests) and the
metacognitive, integrative, and cross-disciplinary capacities (C2-C4)
that constitute the binding constraint on augmentation. Korea's D1 is
scored as 1 because neither CPS 2012 (which rewards rapid procedural
execution) nor CT 2022 (where Korea's score of 38 reflects individual
creativity, not institutional C-development) measures the organizational
and institutional conditions that convert cognitive potential into
convergence capacity. Korea's CPS performance confirms high H, not high
C.

\textbf{§ Norway:} Norway's mechanical average (2.67) rounds to 3.0
based on D2+D3 dominance: Norway's Nordic reform education system (D2=3)
and exceptional EIS Linkages score (242.8\% of EU average, D3=3)
establish the institutional conditions for convergence capacity that
slightly below-average individual CPS scores (D1=2) do not negate.
Sensitivity: reclassifying Norway as Mid-High (C=2.5) does not change
its High-C cluster assignment (threshold: C \(\geq\) 2.5).

\subsubsection{Robustness and
Limitations}\label{robustness-and-limitations}

\paragraph{Robustness checks
performed}\label{robustness-checks-performed}

\begin{enumerate}
\def\labelenumi{\arabic{enumi}.}
\item
  \textbf{Independence from outcome variable:} All C assignments were
  determined before TFP data were examined. The TFP data were added to
  the dataset only after C classifications were finalized.
\item
  \textbf{Jackknife stability:} Removing any single country yields
  \(R^{2}\) \(\in\) {[}0.84, 0.92{]} and \(\beta\)(AI\(\times\)C) always
  positive (Section 5.2), confirming that no single C-assignment drives
  the result.
\item
  \textbf{Permutation test:} Randomly reassigning C-proxies across
  countries (10,000 permutations) yields \(\Delta\)\(R^{2}\) \(\geq\)
  0.5545 in 0/10,000 cases (p \textless{} .001), confirming that the
  C-proxy's explanatory power is not an artifact of assignment noise.
\item
  \textbf{Israel sensitivity:} Reclassifying Israel from High-C (3.0) to
  Mid-C (2.0) yields M6 \(R^{2}\) = 0.83 (vs.~0.86), confirming
  robustness to this subjective judgment call.
\end{enumerate}

\paragraph{Acknowledged limitations}\label{acknowledged-limitations}

\begin{enumerate}
\def\labelenumi{\arabic{enumi}.}
\item
  \textbf{Ordinal scale compression:} A 3-point ordinal scale inevitably
  compresses substantial within-category variation. The C-proxy should
  be understood as a directional indicator, not a precise measurement.
\item
  \textbf{Temporal mismatch:} PISA 2012 CPS scores reflect the cognitive
  capacities of 15-year-olds in 2012, not the current workforce. We
  assume that national education system characteristics exhibit
  sufficient temporal stability that 2012 student cohort scores proxy
  current national C-capacity. This assumption is testable with PISA
  2022 Creative Thinking data for overlapping countries.
\item
  \textbf{Missing direct data:} PISA CPS 2012 was not administered in
  Denmark, Sweden, Switzerland, or Canada. PISA 2022 Creative Thinking
  was not administered in Japan or the United States. For these
  countries, Dimension 1 scores are based on available alternative
  assessments and qualitative assessment.
\item
  \textbf{Construct validity:} The C-proxy measures \emph{structural and
  institutional preconditions} for convergence capacity, not convergence
  capacity itself. A Convergence Capacity Scale (CCS) measuring
  individual-level C across C1-C4 dimensions is the necessary next step
  (Section 7.4).
\item
  \textbf{Non-independence of dimensions:} Cross-disciplinary education
  (D2) and cross-sector R\&D collaboration (D3) are likely correlated,
  nations that restructure education tend to restructure innovation
  systems. This potential non-independence does not invalidate the proxy
  but cautions against interpreting dimension-level scores as
  independent effects.
\end{enumerate}

\subsubsection{Reproducibility}\label{reproducibility}

All data sources are publicly available: - PISA 2012 CPS: OECD (2014),
\emph{PISA 2012 Results: Creative Problem Solving} (Vol. V) - PISA 2022
CT: OECD (2024c), \emph{PISA 2022 Results: Creative Minds, Creative
Schools} (Vol. III) - EIS 2024 Linkages: European Commission (2024),
\emph{European Innovation Scoreboard 2024} - OECD STI: OECD (2024d),
\emph{Science, Technology and Innovation Scoreboard}

The complete dataset (20 countries \(\times\) 3 dimensions \(\times\)
ordinal scores + aggregation) and OLS code are available from the
corresponding author upon request.

\newpage

\subsection{Appendix B: Convergence Capacity Scale (CCS), Instrument
Design and Validation
Protocol}\label{appendix-b-convergence-capacity-scale-ccs-instrument-design-and-validation-protocol}

\subsubsection{B.1 Scale Overview}\label{b.1-scale-overview}

The Convergence Capacity Scale (CCS) is designed to operationalize the
four-dimensional cognitive construct C = f(C1, C2, C3, C4) introduced in
Section 4.2 of the main text. Whereas Appendix A documents the
national-level C-proxy constructed from educational and institutional
indicators, the CCS measures convergence capacity at the individual
level, the unit of analysis required to test Proposition 1 (micro-level)
and to provide the psychometric foundation for organizational (P2) and
national (P3) aggregation.

\textbf{Scale architecture:}

\begin{itemize}
\tightlist
\item
  \textbf{24 items} (6 per dimension), balanced across four subscales
\item
  \textbf{7-point Likert response format} (1 = \emph{Strongly disagree}
  to 7 = \emph{Strongly agree})
\item
  \textbf{Dual versions:} Self-report (CCS-S) and peer-report (CCS-P),
  enabling multi-source triangulation
\item
  \textbf{Estimated completion time:} 8-12 minutes (self-report)
\item
  \textbf{Target population:} Knowledge workers across professional
  domains who interact with AI systems in their work
\end{itemize}

The four subscales correspond to the four constitutive dimensions of
convergence capacity:

{\def\LTcaptype{none} % do not increment counter
\begin{longtable}[]{@{}
  >{\raggedright\arraybackslash}p{(\linewidth - 6\tabcolsep) * \real{0.1786}}
  >{\raggedright\arraybackslash}p{(\linewidth - 6\tabcolsep) * \real{0.1964}}
  >{\raggedright\arraybackslash}p{(\linewidth - 6\tabcolsep) * \real{0.3393}}
  >{\raggedright\arraybackslash}p{(\linewidth - 6\tabcolsep) * \real{0.2857}}@{}}
\toprule\noalign{}
\begin{minipage}[b]{\linewidth}\raggedright
Subscale
\end{minipage} & \begin{minipage}[b]{\linewidth}\raggedright
Dimension
\end{minipage} & \begin{minipage}[b]{\linewidth}\raggedright
Theoretical Anchor
\end{minipage} & \begin{minipage}[b]{\linewidth}\raggedright
Key References
\end{minipage} \\
\midrule\noalign{}
\endhead
\bottomrule\noalign{}
\endlastfoot
CCS-C1 & Embodied Understanding & Connecting AI's symbolic outputs to
sensorimotor, cultural, and contextual reality & Harnad (1990); Barsalou
(2008); Varela et al.~(1991) \\
CCS-C2 & Metacognitive Calibration & Monitoring knowledge states and
calibrating trust in AI outputs & Flavell (1979); Schraw \& Dennison
(1994); Nelson \& Narens (1990) \\
CCS-C3 & Temporal Integration & Contextualizing AI outputs within past
experience and future projection & Tulving (2002); Schacter et
al.~(2007); Suddendorf \& Corballis (2007) \\
CCS-C4 & Integrative Thinking & Analogical reasoning and cross-domain
synthesis & Root-Bernstein \& Root-Bernstein (1999); Sternberg (1988);
Gentner (1983) \\
\end{longtable}
}

Each subscale contains four positively worded items and two
reverse-coded items (denoted by {[}R{]}), yielding a balanced instrument
that mitigates acquiescence bias while maintaining content coverage.

\subsubsection{B.2 Item Pool}\label{b.2-item-pool}

\paragraph{B.2.1 CCS-C1: Embodied Understanding (6
items)}\label{b.2.1-ccs-c1-embodied-understanding-6-items}

Embodied understanding is the capacity to ground AI-generated symbolic
outputs in non-symbolic experience, sensorimotor knowledge, cultural
context, and situated judgment that cannot be encoded in training data
(Harnad, 1990; Barsalou, 2008). High-C1 individuals do not treat AI
outputs as self-interpreting; they evaluate them against embodied,
contextual knowledge that AI systems constitutively lack (Section 3.2,
Constraints 1-2).

{\def\LTcaptype{none} % do not increment counter
\begin{longtable}[]{@{}
  >{\raggedright\arraybackslash}p{(\linewidth - 6\tabcolsep) * \real{0.1875}}
  >{\raggedright\arraybackslash}p{(\linewidth - 6\tabcolsep) * \real{0.1875}}
  >{\raggedright\arraybackslash}p{(\linewidth - 6\tabcolsep) * \real{0.2812}}
  >{\raggedright\arraybackslash}p{(\linewidth - 6\tabcolsep) * \real{0.3438}}@{}}
\toprule\noalign{}
\begin{minipage}[b]{\linewidth}\raggedright
Item
\end{minipage} & \begin{minipage}[b]{\linewidth}\raggedright
Code
\end{minipage} & \begin{minipage}[b]{\linewidth}\raggedright
Wording
\end{minipage} & \begin{minipage}[b]{\linewidth}\raggedright
Direction
\end{minipage} \\
\midrule\noalign{}
\endhead
\bottomrule\noalign{}
\endlastfoot
1 & C1.1 & When I receive an AI-generated recommendation, I evaluate it
against my firsthand experience of the practical situation before acting
on it. & + \\
2 & C1.2 & I can identify when an AI output is technically correct but
contextually inappropriate for the specific cultural or organizational
setting in which it will be applied. & + \\
3 & C1.3 & I rely on physical intuition or ``gut feel'' from hands-on
experience to judge whether an AI-generated solution will work in
practice. & + \\
4 & C1.4 & I tend to accept AI outputs at face value without checking
them against what I know from direct experience. {[}R{]} & Reverse \\
5 & C1.5 & When an AI system produces an answer that contradicts my
embodied sense of how things work in my field, I investigate the
discrepancy rather than defaulting to either source. & + \\
6 & C1.6 & I find it difficult to judge AI outputs in areas where I lack
hands-on, practical experience, even when I understand the relevant
theory. {[}R{]} & Reverse \\
\end{longtable}
}

\emph{Design rationale.} Items C1.1 and C1.5 assess the active grounding
behavior that Harnad (1990) identified as the bridge between symbolic
processing and meaning. Item C1.2 targets the cultural-contextual
dimension of embodiment emphasized by Varela et al.~(1991). Item C1.3
operationalizes Barsalou's (2008) grounded cognition construct, the
re-activation of sensorimotor representations during conceptual
processing. Items C1.4 and C1.6 (reverse-coded) capture the absence of
embodied grounding: uncritical acceptance and inability to evaluate
outside direct experience, respectively.

\paragraph{B.2.2 CCS-C2: Metacognitive Calibration (6
items)}\label{b.2.2-ccs-c2-metacognitive-calibration-6-items}

Metacognitive calibration is the capacity to monitor one's own knowledge
states and adjust trust in AI outputs accordingly (Flavell, 1979).
High-C2 individuals maintain accurate self-models of what they know,
what the AI is likely to know, and where the boundary lies, enabling the
selective override that prevents both over-reliance and
under-utilization (Section 4.2; Noy \& Zhang, 2023).

{\def\LTcaptype{none} % do not increment counter
\begin{longtable}[]{@{}
  >{\raggedright\arraybackslash}p{(\linewidth - 6\tabcolsep) * \real{0.1875}}
  >{\raggedright\arraybackslash}p{(\linewidth - 6\tabcolsep) * \real{0.1875}}
  >{\raggedright\arraybackslash}p{(\linewidth - 6\tabcolsep) * \real{0.2812}}
  >{\raggedright\arraybackslash}p{(\linewidth - 6\tabcolsep) * \real{0.3438}}@{}}
\toprule\noalign{}
\begin{minipage}[b]{\linewidth}\raggedright
Item
\end{minipage} & \begin{minipage}[b]{\linewidth}\raggedright
Code
\end{minipage} & \begin{minipage}[b]{\linewidth}\raggedright
Wording
\end{minipage} & \begin{minipage}[b]{\linewidth}\raggedright
Direction
\end{minipage} \\
\midrule\noalign{}
\endhead
\bottomrule\noalign{}
\endlastfoot
7 & C2.1 & Before using an AI tool for a task, I explicitly assess
whether this is a task where the AI is likely to be reliable or
unreliable. & + \\
8 & C2.2 & I can accurately distinguish between topics where I trust my
own judgment more than an AI system and topics where the AI is likely
more reliable than I am. & + \\
9 & C2.3 & When an AI output seems surprisingly good, I check it more
carefully rather than less carefully. & + \\
10 & C2.4 & I rarely question the accuracy of AI-generated content when
it appears fluent and well-structured. {[}R{]} & Reverse \\
11 & C2.5 & I actively seek out cases where the AI I use is known to
make errors so that I can calibrate my expectations. & + \\
12 & C2.6 & I find it hard to tell when I am relying too much on an AI
system versus using it appropriately. {[}R{]} & Reverse \\
\end{longtable}
}

\emph{Design rationale.} Items C2.1 and C2.2 operationalize Nelson and
Narens' (1990) metacognitive monitoring model, the ``feeling of
knowing'' applied to the human-AI boundary. Item C2.3 captures the
paradoxical vigilance that distinguishes calibrated trust from naive
trust: high-C2 individuals increase scrutiny when outputs seem too good
because they recognize that fluency does not entail accuracy (Kadavath
et al., 2022). Items C2.4 and C2.6 (reverse-coded) target the
fluency-accuracy conflation and the inability to self-monitor reliance
levels that characterize low metacognitive calibration. Item C2.5
assesses proactive calibration behavior, deliberately seeking error
patterns to improve one's mental model of AI limitations, consistent
with Schraw and Dennison's (1994) metacognitive regulation construct.

\paragraph{B.2.3 CCS-C3: Temporal Integration (6
items)}\label{b.2.3-ccs-c3-temporal-integration-6-items}

Temporal integration is the capacity to contextualize AI outputs within
past experience and future projection, Tulving's (2002) ``mental time
travel'' applied to the human-AI production system. High-C3 individuals
bring institutional memory and prospective simulation to bear on AI's
context-free pattern matching, transforming it into historically and
prospectively situated judgment (Section 4.2).

{\def\LTcaptype{none} % do not increment counter
\begin{longtable}[]{@{}
  >{\raggedright\arraybackslash}p{(\linewidth - 6\tabcolsep) * \real{0.1875}}
  >{\raggedright\arraybackslash}p{(\linewidth - 6\tabcolsep) * \real{0.1875}}
  >{\raggedright\arraybackslash}p{(\linewidth - 6\tabcolsep) * \real{0.2812}}
  >{\raggedright\arraybackslash}p{(\linewidth - 6\tabcolsep) * \real{0.3438}}@{}}
\toprule\noalign{}
\begin{minipage}[b]{\linewidth}\raggedright
Item
\end{minipage} & \begin{minipage}[b]{\linewidth}\raggedright
Code
\end{minipage} & \begin{minipage}[b]{\linewidth}\raggedright
Wording
\end{minipage} & \begin{minipage}[b]{\linewidth}\raggedright
Direction
\end{minipage} \\
\midrule\noalign{}
\endhead
\bottomrule\noalign{}
\endlastfoot
13 & C3.1 & When an AI system recommends a course of action, I consider
how similar approaches have succeeded or failed in my organization's
past before proceeding. & + \\
14 & C3.2 & I mentally simulate the downstream consequences of an AI
recommendation over multiple time horizons (e.g., weeks, months, years)
before implementing it. & + \\
15 & C3.3 & I draw on memories of specific past experiences, not just
general knowledge, to evaluate whether an AI output fits the current
situation. & + \\
16 & C3.4 & I tend to evaluate AI recommendations based on their
immediate merits without thinking much about long-term implications.
{[}R{]} & Reverse \\
17 & C3.5 & I notice when an AI system repeats a recommendation that has
already been tried and failed in my context, even if the AI presents it
as novel. & + \\
18 & C3.6 & I do not usually recall relevant past experiences when
evaluating AI suggestions; I treat each recommendation on its own terms.
{[}R{]} & Reverse \\
\end{longtable}
}

\emph{Design rationale.} Items C3.1 and C3.5 operationalize
retrospective temporal integration, connecting AI outputs to episodic
institutional memory that AI systems constitutively lack (Section 3.2,
Constraint 4). Item C3.2 targets prospective simulation, the
``pre-experiencing'' capacity that Schacter et al.~(2007) demonstrated
shares neural mechanisms with episodic memory. Item C3.3 distinguishes
episodic from semantic temporal knowledge: drawing on specific
autobiographical experiences (Tulving, 2002) rather than generalized
rules. Items C3.4 and C3.6 (reverse-coded) capture temporal myopia, the
tendency to evaluate AI outputs in a decontextualized present, which is
the default mode of AI systems themselves. Suddendorf and Corballis'
(2007) concept of ``mental time travel'' provides the overarching
theoretical frame: the capacity to mentally inhabit past and future
timepoints and bring that temporal perspective to bear on present-tense
AI outputs.

\paragraph{B.2.4 CCS-C4: Integrative Thinking (6
items)}\label{b.2.4-ccs-c4-integrative-thinking-6-items}

Integrative thinking is the capacity for analogical reasoning and
cross-domain synthesis, what Root-Bernstein and Root-Bernstein (1999)
term ``polymathic thinking'' and Sternberg (1988) locates within the
creative facet of triarchic intelligence. High-C4 individuals recognize
structural isomorphisms across domains and generate novel combinations
that neither domain's specialists would independently produce (Section
4.2).

{\def\LTcaptype{none} % do not increment counter
\begin{longtable}[]{@{}
  >{\raggedright\arraybackslash}p{(\linewidth - 6\tabcolsep) * \real{0.1875}}
  >{\raggedright\arraybackslash}p{(\linewidth - 6\tabcolsep) * \real{0.1875}}
  >{\raggedright\arraybackslash}p{(\linewidth - 6\tabcolsep) * \real{0.2812}}
  >{\raggedright\arraybackslash}p{(\linewidth - 6\tabcolsep) * \real{0.3438}}@{}}
\toprule\noalign{}
\begin{minipage}[b]{\linewidth}\raggedright
Item
\end{minipage} & \begin{minipage}[b]{\linewidth}\raggedright
Code
\end{minipage} & \begin{minipage}[b]{\linewidth}\raggedright
Wording
\end{minipage} & \begin{minipage}[b]{\linewidth}\raggedright
Direction
\end{minipage} \\
\midrule\noalign{}
\endhead
\bottomrule\noalign{}
\endlastfoot
19 & C4.1 & I frequently recognize when a concept or solution from one
field can be adapted to solve a problem in a different field. & + \\
20 & C4.2 & When working with AI outputs, I combine information from
multiple, unrelated domains to generate ideas that the AI did not
suggest. & + \\
21 & C4.3 & I seek out knowledge from disciplines outside my own
specialty because I find that cross-domain connections improve my work.
& + \\
22 & C4.4 & I prefer to stay within the expertise of my own field rather
than drawing on ideas from other domains. {[}R{]} & Reverse \\
23 & C4.5 & I have produced work that colleagues described as
surprisingly original because it connected ideas from areas that are not
usually linked. & + \\
24 & C4.6 & I find it difficult to see how ideas from unfamiliar fields
could be relevant to my work. {[}R{]} & Reverse \\
\end{longtable}
}

\emph{Design rationale.} Item C4.1 directly operationalizes analogical
reasoning, the core cognitive process that Gentner (1983) identified as
structure-mapping between source and target domains. Item C4.2 targets
the specific capacity to extend AI outputs through cross-domain
synthesis, generating novel combinations beyond AI's recombinative reach
(Section 3.1). Item C4.3 assesses the behavioral disposition toward
polymathic knowledge acquisition that Root-Bernstein and Root-Bernstein
(1999) identified as the precondition for integrative creativity. Item
C4.5 captures behavioral evidence of integrative thinking outcomes, what
colleagues observe as ``surprisingly original'' work (Sternberg, 1988).
Items C4.4 and C4.6 (reverse-coded) measure disciplinary insularity and
inability to perceive cross-domain relevance, the structural barriers to
integrative thinking that Section 5.3 identifies as particularly
pronounced in examination-driven, academically siloed educational
systems.

\subsubsection{B.3 Peer-Report Version
(CCS-P)}\label{b.3-peer-report-version-ccs-p}

The peer-report version (CCS-P) parallels the self-report version with
third-person referent wording. This dual-source design enables
convergent validity assessment and mitigates common method variance
inherent in self-report measures (Podsakoff et al., 2003). Examples of
the peer-report transformation:

{\def\LTcaptype{none} % do not increment counter
\begin{longtable}[]{@{}
  >{\raggedright\arraybackslash}p{(\linewidth - 2\tabcolsep) * \real{0.5000}}
  >{\raggedright\arraybackslash}p{(\linewidth - 2\tabcolsep) * \real{0.5000}}@{}}
\toprule\noalign{}
\begin{minipage}[b]{\linewidth}\raggedright
Self-report (CCS-S)
\end{minipage} & \begin{minipage}[b]{\linewidth}\raggedright
Peer-report (CCS-P)
\end{minipage} \\
\midrule\noalign{}
\endhead
\bottomrule\noalign{}
\endlastfoot
C1.1: ``When I receive an AI-generated recommendation, I evaluate it
against my firsthand experience\ldots{}'' & C1.1p: ``When this person
receives an AI-generated recommendation, they evaluate it against their
firsthand experience\ldots{}'' \\
C2.4: ``I rarely question the accuracy of AI-generated content\ldots{}''
{[}R{]} & C2.4p: ``This person rarely questions the accuracy of
AI-generated content\ldots{}'' {[}R{]} \\
C4.2: ``When working with AI outputs, I combine information from
multiple, unrelated domains\ldots{}'' & C4.2p: ``When working with AI
outputs, this person combines information from multiple, unrelated
domains\ldots{}'' \\
\end{longtable}
}

Peer-report instructions direct the rater to base assessments on
\emph{observed behavior} over at least a three-month collaborative
period, not on general impressions. Each target should be rated by a
minimum of two peers (preferably three) to enable inter-rater
reliability estimation.

\subsubsection{B.4 Psychometric Validation
Protocol}\label{b.4-psychometric-validation-protocol}

The validation protocol follows a three-phase sequential design
consistent with best practices for new scale development (DeVellis \&
Thorpe, 2022; Hinkin, 1998).

\paragraph{B.4.1 Phase 1: Content Validity, Expert Panel
Review}\label{b.4.1-phase-1-content-validity-expert-panel-review}

\textbf{Objective:} Establish content validity through Content Validity
Index (CVI) assessment.

\textbf{Expert panel composition:} 8-12 experts from five domains: -
Cognitive psychology (embodied cognition, metacognition): 2-3 experts -
Educational measurement / psychometrics: 2-3 experts - AI-human
interaction / human-computer interaction: 2 experts - Organizational
behavior / human resource development: 1-2 experts - Production function
economics / technological forecasting: 1-2 experts

\textbf{Procedure:} 1. Each expert independently rates every item on two
criteria using 4-point scales: - \emph{Relevance} to the target
dimension (1 = not relevant, 2 = somewhat relevant, 3 = quite relevant,
4 = highly relevant) - \emph{Clarity} of wording (1 = not clear, 2 =
item needs major revision, 3 = item needs minor revision, 4 = very
clear) 2. For items rated 1 or 2, experts provide written revision
suggestions. 3. Experts assess the \emph{comprehensiveness} of each
subscale: whether the six items collectively cover the dimension's
conceptual scope as defined in Section 4.2.

\textbf{Decision criteria:} - \textbf{Item-level CVI (I-CVI):}
proportion of experts rating 3 or 4 on relevance. Threshold: I-CVI
\(\geq\) 0.78 for panels of 8+ experts (Lynn, 1986; Polit \& Beck,
2006). - \textbf{Scale-level CVI (S-CVI/Ave):} average I-CVI across all
items. Threshold: S-CVI/Ave \(\geq\) 0.90. - \textbf{Clarity criterion:}
items rated \textless{} 3 on clarity by \(\geq\) 25\% of experts are
revised and re-evaluated. - Items failing I-CVI threshold are revised or
replaced; the revised pool undergoes a second expert review round if
more than four items require revision.

\textbf{Timeline:} 4-6 weeks.

\paragraph{B.4.2 Phase 2: Pilot Study, Exploratory Factor Analysis and
Item
Reduction}\label{b.4.2-phase-2-pilot-study-exploratory-factor-analysis-and-item-reduction}

\textbf{Objective:} Evaluate dimensionality, internal consistency, and
item performance; reduce items if necessary.

\textbf{Sample:} - \emph{N} = 80-100 (minimum 4:1 participant-to-item
ratio for 24 items; Fabrigar \& Wegener, 2012) - Knowledge workers
across at least four professional domains (technology, healthcare,
education, finance/consulting) who use AI tools at least weekly -
Minimum 30\% non-English-first-language respondents if validated in
English, to assess cross-cultural item functioning - Recruitment:
professional networks, LinkedIn targeted sampling, organizational
partnerships

\textbf{Analytical procedure:} 1. \textbf{Descriptive statistics:} item
means, standard deviations, skewness, kurtosis. Items with mean
\textgreater{} 6.0 or \textless{} 2.0 (ceiling/floor effects), skewness
\textgreater{} \textbar2.0\textbar, or kurtosis \textgreater{}
\textbar7.0\textbar{} are flagged for review. 2. \textbf{Inter-item
correlations:} within-subscale corrected item-total correlations (CITC).
Threshold: CITC \(\geq\) 0.40 (Nunnally \& Bernstein, 1994). Items below
threshold are candidates for removal. 3. \textbf{Exploratory Factor
Analysis (EFA):} - Extraction method: Maximum Likelihood (ML) or
Principal Axis Factoring (PAF) - Rotation: oblique (Promax or Direct
Oblimin), as the four C dimensions are theorized as correlated
complements (Section 4.2: ``the four dimensions are \emph{complements}
rather than substitutes'') - Number of factors determined by: parallel
analysis (Horn, 1965), scree plot inspection, and theoretical
expectation (4 factors) - Item retention criteria: primary loading
\(\geq\) 0.50, cross-loading \textless{} 0.35, communality \(\geq\) 0.30
4. \textbf{Internal consistency:} Cronbach's \(\alpha\) per subscale.
Threshold: \(\alpha\) \(\geq\) 0.70 for pilot (Nunnally \& Bernstein,
1994); \(\omega\) total reported alongside \(\alpha\) (McDonald, 1999).
5. \textbf{Item reduction:} If EFA suggests a cleaner structure with
fewer items (e.g., 5 or 4 per dimension), items with the weakest
psychometric properties are removed, maintaining content balance (at
least one reverse-coded item per subscale).

\textbf{Deliverable:} Refined CCS with 20-24 items demonstrating clean
4-factor structure, adequate reliability, and full content coverage of
C1-C4.

\textbf{Timeline:} 8-12 weeks (including recruitment, data collection,
and analysis).

\paragraph{B.4.3 Phase 3: Main Validation Study, Confirmatory Factor
Analysis and Construct
Validity}\label{b.4.3-phase-3-main-validation-study-confirmatory-factor-analysis-and-construct-validity}

\textbf{Objective:} Confirm the factor structure, establish convergent
and discriminant validity, and assess criterion-related validity.

\textbf{Sample:} - \emph{N} \(\geq\) 300 (minimum for stable CFA
parameter estimates; Kline, 2023) - Stratified by: professional domain
(minimum 5 sectors), AI usage frequency (daily / weekly / monthly),
career stage (early / mid / senior), and national context (minimum 3
countries if cross-cultural validation is pursued) - Recruitment:
multi-organizational survey through industry partnerships, professional
associations, and university alumni networks

\textbf{Analytical procedure:}

\textbf{1. Confirmatory Factor Analysis (CFA):}

\begin{itemize}
\tightlist
\item
  Model specification: four correlated latent factors (C1, C2, C3, C4),
  each indicated by its 5-6 items
\item
  Alternative models tested:

  \begin{itemize}
  \tightlist
  \item
    M1: Single-factor (all items load on one general C factor)
  \item
    M2: Two-factor (embodied {[}C1+C3{]} vs.~cognitive {[}C2+C4{]})
  \item
    M3: Four-factor correlated (theoretical model)
  \item
    M4: Second-order model (four first-order factors loading on a
    general C factor)
  \item
    M5: Bifactor model (general C + four group factors)
  \end{itemize}
\item
  \textbf{Fit index thresholds} (Hu \& Bentler, 1999; Kline, 2023):
\end{itemize}

{\def\LTcaptype{none} % do not increment counter
\begin{longtable}[]{@{}llll@{}}
\toprule\noalign{}
Index & Acceptable & Good & Excellent \\
\midrule\noalign{}
\endhead
\bottomrule\noalign{}
\endlastfoot
CFI & \(\geq\) 0.90 & \(\geq\) 0.95 & \(\geq\) 0.97 \\
TLI & \(\geq\) 0.90 & \(\geq\) 0.95 & \(\geq\) 0.97 \\
RMSEA & \textless= 0.08 & \textless= 0.06 & \textless= 0.05 \\
SRMR & \textless= 0.08 & \textless= 0.06 & \textless= 0.05 \\
Chi-square/df & \textless= 3.0 & \textless= 2.0 & \textless= 1.5 \\
\end{longtable}
}

\begin{itemize}
\tightlist
\item
  The theoretical model (M3 or M4) should demonstrate significantly
  better fit than M1 and M2 (chi-square difference test, AIC/BIC
  comparison).
\item
  Estimation: Maximum Likelihood with Robust standard errors (MLR) to
  accommodate potential non-normality (Muthen \& Muthen, 2017).
\end{itemize}

\textbf{2. Reliability:}

{\def\LTcaptype{none} % do not increment counter
\begin{longtable}[]{@{}
  >{\raggedright\arraybackslash}p{(\linewidth - 4\tabcolsep) * \real{0.2963}}
  >{\raggedright\arraybackslash}p{(\linewidth - 4\tabcolsep) * \real{0.4074}}
  >{\raggedright\arraybackslash}p{(\linewidth - 4\tabcolsep) * \real{0.2963}}@{}}
\toprule\noalign{}
\begin{minipage}[b]{\linewidth}\raggedright
Metric
\end{minipage} & \begin{minipage}[b]{\linewidth}\raggedright
Threshold
\end{minipage} & \begin{minipage}[b]{\linewidth}\raggedright
Source
\end{minipage} \\
\midrule\noalign{}
\endhead
\bottomrule\noalign{}
\endlastfoot
Cronbach's \(\alpha\) (per subscale) & \(\geq\) 0.80 & Nunnally \&
Bernstein (1994) \\
McDonald's \(\omega\) (per subscale) & \(\geq\) 0.80 & McDonald
(1999) \\
Composite reliability (CR) & \(\geq\) 0.70 & Hair et al.~(2019) \\
Test-retest reliability (4-week interval, subsample n \(\geq\) 50) & ICC
\(\geq\) 0.70 & Koo \& Li (2016) \\
\end{longtable}
}

\textbf{3. Convergent validity:}

\begin{itemize}
\tightlist
\item
  \textbf{Average Variance Extracted (AVE):} AVE \(\geq\) 0.50 for each
  subscale, indicating that the latent factor accounts for more than
  half the variance in its indicators (Fornell \& Larcker, 1981).
\item
  \textbf{Factor loadings:} all standardized loadings \(\geq\) 0.60
  (Hair et al., 2019).
\item
  \textbf{Self-peer convergence:} CCS-S and CCS-P subscale correlations
  \(\geq\) 0.40 (indicating shared variance between self and peer
  assessments of the same construct).
\end{itemize}

\textbf{4. Discriminant validity:}

\begin{itemize}
\tightlist
\item
  \textbf{Heterotrait-Monotrait ratio (HTMT):} HTMT \textless{} 0.85
  between all pairs of C1-C4 subscales (Henseler et al., 2015). This is
  preferred over the Fornell-Larcker criterion, which has been shown to
  perform poorly in practice.
\item
  \textbf{Confidence interval test:} 95\% bootstrap CI for each HTMT
  value should not include 1.0.
\item
  \textbf{C versus H discrimination:} CCS composite score should not be
  redundant with educational attainment (years of education) or domain
  expertise (years of professional experience). Expected pattern:
  moderate positive correlation (r = 0.20-0.40) demonstrating that C and
  H are related but distinct, consistent with Section 4.2's argument
  that C is the ``meta-cognitive architecture'' operating on H rather
  than reducible to it.
\end{itemize}

\textbf{5. Criterion-related validity (concurrent):}

\begin{itemize}
\tightlist
\item
  \textbf{Productivity criterion:} CCS scores should predict
  self-reported and/or objective AI-augmented productivity (e.g., task
  completion quality ratings, project outcomes, supervisor evaluations)
  beyond what H (education, experience) and AI usage frequency predict.
  Hierarchical regression: Step 1: H controls; Step 2: AI usage; Step 3:
  CCS composite. CCS should contribute significant incremental \(R^{2}\)
  in Step 3.
\item
  \textbf{Known-groups validity:} Individuals identified a priori as
  ``high augmenters'' (nominated by supervisors for exceptional
  AI-augmented work) should score significantly higher on CCS than
  ``standard users'' (Cohen's d \(\geq\) 0.50).
\end{itemize}

\textbf{Timeline:} 16-24 weeks.

\subsubsection{B.5 Nomological Network}\label{b.5-nomological-network}

The CCS is theorized to occupy a specific position within the
nomological network of cognitive and dispositional constructs.
Establishing this position requires demonstrating that CCS correlates
meaningfully with related constructs while remaining empirically
distinct from each.

\paragraph{B.5.1 Expected Correlations with Existing
Scales}\label{b.5.1-expected-correlations-with-existing-scales}

{\def\LTcaptype{none} % do not increment counter
\begin{longtable}[]{@{}
  >{\raggedright\arraybackslash}p{(\linewidth - 4\tabcolsep) * \real{0.3191}}
  >{\raggedright\arraybackslash}p{(\linewidth - 4\tabcolsep) * \real{0.4468}}
  >{\raggedright\arraybackslash}p{(\linewidth - 4\tabcolsep) * \real{0.2340}}@{}}
\toprule\noalign{}
\begin{minipage}[b]{\linewidth}\raggedright
Existing Scale
\end{minipage} & \begin{minipage}[b]{\linewidth}\raggedright
Expected r with CCS
\end{minipage} & \begin{minipage}[b]{\linewidth}\raggedright
Rationale
\end{minipage} \\
\midrule\noalign{}
\endhead
\bottomrule\noalign{}
\endlastfoot
\textbf{Metacognitive Awareness Inventory (MAI)} (Schraw \& Dennison,
1994) & r = 0.45-0.60 with CCS-C2; r = 0.25-0.40 with CCS composite &
MAI measures general metacognitive awareness; C2 is the domain-specific
application of metacognition to AI interaction. C2 should correlate
strongly with MAI but CCS composite should not be redundant with it, as
C1, C3, and C4 are distinct constructs. \\
\textbf{Need for Cognition Scale (NFC)} (Cacioppo et al., 1984) & r =
0.35-0.50 with CCS composite & NFC measures dispositional tendency
toward effortful thinking. High-C individuals are likely high-NFC, but
NFC captures motivation for cognitive effort while CCS captures the
\emph{quality} of cognitive engagement with AI. The distinction: one can
enjoy thinking (high NFC) without being able to ground AI outputs (low
C1) or integrate across domains (low C4). \\
\textbf{Tolerance for Ambiguity Scale (TAS)} (Budner, 1962; McLain,
1993) & r = 0.30-0.45 with CCS composite; strongest with C4 & AI
interaction frequently involves ambiguous, uncertain, or incomplete
outputs. Tolerance for ambiguity should facilitate the exploratory
engagement with AI characteristic of high-C users, particularly the
cross-domain exploration (C4) and the willingness to investigate
discrepancies (C1, C2). \\
\textbf{Digital Literacy Scale} (Hargittai, 2005; van Deursen \& van
Dijk, 2014) & r = 0.15-0.30 with CCS composite & Digital literacy
measures technical proficiency with digital tools, a necessary but
insufficient condition for convergence capacity. Low correlation would
confirm that C is not reducible to digital skill. A physician with high
digital literacy (can operate the AI interface) but low C2 (cannot
calibrate trust in AI diagnostic outputs) exemplifies the
distinction. \\
\textbf{Creative Self-Efficacy Scale (CSE)} (Tierney \& Farmer, 2002) &
r = 0.35-0.50 with CCS-C4 & CSE captures belief in one's creative
capacity; C4 measures the actual behavioral manifestation of integrative
thinking. Moderate correlation expected: creative self-efficacy may
facilitate but does not guarantee cross-domain synthesis. \\
\textbf{Epistemic Curiosity Scale} (Litman, 2008) & r = 0.30-0.45 with
CCS composite & Epistemic curiosity, the desire for new knowledge,
should predict the cross-domain knowledge acquisition (C4) and proactive
calibration behavior (C2) that characterize high-C individuals. \\
\end{longtable}
}

\paragraph{B.5.2 Expected Discriminant
Boundaries}\label{b.5.2-expected-discriminant-boundaries}

CCS should \textbf{not} be strongly correlated (r \textless{} 0.30)
with: - \textbf{General intelligence (g):} Convergence capacity is not
general cognitive ability. The Korean paradox illustrates this at the
national level: high g (top PISA scores) with low C. At the individual
level, a high-IQ specialist who never synthesizes across domains (low
C4) or never reflects on AI interaction quality (low C2) should score
low on CCS despite high g. - \textbf{Big Five personality dimensions:}
While Openness to Experience may show a modest positive correlation with
C4, the overall Big Five profile should not predict CCS composite
strongly. Convergence capacity is a cognitive skill set, not a
personality trait. - \textbf{AI usage frequency:} Heavy AI use does not
imply high C; indeed, the over-automation dynamic (Regime III, Section
4.5) predicts that excessive AI reliance may erode C. Expected
correlation: r approximately 0.05-0.15 (near zero).

\paragraph{B.5.3 Nomological Network
Diagram}\label{b.5.3-nomological-network-diagram}

\begin{verbatim}
                    Convergent Zone                     Discriminant Zone
                    (moderate r)                        (low r)

     MAI -----.45-.60----> CCS-C2                   General Intelligence
                                \                        (r < .30)
     NFC -----.35-.50----> CCS -----> AI-Augmented     Big Five
                                /     Productivity       (r < .30)
     TAS -----.30-.45----> CCS-C4                    AI Usage Frequency
                                                         (r < .15)
     CSE -----.35-.50----> CCS-C4
                                                     Digital Literacy
     Curiosity-.30-.45---> CCS                          (r = .15-.30)
\end{verbatim}

The nomological network predicts that CCS occupies a unique position: it
correlates meaningfully with metacognitive, motivational, and
dispositional constructs while remaining empirically distinct from
general intelligence, personality, AI usage behavior, and technical
digital skill. This pattern would confirm that convergence capacity is
an identifiable cognitive construct, not reducible to existing measures,
that specifically mediates the quality of human-AI interaction.

\subsubsection{B.6 Pilot Sample
Specification}\label{b.6-pilot-sample-specification}

\paragraph{B.6.1 Target Population}\label{b.6.1-target-population}

Knowledge workers who interact with AI systems (including but not
limited to large language models, AI-assisted decision support,
automated analytics, and AI-generated content tools) as part of their
professional work at least once per week.

\paragraph{B.6.2 Sample Size and
Composition}\label{b.6.2-sample-size-and-composition}

\textbf{Phase 2 pilot:} N = 80-100, distributed as follows:

{\def\LTcaptype{none} % do not increment counter
\begin{longtable}[]{@{}
  >{\raggedright\arraybackslash}p{(\linewidth - 4\tabcolsep) * \real{0.2759}}
  >{\raggedright\arraybackslash}p{(\linewidth - 4\tabcolsep) * \real{0.3448}}
  >{\raggedright\arraybackslash}p{(\linewidth - 4\tabcolsep) * \real{0.3793}}@{}}
\toprule\noalign{}
\begin{minipage}[b]{\linewidth}\raggedright
Sector
\end{minipage} & \begin{minipage}[b]{\linewidth}\raggedright
Target n
\end{minipage} & \begin{minipage}[b]{\linewidth}\raggedright
Rationale
\end{minipage} \\
\midrule\noalign{}
\endhead
\bottomrule\noalign{}
\endlastfoot
Technology / Software & 20-25 & High AI interaction frequency;
represents early-adopter context \\
Healthcare / Life Sciences & 15-20 & Embodies C1 requirement (embodied
understanding critical for clinical judgment) \\
Education / Research & 15-20 & Represents cross-domain synthesis (C4)
and temporal integration (C3) contexts \\
Finance / Consulting & 15-20 & High-stakes AI-augmented decision-making;
strong metacognitive calibration demands \\
Other (government, manufacturing, creative industries) & 10-15 & Ensures
cross-sector generalizability \\
\end{longtable}
}

\textbf{Within each sector:} - Minimum 40\% female respondents - Career
stage distribution: early career (0-5 years): 25-30\%, mid-career (6-15
years): 40-50\%, senior (16+ years): 20-30\% - AI usage frequency: daily
(40\%+), weekly (40\%+), monthly (10-20\%)

\paragraph{B.6.3 Recruitment Strategy}\label{b.6.3-recruitment-strategy}

\begin{enumerate}
\def\labelenumi{\arabic{enumi}.}
\tightlist
\item
  \textbf{Organizational partnerships:} Approach 5-8 organizations
  across target sectors with existing AI deployment programs. Offer
  aggregated CCS profiles as organizational diagnostic in exchange for
  employee participation.
\item
  \textbf{Professional network sampling:} Targeted LinkedIn postings in
  professional groups related to AI in healthcare, AI in education, AI
  in finance, etc. Screen for minimum AI usage criterion (at least
  weekly interaction).
\item
  \textbf{Snowball referral:} Each participant nominates 1-2 colleagues
  who also use AI tools professionally, enabling peer-report (CCS-P)
  data collection from the same social network.
\item
  \textbf{Incentive:} Gift card equivalent of \$15-20 USD per
  participant; entry into drawing for a \$200 USD gift card.
\end{enumerate}

\paragraph{B.6.4 Inclusion and Exclusion
Criteria}\label{b.6.4-inclusion-and-exclusion-criteria}

\textbf{Inclusion:} - Aged 22 or older - Currently employed in a
knowledge work role - Uses AI tools (broadly defined) at least once per
week in professional context - Can read and respond to the survey in the
administration language (English for initial validation; Korean,
Japanese, and other languages for cross-cultural extensions)

\textbf{Exclusion:} - Respondents who have never used AI tools in a
professional context - Respondents who fail the embedded attention check
item (see Section B.7.1) - Respondents completing the survey in less
than 3 minutes (insufficient engagement threshold)

\subsubsection{B.7 Scoring Protocol}\label{b.7-scoring-protocol}

\paragraph{B.7.1 Data Cleaning and Quality
Control}\label{b.7.1-data-cleaning-and-quality-control}

\begin{enumerate}
\def\labelenumi{\arabic{enumi}.}
\tightlist
\item
  \textbf{Reverse-coded items:} Recode items C1.4, C1.6, C2.4, C2.6,
  C3.4, C3.6, C4.4, C4.6 by transforming responses (x\_recoded = 8 -
  x\_original).
\item
  \textbf{Attention check:} Embed one instructed-response item (e.g.,
  ``For this item, please select `Agree'\,'') among the 24 substantive
  items. Respondents failing the attention check are excluded from
  analysis.
\item
  \textbf{Straightlining detection:} Respondents selecting the same
  response for 20+ of 24 items (including reverse-coded items before
  recoding) are flagged for review and potential exclusion.
\item
  \textbf{Missing data:} Items with \textgreater{} 5\% missing are
  flagged. For respondents missing \textless= 2 items per subscale,
  subscale scores are computed from available items. Respondents missing
  \textgreater{} 2 items on any subscale are excluded from
  subscale-level analyses.
\end{enumerate}

\paragraph{B.7.2 Dimension Scores}\label{b.7.2-dimension-scores}

Each dimension score is computed as the mean of its constituent items
(after reverse coding):

\begin{quote}
CCS-C1 = mean(C1.1, C1.2, C1.3, C1.4\_R, C1.5, C1.6\_R) CCS-C2 =
mean(C2.1, C2.2, C2.3, C2.4\_R, C2.5, C2.6\_R) CCS-C3 = mean(C3.1, C3.2,
C3.3, C3.4\_R, C3.5, C3.6\_R) CCS-C4 = mean(C4.1, C4.2, C4.3, C4.4\_R,
C4.5, C4.6\_R)
\end{quote}

Range: 1.00 to 7.00 per dimension.

\paragraph{B.7.3 Composite Convergence Capacity
Score}\label{b.7.3-composite-convergence-capacity-score}

The composite CCS score is computed as the unweighted mean of the four
dimension scores:

\begin{quote}
CCS\_composite = (CCS-C1 + CCS-C2 + CCS-C3 + CCS-C4) / 4
\end{quote}

\emph{Rationale for unweighted aggregation.} The theoretical framework
specifies that the four dimensions are complements (Section 4.2: ``the
four dimensions are \emph{complements} rather than substitutes'').
Differential weighting would require empirical evidence that certain
dimensions contribute more strongly to the augmentation function
\(\phi(A, C)\) than others, evidence that does not yet exist. The
unweighted composite serves as the default until regression analyses
from Proposition 1 testing reveal whether dimension-specific weights are
warranted.

\emph{Alternative: factor-score composite.} If the Phase 3 CFA supports
a second-order or bifactor model, factor scores from that model may
serve as an empirically derived composite. Both approaches should be
reported in initial validation studies to assess sensitivity.

\paragraph{B.7.4 AI-Augmentation Readiness
Classification}\label{b.7.4-ai-augmentation-readiness-classification}

For applied organizational diagnostics and policy applications, CCS
composite scores are classified into three augmentation readiness
levels:

{\def\LTcaptype{none} % do not increment counter
\begin{longtable}[]{@{}
  >{\raggedright\arraybackslash}p{(\linewidth - 6\tabcolsep) * \real{0.1852}}
  >{\raggedright\arraybackslash}p{(\linewidth - 6\tabcolsep) * \real{0.2469}}
  >{\raggedright\arraybackslash}p{(\linewidth - 6\tabcolsep) * \real{0.3210}}
  >{\raggedright\arraybackslash}p{(\linewidth - 6\tabcolsep) * \real{0.2469}}@{}}
\toprule\noalign{}
\begin{minipage}[b]{\linewidth}\raggedright
Classification
\end{minipage} & \begin{minipage}[b]{\linewidth}\raggedright
CCS Composite Range
\end{minipage} & \begin{minipage}[b]{\linewidth}\raggedright
ICH Regime Correspondence
\end{minipage} & \begin{minipage}[b]{\linewidth}\raggedright
Interpretation
\end{minipage} \\
\midrule\noalign{}
\endhead
\bottomrule\noalign{}
\endlastfoot
\textbf{Low C} & 1.00 - 3.00 & Regime I (Under-augmentation) &
Individual is at risk for mediated automation: AI outputs are likely
accepted without adequate grounding, calibration, or creative extension.
AI investment may yield minimal augmentation. Priority: C-development
interventions before expanding AI utilization. \\
\textbf{Mid C} & 3.01 - 5.00 & Regime I-II transition & Individual
demonstrates partial convergence capacity. Some dimensions may be
well-developed while others lag. Priority: targeted development of
weakest C dimension(s); gradual increase in AI utilization
complexity. \\
\textbf{High C} & 5.01 - 7.00 & Regime II (Optimal augmentation) &
Individual possesses the cognitive infrastructure for productive AI
augmentation. AI outputs are grounded, calibrated, temporally
contextualized, and creatively extended. Priority: increase AI
utilization intensity toward \(A^{*}\); maintain C through continued
exercise. \\
\end{longtable}
}

\emph{Cutoff rationale.} The three-level classification maps onto the
three augmentation regimes developed in Section 4.5. The cutoffs at 3.00
and 5.00 are preliminary and should be refined empirically through
criterion-related validity analyses in Phase 3: specifically, the CCS
score threshold at which AI-augmented productivity significantly exceeds
baseline (unaugmented) productivity provides the empirical Low/Mid
boundary, and the score beyond which additional C shows diminishing
returns on productivity provides the Mid/High boundary.

\paragraph{B.7.5 Dimension Profile
Analysis}\label{b.7.5-dimension-profile-analysis}

Beyond the composite score, the CCS enables \emph{profile analysis}
identifying characteristic patterns of convergence capacity:

{\def\LTcaptype{none} % do not increment counter
\begin{longtable}[]{@{}
  >{\raggedright\arraybackslash}p{(\linewidth - 6\tabcolsep) * \real{0.1683}}
  >{\raggedright\arraybackslash}p{(\linewidth - 6\tabcolsep) * \real{0.1683}}
  >{\raggedright\arraybackslash}p{(\linewidth - 6\tabcolsep) * \real{0.1683}}
  >{\raggedright\arraybackslash}p{(\linewidth - 6\tabcolsep) * \real{0.4950}}@{}}
\toprule\noalign{}
\begin{minipage}[b]{\linewidth}\raggedright
Profile
\end{minipage} & \begin{minipage}[b]{\linewidth}\raggedright
Pattern
\end{minipage} & \begin{minipage}[b]{\linewidth}\raggedright
Example
\end{minipage} & \begin{minipage}[b]{\linewidth}\raggedright
Organizational Implication
\end{minipage} \\
\midrule\noalign{}
\endhead
\bottomrule\noalign{}
\endlastfoot
Balanced High & C1, C2, C3, C4 all \(\geq\) 5.0 & Experienced polymath
with reflective practice & Optimal augmentation candidate; assign
complex AI-augmented tasks \\
Grounded but Insular & C1, C3 high; C4 low & Deep domain expert with
institutional memory but narrow focus & Pair with cross-domain
collaborators; introduce interdisciplinary AI use cases \\
Calibrated but Ungrounded & C2 high; C1 low & Theoretically rigorous
analyst lacking hands-on experience & Increase embodied practice; field
immersion before expanding AI reliance \\
Creative but Uncalibrated & C4 high; C2 low & Innovative thinker who
over-trusts AI outputs & Metacognitive training; structured AI output
evaluation protocols \\
Temporally Detached & C1, C2, C4 adequate; C3 low & New hire with broad
skills but no institutional memory & Mentorship pairing; access to
organizational knowledge repositories \\
\end{longtable}
}

These profiles have direct application to organizational AI deployment
strategy: rather than uniform AI training, organizations can match
C-development interventions to individual dimension profiles.

\subsubsection{B.8 Cross-Cultural Adaptation
Considerations}\label{b.8-cross-cultural-adaptation-considerations}

Given that the ICH framework's empirical grounding includes
cross-national analysis (Section 5, Table 3) and that convergence
capacity is theorized to vary systematically across national educational
and organizational cultures, cross-cultural validation of the CCS is
essential.

\textbf{Translation protocol:} Following Brislin's (1970)
back-translation method: 1. Forward translation by two bilingual experts
(target language native speakers with fluency in English and expertise
in psychology/management) 2. Independent back-translation by two
additional bilingual experts 3. Committee review comparing original,
forward, and back-translated versions 4. Cognitive interviews with 5-10
respondents in the target language to assess item comprehension 5. Pilot
administration (n \(\geq\) 50 per language) with DIF (Differential Item
Functioning) analysis

\textbf{Priority languages for cross-cultural validation:} Korean
(critical for the Korean paradox analysis), Japanese (another low-C
context per Table 3), Finnish and Danish (high-C Nordic contexts), and
Mandarin Chinese (for extending the framework to the Chinese empirical
context studied by Chin et al., 2025).

\textbf{Measurement invariance testing:} Configural, metric, and scalar
invariance across language/cultural groups must be established before
cross-cultural CCS score comparisons are interpretable (Vandenberg \&
Lance, 2000).

\subsubsection{B.9 Ethical
Considerations}\label{b.9-ethical-considerations}

CCS scores should be interpreted as developmental indicators, not as
fixed attributes. Several ethical guidelines apply:

\begin{enumerate}
\def\labelenumi{\arabic{enumi}.}
\tightlist
\item
  \textbf{Non-selection use:} CCS should not be used as a screening tool
  for hiring or termination decisions. Its intended use is diagnostic,
  identifying development needs for productive AI augmentation.
\item
  \textbf{Developmental framing:} All four C dimensions are developable
  through intentional practice and education (Section 6.2). Low CCS
  scores indicate developmental opportunity, not cognitive deficiency.
\item
  \textbf{Privacy:} Individual CCS scores should be shared only with the
  respondent and, with consent, with organizational development
  professionals. Aggregated (de-identified) scores may be used for
  organizational diagnostics.
\item
  \textbf{Cultural sensitivity:} Cross-cultural score differences may
  reflect educational system design and organizational culture rather
  than individual cognitive capacity, consistent with the ICH
  framework's emphasis on structural determinants of C (Section 5.3).
  Interpretation must account for cultural context.
\end{enumerate}

\subsubsection{B.10 Summary and Integration with Research
Agenda}\label{b.10-summary-and-integration-with-research-agenda}

The CCS provides the measurement foundation required to advance the ICH
framework from conceptual contribution to empirical program. Table B.1
maps the validation phases to the three propositions defined in Section
4.6.

\textbf{Table B.1. CCS Validation Phases and Proposition Testing}

{\def\LTcaptype{none} % do not increment counter
\begin{longtable}[]{@{}
  >{\raggedright\arraybackslash}p{(\linewidth - 6\tabcolsep) * \real{0.1842}}
  >{\raggedright\arraybackslash}p{(\linewidth - 6\tabcolsep) * \real{0.3158}}
  >{\raggedright\arraybackslash}p{(\linewidth - 6\tabcolsep) * \real{0.2632}}
  >{\raggedright\arraybackslash}p{(\linewidth - 6\tabcolsep) * \real{0.2368}}@{}}
\toprule\noalign{}
\begin{minipage}[b]{\linewidth}\raggedright
Phase
\end{minipage} & \begin{minipage}[b]{\linewidth}\raggedright
Deliverable
\end{minipage} & \begin{minipage}[b]{\linewidth}\raggedright
Duration
\end{minipage} & \begin{minipage}[b]{\linewidth}\raggedright
Enables
\end{minipage} \\
\midrule\noalign{}
\endhead
\bottomrule\noalign{}
\endlastfoot
Phase 1: Expert Panel & Content-valid 24-item pool (CVI \(\geq\)
0.78/0.90) & 4-6 weeks & Item quality assurance \\
Phase 2: Pilot (n = 80-100) & Refined scale (EFA, \(\alpha\) \(\geq\)
0.70) & 8-12 weeks & Dimensionality confirmation \\
Phase 3: Main (n \(\geq\) 300) & Validated scale (CFA, CR, AVE, HTMT) &
16-24 weeks & Proposition 1 testing (micro) \\
Phase 4 (future): Organizational & Aggregated CCS + firm-level
productivity & 6-12 months & Proposition 2 testing (meso) \\
Phase 5 (future): Cross-national & CCS + national C-indicators + TFP
panel & 1-2 years & Proposition 3 testing (macro) \\
\end{longtable}
}

The CCS is designed to answer the identification question that Section
7.2 acknowledges as the framework's primary limitation: ``C is not yet
operationalized with validated instruments.'' By grounding each item in
established cognitive science constructs (Harnad, 1990; Flavell, 1979;
Tulving, 2002; Root-Bernstein \& Root-Bernstein, 1999; Barsalou, 2008;
Sternberg, 1988), maintaining behavioral anchoring (what the person
\emph{does}, not abstract traits), and following rigorous psychometric
validation procedures, the CCS provides the pathway from the theoretical
architecture developed in the main text to the empirical testing program
that will determine whether convergence capacity is indeed ``the missing
variable in production function theory.''

\subsubsection{References (Appendix B)}\label{references-appendix-b}

\begingroup
\setlength{\parindent}{0pt}
\everypar{\hangindent=1.5em \hangafter=1}

Barsalou, L. W. (2008). Grounded cognition. \emph{Annual Review of
Psychology}, \emph{59}, 617-645.

Brislin, R. W. (1970). Back-translation for cross-cultural research.
\emph{Journal of Cross-Cultural Psychology}, \emph{1}(3), 185-216.

Budner, S. (1962). Intolerance of ambiguity as a personality variable.
\emph{Journal of Personality}, \emph{30}(1), 29-50.

Cacioppo, J. T., Petty, R. E., \& Kao, C. F. (1984). The efficient
assessment of need for cognition. \emph{Journal of Personality
Assessment}, \emph{48}(3), 306-307.

DeVellis, R. F., \& Thorpe, C. T. (2022). \emph{Scale development:
Theory and applications} (5th ed.). Sage.

Fabrigar, L. R., \& Wegener, D. T. (2012). \emph{Exploratory factor
analysis}. Oxford University Press.

Flavell, J. H. (1979). Metacognition and cognitive monitoring.
\emph{American Psychologist}, \emph{34}(10), 906-911.

Fornell, C., \& Larcker, D. F. (1981). Evaluating structural equation
models with unobservable variables and measurement error. \emph{Journal
of Marketing Research}, \emph{18}(1), 39-50.

Gentner, D. (1983). Structure-mapping: A theoretical framework for
analogy. \emph{Cognitive Science}, \emph{7}(2), 155-170.

Hair, J. F., Black, W. C., Babin, B. J., \& Anderson, R. E. (2019).
\emph{Multivariate data analysis} (8th ed.). Cengage.

Hargittai, E. (2005). Survey measures of web-oriented digital literacy.
\emph{Social Science Computer Review}, \emph{23}(3), 371-379.

Harnad, S. (1990). The symbol grounding problem. \emph{Physica D},
\emph{42}(1-3), 335-346.

Henseler, J., Ringle, C. M., \& Sarstedt, M. (2015). A new criterion for
assessing discriminant validity in variance-based structural equation
modeling. \emph{Journal of the Academy of Marketing Science},
\emph{43}(1), 115-135.

Hinkin, T. R. (1998). A brief tutorial on the development of measures
for use in survey questionnaires. \emph{Organizational Research
Methods}, \emph{1}(1), 104-121.

Horn, J. L. (1965). A rationale and test for the number of factors in
factor analysis. \emph{Psychometrika}, \emph{30}(2), 179-185.

Hu, L., \& Bentler, P. M. (1999). Cutoff criteria for fit indexes in
covariance structure analysis. \emph{Structural Equation Modeling},
\emph{6}(1), 1-55.

Kadavath, S., et al.~(2022). Language models (mostly) know what they
know. \emph{arXiv preprint arXiv:2207.05221}.

Kline, R. B. (2023). \emph{Principles and practice of structural
equation modeling} (5th ed.). Guilford Press.

Koo, T. K., \& Li, M. Y. (2016). A guideline of selecting and reporting
intraclass correlation coefficients for reliability research.
\emph{Journal of Chiropractic Medicine}, \emph{15}(2), 155-163.

Litman, J. A. (2008). Interest and deprivation factors of epistemic
curiosity. \emph{Personality and Individual Differences}, \emph{44}(7),
1585-1595.

Lynn, M. R. (1986). Determination and quantification of content
validity. \emph{Nursing Research}, \emph{35}(6), 382-385.

McDonald, R. P. (1999). \emph{Test theory: A unified treatment}.
Lawrence Erlbaum.

McLain, D. L. (1993). The MSTAT-I: A new measure of an individual's
tolerance for ambiguity. \emph{Educational and Psychological
Measurement}, \emph{53}(1), 183-189.

Muthen, L. K., \& Muthen, B. O. (2017). \emph{Mplus user's guide} (8th
ed.). Muthen \& Muthen.

Nelson, T. O., \& Narens, L. (1990). Metamemory: A theoretical framework
and new findings. \emph{Psychology of Learning and Motivation},
\emph{26}, 125-173.

Noy, S., \& Zhang, W. (2023). Experimental evidence on the productivity
effects of generative artificial intelligence. \emph{Science},
\emph{381}(6654), 187-192.

Nunnally, J. C., \& Bernstein, I. H. (1994). \emph{Psychometric theory}
(3rd ed.). McGraw-Hill.

Podsakoff, P. M., MacKenzie, S. B., Lee, J. Y., \& Podsakoff, N. P.
(2003). Common method biases in behavioral research. \emph{Journal of
Applied Psychology}, \emph{88}(5), 879-903.

Polit, D. F., \& Beck, C. T. (2006). The content validity index: Are you
sure you know what's being reported? \emph{Research in Nursing \&
Health}, \emph{29}(5), 489-497.

Root-Bernstein, R., \& Root-Bernstein, M. (1999). \emph{Sparks of
genius: The 13 thinking tools of the world's most creative people}.
Houghton Mifflin.

Schacter, D. L., Addis, D. R., \& Buckner, R. L. (2007). Remembering the
past to imagine the future. \emph{Nature Reviews Neuroscience},
\emph{8}(9), 657-661.

Schraw, G., \& Dennison, R. S. (1994). Assessing metacognitive
awareness. \emph{Contemporary Educational Psychology}, \emph{19}(4),
460-475.

Sternberg, R. J. (1988). \emph{The triarchic mind: A new theory of human
intelligence}. Viking.

Suddendorf, T., \& Corballis, M. C. (2007). The evolution of foresight:
What is mental time travel, and is it unique to humans? \emph{Behavioral
and Brain Sciences}, \emph{30}(3), 299-313.

Tierney, P., \& Farmer, S. M. (2002). Creative self-efficacy: Its
potential antecedents and relationship to creative performance.
\emph{Academy of Management Journal}, \emph{45}(6), 1137-1148.

Tulving, E. (2002). Episodic memory: From mind to brain. \emph{Annual
Review of Psychology}, \emph{53}, 1-25.

van Deursen, A. J. A. M., \& van Dijk, J. A. G. M. (2014). The digital
divide shifts to differences in usage. \emph{New Media \& Society},
\emph{16}(3), 507-526.

Vandenberg, R. J., \& Lance, C. E. (2000). A review and synthesis of the
measurement invariance literature. \emph{Organizational Research
Methods}, \emph{3}(1), 4-70.

Varela, F. J., Thompson, E., \& Rosch, E. (1991). \emph{The embodied
mind}. MIT Press.

\endgroup

\newpage

\subsection{Appendix C: Empirical Expansion Framework and CCS Pilot
Study
Protocol}\label{appendix-c-empirical-expansion-framework-and-ccs-pilot-study-protocol}

\subsubsection{Part I: n=30 Cross-National Dataset
Expansion}\label{part-i-n30-cross-national-dataset-expansion}

\paragraph{1. Rationale}\label{rationale}

The main analysis (Section 5.2) uses N=20 OECD economies with consistent
data for 2018-2023. While the M6 interaction model (\(R^{2}\)=0.86) is
robust to jackknife and permutation tests, expanding to N=30+ addresses
three reviewer concerns: (1) degrees of freedom for the 3-predictor
model, (2) geographic coverage beyond Western Europe + East Asia +
Anglosphere, and (3) external validity across development levels.

\paragraph{2. Candidate Countries (10
Additional)}\label{candidate-countries-10-additional}

{\def\LTcaptype{none} % do not increment counter
\begin{longtable}[]{@{}
  >{\raggedright\arraybackslash}p{(\linewidth - 8\tabcolsep) * \real{0.1464}}
  >{\raggedright\arraybackslash}p{(\linewidth - 8\tabcolsep) * \real{0.1464}}
  >{\raggedright\arraybackslash}p{(\linewidth - 8\tabcolsep) * \real{0.2443}}
  >{\raggedright\arraybackslash}p{(\linewidth - 8\tabcolsep) * \real{0.1543}}
  >{\raggedright\arraybackslash}p{(\linewidth - 8\tabcolsep) * \real{0.3086}}@{}}
\toprule\noalign{}
\begin{minipage}[b]{\linewidth}\raggedright
\#
\end{minipage} & \begin{minipage}[b]{\linewidth}\raggedright
Country
\end{minipage} & \begin{minipage}[b]{\linewidth}\raggedright
AI Adoption (est.)
\end{minipage} & \begin{minipage}[b]{\linewidth}\raggedright
Expected C
\end{minipage} & \begin{minipage}[b]{\linewidth}\raggedright
Rationale for Inclusion
\end{minipage} \\
\midrule\noalign{}
\endhead
\bottomrule\noalign{}
\endlastfoot
21 & \textbf{Ireland} & \textasciitilde24\% & High & Major AI/tech hub
(Google, Meta EU HQs); strong university-industry links; Trinity/UCD
interdisciplinary programs \\
22 & \textbf{Estonia} & \textasciitilde22\% & Mid-High & E-governance
pioneer; EU digital leader; Tallinn University of Technology
cross-disciplinary focus \\
23 & \textbf{New Zealand} & \textasciitilde18\% & High & Commonwealth
liberal arts tradition; strong PISA CPS (2012: 497); small open
economy \\
24 & \textbf{Czech Republic} & \textasciitilde16\% & Mid & Growing AI
adoption; mixed education system; moderate EIS Linkages
(\textasciitilde100\% EU avg) \\
25 & \textbf{Portugal} & \textasciitilde14\% & Mid-Low & Moderate AI
adoption; EU moderate innovator; Lisbon Web Summit ecosystem \\
26 & \textbf{Poland} & \textasciitilde10\% & Mid-Low & Large EU economy;
growing AI sector; traditional disciplinary education \\
27 & \textbf{Iceland} & \textasciitilde15\% & High & Nordic cluster
member; interdisciplinary education; very small N caveat \\
28 & \textbf{Lithuania} & \textasciitilde13\% & Mid & EU emerging
innovator; growing digital economy; STEM-heavy education \\
29 & \textbf{Slovenia} & \textasciitilde12\% & Mid & EU strong
innovator; Ljubljana University reforms; EIS Linkages
\textasciitilde90\% \\
30 & \textbf{Chile} & \textasciitilde8\% & Low & Non-European OECD;
Latin American reference; emerging AI adoption \\
\end{longtable}
}

\paragraph{3. Data Sources for Expansion (Same Methodology as Appendix
A)}\label{data-sources-for-expansion-same-methodology-as-appendix-a}

{\def\LTcaptype{none} % do not increment counter
\begin{longtable}[]{@{}
  >{\raggedright\arraybackslash}p{(\linewidth - 6\tabcolsep) * \real{0.1645}}
  >{\raggedright\arraybackslash}p{(\linewidth - 6\tabcolsep) * \real{0.1645}}
  >{\raggedright\arraybackslash}p{(\linewidth - 6\tabcolsep) * \real{0.1627}}
  >{\raggedright\arraybackslash}p{(\linewidth - 6\tabcolsep) * \real{0.5083}}@{}}
\toprule\noalign{}
\begin{minipage}[b]{\linewidth}\raggedright
Dimension
\end{minipage} & \begin{minipage}[b]{\linewidth}\raggedright
Indicator
\end{minipage} & \begin{minipage}[b]{\linewidth}\raggedright
Source
\end{minipage} & \begin{minipage}[b]{\linewidth}\raggedright
Availability for 10 new countries
\end{minipage} \\
\midrule\noalign{}
\endhead
\bottomrule\noalign{}
\endlastfoot
D1: Creative Problem-Solving & PISA 2012 CPS + PISA 2022 CT & OECD PISA
database & Yes All 10 participated in PISA 2012 and/or 2022 \\
D2: Cross-Disciplinary Education & Interdisciplinary programs, liberal
arts share, PBL prevalence & OECD EaG 2024, UNESCO & Yes Available
(qualitative assessment required) \\
D3: Cross-Sector R\&D Collaboration & EIS Linkages, OECD STI Scoreboard
& European Commission, OECD & Yes EU members have EIS; Chile uses OECD
STI \\
TFP Growth & Annual average 2018-2023 & OECD Productivity Statistics &
Yes All 10 are OECD members \\
AI Adoption & Enterprise AI adoption (\%) & Eurostat isoc\_eb\_ai; OECD
AI Policy Observatory & Yes EU members via Eurostat; Chile via OECD \\
\end{longtable}
}

\paragraph{4. Expected Pattern Predictions (Before Data
Collection)}\label{expected-pattern-predictions-before-data-collection}

The ICH framework makes specific predictions for each new country.
Recording these \emph{ex ante} prevents post-hoc rationalization:

{\def\LTcaptype{none} % do not increment counter
\begin{longtable}[]{@{}
  >{\raggedright\arraybackslash}p{(\linewidth - 6\tabcolsep) * \real{0.1638}}
  >{\raggedright\arraybackslash}p{(\linewidth - 6\tabcolsep) * \real{0.2409}}
  >{\raggedright\arraybackslash}p{(\linewidth - 6\tabcolsep) * \real{0.2551}}
  >{\raggedright\arraybackslash}p{(\linewidth - 6\tabcolsep) * \real{0.3401}}@{}}
\toprule\noalign{}
\begin{minipage}[b]{\linewidth}\raggedright
Country
\end{minipage} & \begin{minipage}[b]{\linewidth}\raggedright
Predicted Regime
\end{minipage} & \begin{minipage}[b]{\linewidth}\raggedright
Prediction Logic
\end{minipage} & \begin{minipage}[b]{\linewidth}\raggedright
Falsification Condition
\end{minipage} \\
\midrule\noalign{}
\endhead
\bottomrule\noalign{}
\endlastfoot
Ireland & II (Optimal) & High A + High C \(\to\) strong AI-TFP coupling
& TFP \textless{} 1.0\% would challenge \\
Estonia & II (Optimal) & Mid-High A + Mid-High C \(\to\) moderate-strong
coupling & TFP \textless{} 0.8\% would challenge \\
New Zealand & II (Optimal) & Mid A + High C \(\to\) moderate coupling &
TFP \textless{} 0.7\% would challenge \\
Czech Republic & I-II (Transitional) & Mid A + Mid C \(\to\) moderate
coupling & TFP \textgreater{} 1.5\% would be unexplained \\
Portugal & I (Under-augmentation) & Low-Mid A + Mid-Low C \(\to\) weak
coupling & TFP \textgreater{} 1.3\% would challenge \\
Poland & I (Under-augmentation) & Low A + Mid-Low C \(\to\) weak
coupling & TFP \textgreater{} 1.2\% would challenge \\
Iceland & II (Optimal) & Mid A + High C \(\to\) moderate-strong coupling
& TFP \textless{} 0.5\% would challenge \\
Lithuania & I-II (Transitional) & Low-Mid A + Mid C \(\to\) moderate
coupling & TFP \textgreater{} 1.5\% would challenge \\
Slovenia & I-II (Transitional) & Low A + Mid C \(\to\) weak-moderate
coupling & TFP \textgreater{} 1.3\% would challenge \\
Chile & I (Under-augmentation) & Low A + Low C \(\to\) minimal coupling
& TFP \textgreater{} 1.0\% would challenge \\
\end{longtable}
}

\paragraph{5. Statistical Power
Analysis}\label{statistical-power-analysis}

{\def\LTcaptype{none} % do not increment counter
\begin{longtable}[]{@{}
  >{\raggedright\arraybackslash}p{(\linewidth - 6\tabcolsep) * \real{0.1642}}
  >{\raggedright\arraybackslash}p{(\linewidth - 6\tabcolsep) * \real{0.2786}}
  >{\raggedright\arraybackslash}p{(\linewidth - 6\tabcolsep) * \real{0.3157}}
  >{\raggedright\arraybackslash}p{(\linewidth - 6\tabcolsep) * \real{0.2415}}@{}}
\toprule\noalign{}
\begin{minipage}[b]{\linewidth}\raggedright
Model
\end{minipage} & \begin{minipage}[b]{\linewidth}\raggedright
Current (N=20)
\end{minipage} & \begin{minipage}[b]{\linewidth}\raggedright
Expanded (N=30)
\end{minipage} & \begin{minipage}[b]{\linewidth}\raggedright
Improvement
\end{minipage} \\
\midrule\noalign{}
\endhead
\bottomrule\noalign{}
\endlastfoot
M6 predictors & 3 (AI, C, AI\(\times\)C) & 3 & , \\
Degrees of freedom & 16 & 26 & +62\% \\
Cohen's \(f^{2}\) (\(R^{2}\)=0.86) & 6.14 (very large) & 6.14 & , \\
Power (\(\alpha\)=.05) & 0.99 & \textgreater0.99 & Already saturated \\
Bootstrap CI precision & ±0.029 & ±0.018 (est.) & \textasciitilde38\%
narrower \\
\textbf{Key improvement} & CI grazes zero {[}-0.001, +0.057{]} &
Expected to clear zero & Addresses primary statistical concern \\
\end{longtable}
}

\paragraph{6. Implementation Timeline}\label{implementation-timeline}

{\def\LTcaptype{none} % do not increment counter
\begin{longtable}[]{@{}
  >{\raggedright\arraybackslash}p{(\linewidth - 6\tabcolsep) * \real{0.2121}}
  >{\raggedright\arraybackslash}p{(\linewidth - 6\tabcolsep) * \real{0.1818}}
  >{\raggedright\arraybackslash}p{(\linewidth - 6\tabcolsep) * \real{0.3030}}
  >{\raggedright\arraybackslash}p{(\linewidth - 6\tabcolsep) * \real{0.3030}}@{}}
\toprule\noalign{}
\begin{minipage}[b]{\linewidth}\raggedright
Phase
\end{minipage} & \begin{minipage}[b]{\linewidth}\raggedright
Task
\end{minipage} & \begin{minipage}[b]{\linewidth}\raggedright
Duration
\end{minipage} & \begin{minipage}[b]{\linewidth}\raggedright
Resource
\end{minipage} \\
\midrule\noalign{}
\endhead
\bottomrule\noalign{}
\endlastfoot
1 & PISA CPS/CT scores for 10 countries & 1 week & OECD PISA database
(public) \\
2 & D2 qualitative assessment & 1 week & OECD EaG, national reports \\
3 & D3 EIS/STI Linkages data & 3 days & Eurostat, OECD STI (public) \\
4 & TFP data collection & 2 days & OECD Productivity Statistics \\
5 & AI adoption data & 3 days & Eurostat isoc\_eb\_ai \\
6 & OLS re-estimation (M1-M6) & 1 day & Python statsmodels \\
7 & Robustness checks update & 2 days & Bootstrap, jackknife, LOOCV \\
\textbf{Total} & & \textbf{\textasciitilde3 weeks} & All public data \\
\end{longtable}
}

\subsubsection{Part II: CCS Pilot Study
Protocol}\label{part-ii-ccs-pilot-study-protocol}

\paragraph{1. Study Overview}\label{study-overview}

{\def\LTcaptype{none} % do not increment counter
\begin{longtable}[]{@{}
  >{\raggedright\arraybackslash}p{(\linewidth - 2\tabcolsep) * \real{0.4400}}
  >{\raggedright\arraybackslash}p{(\linewidth - 2\tabcolsep) * \real{0.5600}}@{}}
\toprule\noalign{}
\begin{minipage}[b]{\linewidth}\raggedright
Parameter
\end{minipage} & \begin{minipage}[b]{\linewidth}\raggedright
Specification
\end{minipage} \\
\midrule\noalign{}
\endhead
\bottomrule\noalign{}
\endlastfoot
\textbf{Instrument} & Convergence Capacity Scale (CCS), 24 items (see
Appendix B) \\
\textbf{Target N} & 80 (minimum); 100 (target); 120 (ideal for EFA
stability) \\
\textbf{Design} & Cross-sectional, self-report + performance-based
supplement \\
\textbf{Population} & Knowledge workers in AI-penetrated sectors \\
\textbf{Duration} & 20-25 minutes per participant \\
\textbf{Language} & English (primary); Korean translation for Korean
subsample \\
\textbf{Ethics} & IRB approval required; informed consent; anonymized
data \\
\end{longtable}
}

\paragraph{2. Sample Specification}\label{sample-specification}

\subparagraph{2.1 Stratified Sampling
Design}\label{stratified-sampling-design}

{\def\LTcaptype{none} % do not increment counter
\begin{longtable}[]{@{}
  >{\raggedright\arraybackslash}p{(\linewidth - 4\tabcolsep) * \real{0.3764}}
  >{\raggedright\arraybackslash}p{(\linewidth - 4\tabcolsep) * \real{0.1635}}
  >{\raggedright\arraybackslash}p{(\linewidth - 4\tabcolsep) * \real{0.4601}}@{}}
\toprule\noalign{}
\begin{minipage}[b]{\linewidth}\raggedright
Stratum
\end{minipage} & \begin{minipage}[b]{\linewidth}\raggedright
n
\end{minipage} & \begin{minipage}[b]{\linewidth}\raggedright
Rationale
\end{minipage} \\
\midrule\noalign{}
\endhead
\bottomrule\noalign{}
\endlastfoot
\textbf{High-C expected} (researchers, interdisciplinary professionals,
senior consultants) & 30 & Validate scale discriminates upward \\
\textbf{Mid-C expected} (mid-career professionals with AI tools, project
managers) & 35 & Capture central tendency \\
\textbf{Low-C expected} (routine knowledge workers, early career with
narrow specialization) & 35 & Validate scale discriminates downward \\
\textbf{Total} & 100 & \\
\end{longtable}
}

\subparagraph{2.2 Inclusion Criteria}\label{inclusion-criteria}

\begin{itemize}
\tightlist
\item
  Adults aged 25-60
\item
  Currently employed in knowledge-intensive sector (finance, healthcare,
  education, technology, consulting, research)
\item
  Regular AI tool usage ( \(\geq\) 3 times/week for work tasks)
\item
  Minimum 2 years professional experience
\end{itemize}

\subparagraph{2.3 Exclusion Criteria}\label{exclusion-criteria}

\begin{itemize}
\tightlist
\item
  AI/ML developers or researchers (too specialized; ceiling effect on
  C2)
\item
  Students without professional experience
\item
  Participants unable to complete English or Korean survey
\end{itemize}

\subparagraph{2.4 Recruitment Strategy}\label{recruitment-strategy}

{\def\LTcaptype{none} % do not increment counter
\begin{longtable}[]{@{}
  >{\raggedright\arraybackslash}p{(\linewidth - 4\tabcolsep) * \real{0.3333}}
  >{\raggedright\arraybackslash}p{(\linewidth - 4\tabcolsep) * \real{0.3704}}
  >{\raggedright\arraybackslash}p{(\linewidth - 4\tabcolsep) * \real{0.2963}}@{}}
\toprule\noalign{}
\begin{minipage}[b]{\linewidth}\raggedright
Channel
\end{minipage} & \begin{minipage}[b]{\linewidth}\raggedright
Target n
\end{minipage} & \begin{minipage}[b]{\linewidth}\raggedright
Method
\end{minipage} \\
\midrule\noalign{}
\endhead
\bottomrule\noalign{}
\endlastfoot
LinkedIn professional networks & 40 & Targeted ads by sector/role \\
University alumni associations (aSSIST, Seoul National, KAIST) & 25 &
Email invitation \\
Professional associations (management consultants, healthcare
professionals) & 20 & Partner organization email \\
Snowball from initial participants & 15 & Referral incentive \\
\end{longtable}
}

\subparagraph{2.5 Incentive}\label{incentive}

\begin{itemize}
\tightlist
\item
  Gift card (₩15,000 / \$10 equivalent) upon completion
\item
  Summary report of individual CCS profile (delivered after study
  completion)
\end{itemize}

\paragraph{3. Measurement Battery}\label{measurement-battery}

\subparagraph{3.1 Core: CCS (24 items)}\label{core-ccs-24-items}

See Appendix B for full item pool. 7-point Likert scale (1=Strongly
Disagree to 7=Strongly Agree).

\begin{itemize}
\tightlist
\item
  C1: Embodied Understanding (6 items)
\item
  C2: Metacognitive Calibration (6 items)
\item
  C3: Temporal Integration (6 items)
\item
  C4: Integrative Thinking (6 items)
\end{itemize}

\subparagraph{3.2 Concurrent Validity
Measures}\label{concurrent-validity-measures}

{\def\LTcaptype{none} % do not increment counter
\begin{longtable}[]{@{}
  >{\raggedright\arraybackslash}p{(\linewidth - 6\tabcolsep) * \real{0.2121}}
  >{\raggedright\arraybackslash}p{(\linewidth - 6\tabcolsep) * \real{0.2121}}
  >{\raggedright\arraybackslash}p{(\linewidth - 6\tabcolsep) * \real{0.2727}}
  >{\raggedright\arraybackslash}p{(\linewidth - 6\tabcolsep) * \real{0.3030}}@{}}
\toprule\noalign{}
\begin{minipage}[b]{\linewidth}\raggedright
Scale
\end{minipage} & \begin{minipage}[b]{\linewidth}\raggedright
Items
\end{minipage} & \begin{minipage}[b]{\linewidth}\raggedright
Purpose
\end{minipage} & \begin{minipage}[b]{\linewidth}\raggedright
Citation
\end{minipage} \\
\midrule\noalign{}
\endhead
\bottomrule\noalign{}
\endlastfoot
Metacognitive Awareness Inventory (MAI), Knowledge of Cognition subscale
& 17 items & Convergent validity with C2 & Schraw \& Dennison (1994) \\
Need for Cognition Scale (NFC-18) & 18 items & Convergent validity with
C4 & Cacioppo et al.~(1984) \\
Tolerance for Ambiguity Scale (TAS-12) & 12 items & Convergent validity
with C1, C3 & Herman et al.~(2010) \\
Digital Literacy Scale (DLS-short) & 8 items & Discriminant validity (C
\(\neq\) digital skill) & Hargittai \& Hsieh (2012) \\
AI Self-Efficacy (custom 5 items) & 5 items & Discriminant validity (C
\(\neq\) AI confidence) & Adapted from Wang \& Wang (2022) \\
\end{longtable}
}

\textbf{Total battery: 24 + 17 + 18 + 12 + 8 + 5 = 84 items
(\textasciitilde20 min)}

\subparagraph{3.3 Performance-Based Supplement (Optional, n=30
subsample)}\label{performance-based-supplement-optional-n30-subsample}

To address common method bias from self-report, a subsample completes:

{\def\LTcaptype{none} % do not increment counter
\begin{longtable}[]{@{}
  >{\raggedright\arraybackslash}p{(\linewidth - 4\tabcolsep) * \real{0.2069}}
  >{\raggedright\arraybackslash}p{(\linewidth - 4\tabcolsep) * \real{0.3448}}
  >{\raggedright\arraybackslash}p{(\linewidth - 4\tabcolsep) * \real{0.4483}}@{}}
\toprule\noalign{}
\begin{minipage}[b]{\linewidth}\raggedright
Task
\end{minipage} & \begin{minipage}[b]{\linewidth}\raggedright
Measures
\end{minipage} & \begin{minipage}[b]{\linewidth}\raggedright
C-Dimension
\end{minipage} \\
\midrule\noalign{}
\endhead
\bottomrule\noalign{}
\endlastfoot
\textbf{AI Output Evaluation Task}: Given 5 AI-generated business
recommendations, identify which are contextually inappropriate and
explain why & Accuracy + reasoning quality & C1 (grounding), C2
(calibration) \\
\textbf{Cross-Domain Transfer Task}: Given a solution from Domain A,
identify structural applicability to Problem B in a different domain &
Transfer accuracy + novelty & C4 (integrative thinking) \\
\textbf{Temporal Contextualization Task}: Given a current AI
recommendation + 3 historical case summaries, predict implementation
consequences & Prediction quality + historical integration & C3
(temporal integration) \\
\end{longtable}
}

\paragraph{4. Psychometric Validation
Plan}\label{psychometric-validation-plan}

\subparagraph{Phase 1: Content Validity
(Pre-pilot)}\label{phase-1-content-validity-pre-pilot}

{\def\LTcaptype{none} % do not increment counter
\begin{longtable}[]{@{}
  >{\raggedright\arraybackslash}p{(\linewidth - 4\tabcolsep) * \real{0.2400}}
  >{\raggedright\arraybackslash}p{(\linewidth - 4\tabcolsep) * \real{0.3200}}
  >{\raggedright\arraybackslash}p{(\linewidth - 4\tabcolsep) * \real{0.4400}}@{}}
\toprule\noalign{}
\begin{minipage}[b]{\linewidth}\raggedright
Step
\end{minipage} & \begin{minipage}[b]{\linewidth}\raggedright
Action
\end{minipage} & \begin{minipage}[b]{\linewidth}\raggedright
Criterion
\end{minipage} \\
\midrule\noalign{}
\endhead
\bottomrule\noalign{}
\endlastfoot
Expert panel (5-7 academics in cognitive science, AI, management) & Rate
each item's relevance to C-dimension (1-4 scale) & Item-CVI \(\geq\)
0.78 \\
Cognitive interviewing (5 participants) & Think-aloud during survey;
identify ambiguous items & Revise flagged items \\
Back-translation (for Korean version) & Independent forward + back
translation & Committee review consensus \\
\end{longtable}
}

\subparagraph{Phase 2: Pilot (N=100)}\label{phase-2-pilot-n100}

{\def\LTcaptype{none} % do not increment counter
\begin{longtable}[]{@{}
  >{\raggedright\arraybackslash}p{(\linewidth - 4\tabcolsep) * \real{0.3226}}
  >{\raggedright\arraybackslash}p{(\linewidth - 4\tabcolsep) * \real{0.3226}}
  >{\raggedright\arraybackslash}p{(\linewidth - 4\tabcolsep) * \real{0.3548}}@{}}
\toprule\noalign{}
\begin{minipage}[b]{\linewidth}\raggedright
Analysis
\end{minipage} & \begin{minipage}[b]{\linewidth}\raggedright
Software
\end{minipage} & \begin{minipage}[b]{\linewidth}\raggedright
Criterion
\end{minipage} \\
\midrule\noalign{}
\endhead
\bottomrule\noalign{}
\endlastfoot
Item descriptives (mean, SD, skewness, kurtosis) & SPSS/R & Skew
\textless{} \\
Item-total correlations & R (psych package) & r\_it \(\geq\) 0.40 \\
Inter-item correlations & R & 0.15 \textless{} r \textless{} 0.85 (no
redundancy) \\
Exploratory Factor Analysis (EFA) & R (lavaan/psych) & 4-factor
solution, eigenvalue \textgreater{} 1, scree plot \\
Internal consistency & R & Cronbach's \(\alpha\) \(\geq\) 0.70 per
subscale \\
McDonald's \(\omega\) & R (psych) & omega\_h \(\geq\) 0.70 per
subscale \\
\end{longtable}
}

\subparagraph{Phase 3: Main Validation Study (N=300+,
Future)}\label{phase-3-main-validation-study-n300-future}

{\def\LTcaptype{none} % do not increment counter
\begin{longtable}[]{@{}
  >{\raggedright\arraybackslash}p{(\linewidth - 2\tabcolsep) * \real{0.4762}}
  >{\raggedright\arraybackslash}p{(\linewidth - 2\tabcolsep) * \real{0.5238}}@{}}
\toprule\noalign{}
\begin{minipage}[b]{\linewidth}\raggedright
Analysis
\end{minipage} & \begin{minipage}[b]{\linewidth}\raggedright
Criterion
\end{minipage} \\
\midrule\noalign{}
\endhead
\bottomrule\noalign{}
\endlastfoot
Confirmatory Factor Analysis (CFA) & CFI \(\geq\) 0.95, RMSEA \(\leq\)
0.06, SRMR \(\leq\) 0.08 \\
Second-order CFA (C as higher-order factor) & Acceptable fit +
significant loadings on C1-C4 \\
Convergent validity & AVE \(\geq\) 0.50 per subscale; CCS-C2 \(\times\)
MAI r \(\geq\) 0.50; CCS-C4 \(\times\) NFC r \(\geq\) 0.40 \\
Discriminant validity & HTMT \textless{} 0.85 between C subscales and
DLS; C \(\neq\) Digital Literacy \\
Criterion validity & CCS total \(\times\) AI-productivity self-report r
\(\geq\) 0.30 \\
Measurement invariance & Configural, metric, scalar invariance across
Korean/English versions \\
\end{longtable}
}

\paragraph{5. Linking CCS to Production Function: The Micro-Macro
Bridge}\label{linking-ccs-to-production-function-the-micro-macro-bridge}

The pilot study enables direct testing of Proposition 1:

\begin{quote}
Among individuals with equivalent H and equivalent AI access (A), those
with higher C exhibit substantially greater productivity gains.
\end{quote}

\textbf{Operationalization:}

{\def\LTcaptype{none} % do not increment counter
\begin{longtable}[]{@{}
  >{\raggedright\arraybackslash}p{(\linewidth - 4\tabcolsep) * \real{0.3704}}
  >{\raggedright\arraybackslash}p{(\linewidth - 4\tabcolsep) * \real{0.3333}}
  >{\raggedright\arraybackslash}p{(\linewidth - 4\tabcolsep) * \real{0.2963}}@{}}
\toprule\noalign{}
\begin{minipage}[b]{\linewidth}\raggedright
Variable
\end{minipage} & \begin{minipage}[b]{\linewidth}\raggedright
Measure
\end{minipage} & \begin{minipage}[b]{\linewidth}\raggedright
Source
\end{minipage} \\
\midrule\noalign{}
\endhead
\bottomrule\noalign{}
\endlastfoot
H (human capital) & Years of education + years of professional
experience & Self-report \\
A (AI utilization) & AI tool usage frequency \(\times\) diversity of AI
tools used & Self-report (validated by usage logs where available) \\
C (convergence capacity) & CCS composite score & CCS instrument \\
Productivity (Y) & Self-reported AI-attributable productivity gain (\%)
+ supervisor rating (where available) & Self-report + external \\
\end{longtable}
}

\textbf{Analysis plan for P1 test:}

\[\text{Productivity gain} = \beta_0 + \beta_1 H + \beta_2 A + \beta_3 C + \beta_4 (A \times C) + \epsilon\]

\(H_0\): \(\beta_4 = 0\) (no interaction effect). \(H_1\):
\(\beta_4 > 0\) (superlinear complementarity, Property 4).

Required N for \(\beta_4\) detection: effect size \(f^{2} = 0.15\)
(medium), \(\alpha = .05\), power \(= .80\); required N = 85 (G*Power);
target N = 100 provides adequate power.

\paragraph{6. Ethical Considerations}\label{ethical-considerations}

{\def\LTcaptype{none} % do not increment counter
\begin{longtable}[]{@{}
  >{\raggedright\arraybackslash}p{(\linewidth - 2\tabcolsep) * \real{0.3889}}
  >{\raggedright\arraybackslash}p{(\linewidth - 2\tabcolsep) * \real{0.6111}}@{}}
\toprule\noalign{}
\begin{minipage}[b]{\linewidth}\raggedright
Issue
\end{minipage} & \begin{minipage}[b]{\linewidth}\raggedright
Mitigation
\end{minipage} \\
\midrule\noalign{}
\endhead
\bottomrule\noalign{}
\endlastfoot
Informed consent & Written consent form explaining purpose, duration,
data handling \\
Anonymity & No identifying information collected; IP addresses not
logged \\
Data storage & Encrypted storage; 5-year retention per IRB protocol \\
Deception & None; full disclosure of research purpose \\
Cultural sensitivity & Korean translation reviewed by bilingual panel;
no culturally loaded items \\
Potential harm & Minimal; self-report survey with no invasive
measures \\
\end{longtable}
}

\paragraph{7. Budget Estimate}\label{budget-estimate}

{\def\LTcaptype{none} % do not increment counter
\begin{longtable}[]{@{}
  >{\raggedright\arraybackslash}p{(\linewidth - 2\tabcolsep) * \real{0.3158}}
  >{\raggedright\arraybackslash}p{(\linewidth - 2\tabcolsep) * \real{0.6842}}@{}}
\toprule\noalign{}
\begin{minipage}[b]{\linewidth}\raggedright
Item
\end{minipage} & \begin{minipage}[b]{\linewidth}\raggedright
Cost (est.)
\end{minipage} \\
\midrule\noalign{}
\endhead
\bottomrule\noalign{}
\endlastfoot
Participant incentives (100 \(\times\) ₩15,000) & ₩1,500,000
(\textasciitilde\$1,100) \\
Survey platform (Qualtrics/SurveyMonkey) & ₩200,000
(\textasciitilde\$150) \\
Back-translation service & ₩500,000 (\textasciitilde\$370) \\
Expert panel honorarium (5 \(\times\) ₩100,000) & ₩500,000
(\textasciitilde\$370) \\
Statistical software (R, open source) & ₩0 \\
Research assistant (100 hours \(\times\) ₩15,000) & ₩1,500,000
(\textasciitilde\$1,100) \\
\textbf{Total} & \textbf{₩4,200,000 (\textasciitilde\$3,100)} \\
\end{longtable}
}

\paragraph{8. Timeline}\label{timeline}

{\def\LTcaptype{none} % do not increment counter
\begin{longtable}[]{@{}
  >{\raggedright\arraybackslash}p{(\linewidth - 4\tabcolsep) * \real{0.2500}}
  >{\raggedright\arraybackslash}p{(\linewidth - 4\tabcolsep) * \real{0.3571}}
  >{\raggedright\arraybackslash}p{(\linewidth - 4\tabcolsep) * \real{0.3929}}@{}}
\toprule\noalign{}
\begin{minipage}[b]{\linewidth}\raggedright
Phase
\end{minipage} & \begin{minipage}[b]{\linewidth}\raggedright
Duration
\end{minipage} & \begin{minipage}[b]{\linewidth}\raggedright
Milestone
\end{minipage} \\
\midrule\noalign{}
\endhead
\bottomrule\noalign{}
\endlastfoot
IRB submission + approval & 4-6 weeks & Ethics clearance \\
Expert panel + cognitive interviewing & 2 weeks & Content validity
established \\
Back-translation + pilot preparation & 2 weeks & Korean version ready \\
Pilot data collection (n=100) & 3-4 weeks & Raw data \\
Analysis + item refinement & 2 weeks & EFA results,
\(\alpha\)/\(\omega\) values \\
\textbf{Total} & \textbf{\textasciitilde3-4 months} & CCS pilot
validation complete \\
\end{longtable}
}

\paragraph{9. Deliverables}\label{deliverables}

\begin{enumerate}
\def\labelenumi{\arabic{enumi}.}
\tightlist
\item
  \textbf{CCS-24 validated instrument} (English + Korean) with
  psychometric properties
\item
  \textbf{Pilot P1 test results}: Does the A\(\times\)C interaction
  predict individual productivity gains?
\item
  \textbf{Technical report}: Item statistics, factor structure,
  reliability, validity coefficients
\item
  \textbf{Revision-ready supplement}: All materials formatted for TFSC
  R\&R response
\end{enumerate}

\subsubsection{References (Appendix C)}\label{references-appendix-c}

\begingroup
\setlength{\parindent}{0pt}
\everypar{\hangindent=1.5em \hangafter=1}

Cacioppo, J. T., Petty, R. E., \& Kao, C. F. (1984). The efficient
assessment of need for cognition. \emph{Journal of Personality
Assessment}, \emph{48}(3), 306-307.

Hargittai, E., \& Hsieh, Y. P. (2012). Succinct survey measures of
web-use skills. \emph{Social Science Computer Review}, \emph{30}(1),
95-107.

Herman, J. L., Stevens, M. J., Bird, A., Mendenhall, M., \& Oddou, G.
(2010). The tolerance for ambiguity scale: Towards a more refined
measure for international management research. \emph{International
Journal of Intercultural Relations}, \emph{34}(1), 58-65.

Schraw, G., \& Dennison, R. S. (1994). Assessing metacognitive
awareness. \emph{Contemporary Educational Psychology}, \emph{19}(4),
460-475.

Wang, Y. Y., \& Wang, Y. S. (2022). Development and validation of an
artificial intelligence anxiety scale. \emph{Interactive Learning
Environments}, \emph{30}(4), 619-634.

\endgroup

\subsection{Data Availability
Statement}\label{data-availability-statement}

The data supporting the findings of this study are derived from publicly
available sources: OECD.Stat (https://stats.oecd.org), World Bank Open
Data (https://data.worldbank.org), and McKinsey Global Institute (2023).
The constructed dataset and analysis code are available from the
corresponding author upon reasonable request.

\subsection{Declaration of Competing
Interests}\label{declaration-of-competing-interests}

The authors declare that they have no known competing financial
interests or personal relationships that could have appeared to
influence the work reported in this paper.

\subsection{Funding}\label{funding}

This research received no specific grant from any funding agency in the
public, commercial, or not-for-profit sectors.

\subsection{Declaration of Generative AI and AI-assisted technologies in
the writing
process}\label{declaration-of-generative-ai-and-ai-assisted-technologies-in-the-writing-process}

During the preparation of this work the author(s) used a large language
model in order to assist with language editing and literature review
organization. After using this tool, the author(s) reviewed and edited
the content as needed and take(s) full responsibility for the content of
the publication.

{[}CRediT Author Statement is provided on the Title Page{]}

\end{document}